# University of Warsaw
## Faculty of Mathematics, Informatics and Mechanics

**Jan Mirosław Kopański**

Student no. 359785

# Optimisation of job scheduling for supercomputers with burst buffers

**Master's thesis**
**in COMPUTER SCIENCE**

Supervisor:
**dr hab. Krzysztof Rządca**
Institute of Informatics

Warsaw, December 2020

# Abstract

The ever-increasing gap between compute and I/O performance in HPC platforms, together with the development of novel NVMe storage devices (NVRAM), led to the emergence of the burst buffer concept - an intermediate persistent storage layer logically positioned between random-access main memory and a parallel file system. Since the appearance of this technology, numerous supercomputers have been equipped with burst buffers exploring various architectures. Despite the development of real-world architectures as well as research concepts, Resource and Job Management Systems, such as Slurm, provide only marginal support for scheduling jobs with burst buffer requirements. This research is primarily motivated by the alerting observation that burst buffers are omitted from reservations in the procedure of backfilling in existing job schedulers.

In this dissertation, we forge a detailed supercomputer simulator based on Batsim and SimGrid, which is capable of simulating I/O contention and I/O congestion effects. Due to the lack of publicly available workloads with burst buffer requests, we create a burst buffer request distribution model derived from Parallel Workload Archive logs. We investigate the impact of burst buffer reservations on the overall efficiency of online job scheduling for canonical algorithms: First-Come-First-Served (FCFS) and Shortest-Job-First (SJF) EASY-backfilling.

Our results indicate that the lack of burst buffer reservations in backfilling may significantly deteriorate the performance of scheduling. Backfilling without future burst buffer reservations can result in almost identical mean waiting time as FCFS without any backfilling and have enormously high waiting time dispersion. Furthermore, this lack of reservations may cause the starvation of medium-size and wide jobs. Finally, we propose a burst-buffer–aware plan-based scheduling algorithm with simulated annealing optimisation, which improves the mean waiting time by over 20% and mean bounded slowdown by 27% compared to the SJF EASY-backfilling.

## Keywords

High performance computing (HPC); burst buffer; online job scheduling; EASY backfilling; multi-resource scheduling; plan-based scheduling; simulated annealing; nonvolatile memory; simulation; supercomputer

## Thesis domain (Socrates-Erasmus subject area codes)

11.3 Informatics, Computer Science

## Subject classification

- Software and its engineering—Software organization and properties—Contextual software domains—Operating systems—Process management—Scheduling
- Theory of computation—Design and analysis of algorithms—Online algorithms—Online learning algorithms—Scheduling algorithms
- Computing methodologies—Modeling and simulation—Simulation types and techniques—Massively parallel and high-performance simulations
- Information systems—Information storage systems—Storage management—Hierarchical storage management

## Tytuł pracy w języku polskim

Optymalizacja szeregowania zadań na superkomputerach z uwzględnieniem buforów impulsowych

# Contents

# Introduction

**Background**   With the deployment of Fugaku [Don20], supercomputing has already exceeded the threshold of exascale computing. Nevertheless, High-performance computing (HPC) is still facing the challenge of a constantly growing performance gap between compute resources and I/O capabilities. HPC applications typically alternate between compute-intensive and I/O-intensive execution phases, where the latter is characterised by emitting bursty I/O requests. Those I/O spikes produced by multiple parallel jobs can saturate network bandwidth and lead to I/O congestion, which effectively stretches the I/O phases of applications resulting in higher turnaround time. However, the development of novel storage technologies, such as NVRAM, pave the way to solve the issue of bursty I/O by the introduction of burst buffers [Liu+12].

Burst buffer is an intermediate fast persistent storage layer, which is logically positioned between random-access main memory in compute nodes and a parallel file system (PFS). Burst buffers enable to immediately absorb bursty I/O requests and gradually flush them to the PFS. This application provides the possibility to implement checkpointing in HPC platforms efficiently. There are, however, several other applications of burst buffers, among others, data staging, write-through cache or in-situ analysis.

Since the introduction of the burst buffer concept, many supercomputers have been equipped with high-performance storage devices, mostly in the form of Non-Volatile Memory Express (NVMe) Solid State Drives (SSD). They have also been configured in various architectures, where the most prominent are node-local burst buffers as in Summit [Vaz+18] and remote shared burst buffers as in Cori [Bhi+16]. Simultaneously, numerous research publications on I/O-aware scheduling has been released, which span from methods of avoiding I/O contentions [Zho+15; Her+16], through deciding whether or not to use burst buffers [LFW19] and ending on improving the efficiency of high-level job scheduling [Fan+19].

Despite the intensive development of real supercomputing platforms and appearance of novel research algorithms, Resources and Jobs Management Systems (RJMS) provide only basic software support for burst buffer utilisation. RJMSs, such as Slurm, usually implement some sort of backfilling algorithm for online job scheduling (online batch scheduling). We observed that even though HPC platforms have been equipped with the new kind of resources - burst buffers, the process of allocating compute resources for future reservations in backfilling has not been extended with storage resources. This issue may potentially be a cause of starvation and deteriorate the performance of scheduling. Therefore, we dedicate this dissertation to study burst-buffer–aware scheduling algorithms and evaluate their efficiency in a simulated environment.

**Simulation**   In order to test and compare scheduling algorithm, we have created a detailed supercomputer simulator, based on Batsim [Dut+16] and SimGrid [Cas+14], which is capable of simulating I/O contention and I/O congestion effects. In fact, we developed two simulation models: one focused on simulating allocation of resources, and the other that extends it with I/O side effects. We modelled the SimGrid platform network to resemble a relatively small HPC cluster in a Dragonfly topology [Kim+08]. The chosen size of the cluster enabled us to perform an in-depth analysis of scheduling traces. This work is confined to research the job scheduling in the shared burst buffer architecture, which we modelled to resemble the architecture utilised in Fugaku.

One of the challenges was the availability of workload logs. The Parallel Workload Archive [FTK14] is known from a grand collection of computer workload logs gathered over that last two decades, which is an indispensable data source for scheduling research. Nevertheless, it does not contain information about burst buffer requests associated with parallel jobs, neither does any well-know publicly available dataset. Hence, we created a model of burst buffer request size distribution using the information about the requested main memory from the workload logs.

**Online job scheduling**   Our first major contribution is the study of the impact of future burst buffer reservations on the overall efficiency of job scheduling. For this part, we experiment with four variants of First-Come-First-Served (FCFS) scheduling characterised with different approaches to resource reservations in aggressive backfilling (EASY backfilling). Namely, we compare FCFS (1) without backfilling, (2) with backfilling and future reservations for only compute resources (canonical backfilling), (3) with backfilling and reservations for both compute and storage resources, and lastly a (4) greedy filling, which may be perceived as a pure backfilling without any reservations. As we tackle the online scheduling, we study all the results based on four well-known user-centric metrics: waiting time, turnaround time (response time), slowdown (stretch) and bounded slowdown.

The next major contribution is the proposal of new burst-buffer–aware scheduling algorithms, out of which the best performing is plan-based scheduling with simulated annealing optimisation. Based on real workload logs, we prove that it achieves a significant reduction of mean waiting time and mean bounded slowdown while maintaining a comparable statistical dispersion of waiting times. We collocate its results with FCFS backfilling and Shortest-Jobs-First (SJF) backfilling.

Additionally, we study the influence of backfilling reservation depth - the number of jobs at the front of a waiting queue that is prioritised in the backfilling and assigned with future reservations of resources.

**Thesis organisation**   The reminder of this dissertation is organised as follows. In Chapter 1, we present an in-depth description of related work. We discuss, use cases of burst buffers, architectures of burst buffers in supercomputers, hardware implementation and provide a detailed overview of job scheduling terminology. In Chapter 2, we present the simulation model, its assumptions and the created workload. Chapter 3, defines the canonical scheduling algorithms and also introduce three novel burst-buffer–aware scheduling algorithms. All described algorithms are followed with their evaluation in Chapter 4, where we present our analysis of experiments. In Chapter 5, we form all conclusions arising from this dissertation. In Chapter 6, we propose several future research directions. Lastly, in Appendix A, we gathered all online resources which were immensely helpful for us. We decided to share them as we find that they may be useful for related research.

**All code sources associated with this dissertation are available at:**
`https://github.com/jankopanski/Burst-Buffer-Scheduling`

# Chapter 1

# Related work

## 1.1. Definition of burst buffers

In the domain of high-performance computing (HPC), a burst buffer is a fast, non-volatile, intermediate storage layer positioned between computing resources and backend storage systems. It emerges as a solution to bridge the ever-increasing performance gap between processing speed of compute nodes and input/output (I/O) bandwidth of the storage systems. It typically offers from one to two orders of magnitude higher I/O bandwidth than the backend storage systems, albeit at a much lower capacity.

The term burst buffer is used both in the context of the intermediate storage layer and non-volatile storage resource assigned to a job at the scheduling. Two modes of the burst buffer allocation may be distinguished: ephemeral and persistent. The ephemeral allocation, also called a per-job instance, is associated with a given job and exists only for the lifetime of this job. The persistent burst buffer instance is manually created by a user and may be used by multiple jobs.

The placement of burst buffers in the memory hierarchy of a modern HPC system is presented in Figure 1.1. The directions of arrows represent increasing values of given properties. Explanation of Abbreviations in the figure: HBM - High Bandwidth Memory, DRAM - Dynamic Random Access Memory, SSD - Solid State Drive, NVRAM - Non-Volatile Random Access Memory, HDD - Hard Disk Drive.

## 1.2. Far storage in a Parallel File System

A parallel file system (PFS), also known as a clustered file system, is a type of file system designed to store data across multiple networked servers and to facilitate high-performance access through simultaneous, coordinated input/output operations between parallel applications running on compute nodes and storage nodes. PFS is usually deployed on the backend storage (far storage, long-term storage) of supercomputers and HPC clusters.

PFS breaks up a dataset and distributes/stripes blocks to multiple storage drives, which can be located in local or remote servers. Users do not need to know the physical location of the data blocks to retrieve a file, as the system uses a global namespace to facilitate data access. Data is read and written to the storage devices using multiple I/O paths concurrently, which provides a significant performance benefit. Storage capacity and bandwidth can be scaled to accommodate enormous quantities of data. Features of PFS may include high availability, mirroring, replication and snapshots.

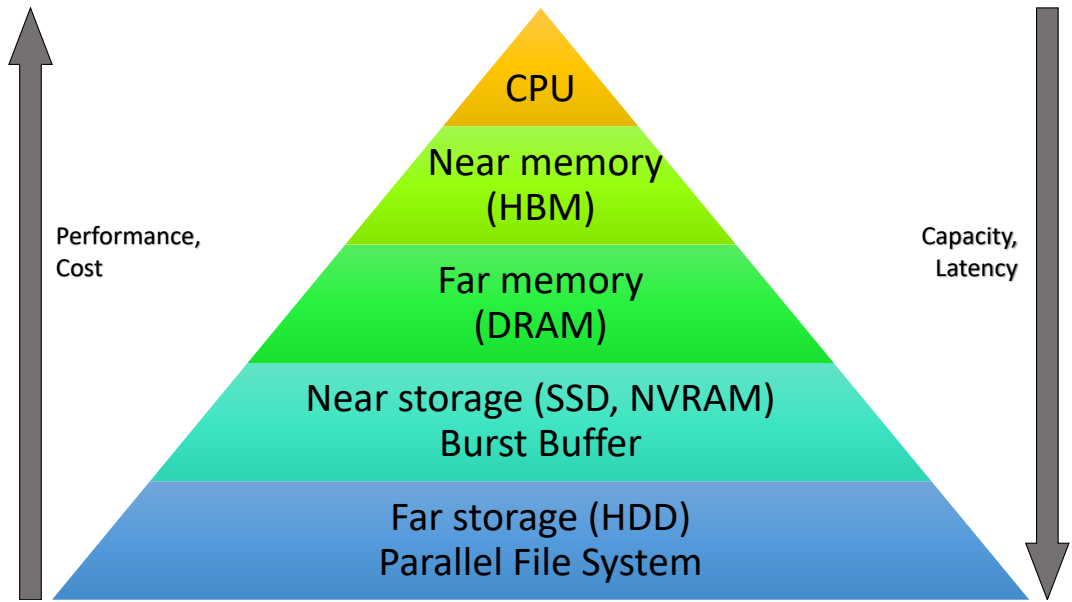

Figure 1.1: Memory hierarchy of a modern supercomputer with burst buffers

Throughout this work, we will use the term PFS interchangeably with the backend storage system.

## 1.3. Burst buffer use cases

The primary motivation for the existence of burst buffers was to absorb the periodic I/O bursts of HPC applications, especially increasing the performance of periodic checkpoints. However, burst buffers bring several additional use cases that may increase the performance of individual applications or a whole HPC system. Based on the gathered literature [Har+16; Bhi+16; Liu+12; Khe+19; CSB17; RRP15; Loc17] we selected common use cases of burst buffers.

**Periodic I/O bursts**   Scientific applications' I/O patterns are often characterised by enormous spikes in I/O traffic, which is due to their life cycles that typically alternate between computation phases and I/O phases. In other words, after each round of computation, all computing processes concurrently write their intermediate data to backend storage systems. This pattern is then repeated with another round of computation and data movement operations.

Traffic to PFS is often mostly limited by slow hard drives as well as bandwidth and contention of storage network. As a consequence, the significant I/O bursts are causing high I/O latency of application and under-utilisation of computing resources.

With the deployment of burst buffers, applications can quickly write their data to a burst buffer after a round of computations instead of writing to the slow hard disk–based storage systems. Therefore, they receive a quick confirmation of data being saved in persistent memory and may proceed to the next round of computations without waiting for the data to be moved to the slow backend storage systems. The data stored in burst buffers are then asynchronously flushed to the PFS, which overlap with the next round of computation. In this way, the extended I/O time spent in moving data to the storage systems is hidden by the computation time.

**Checkpoint/restart** One complexity in HPC is that an interruption of any process of a parallel application destroys the distributed state of the entire application. As the number of nodes in supercomputers constantly increases, the size of the distributed state grows. To ensure forward progress parallel applications use checkpoints to allow a job to restart from an intermediate state in case of a failure. That is a conceptually simple technique of persisting their distributed state at regular time intervals and avoiding repetition of all computations.

**Data staging** Stage-in and stage-out is a technique of moving input and output data of an application from PFS to burst buffers. The key concept is to make the input data available in the fast storage, close to compute resources, rather than slow backend storage. In the naive approach, data staging could take place immediately when a job starts. However, as compute resources are not necessary for data moving, the stage-in phase of one application could be overlapped with computations of another application that was previously allocated with a set of computing resources shared with the first application. Similarly, the stage-out from a burst buffer could allow flushing of output data to PFS after freeing of computing resources. In result, this technique could improve the overall utilisation of a supercomputing system. According to [Dal+17] stage-in and stage-out with burst buffers are used in the following scientific applications: SWarp and CAMP (Community Access MODIS Pipeline).

**Write-through cache** A write-through cache is a caching technique in which data is simultaneously copied to cache and to backend storage systems. It allows to increase the read performance as data is already available in fast storage, which may be gradually synchronised with far persistence storage. The urge to make the caching as easy as possible from the application perspective implies attempts to implement write-through cache in the form of the transparent caching mode. For instance, both Data Direct Network's Infinite Memory Engine and Cray DataWarp systems implement the transparent caching mode [Loc17].

**Data sharing** Burst buffers may be applied to enable data sharing within and across applications. An application could utilise burst buffers to create a shared file and use it to communicate among compute nodes. Different I/O models may be applied for this purpose, such as one-file–per-job (N-1), file-per-process (N-N) or more complex patterns (N-M). These models were discussed in [Wan+16a; Ben+09].

Burst buffers may also be used to enable data sharing between consecutive jobs. A scientific workflow may be divided into several stages, each performed as a separate job. In this scenario, intermediate data passed between jobs may be persisted in a burst buffer instead of sending it back and forth to PFS.

**In-situ analysis** An in-situ analysis is a type of data analysis and visualisation that happen concurrently to the main computational task. Burst buffers are utilised here to store data and perform analysis without data movement to PFS. The kind of data subject to analysis may be an application checkpoint.

**Out-of-core memory** For applications for which available main memory (DRAM) is not enough, non-volatile devices may be used to extend the application memory. NVDIMM is the type of NVM hardware that could provide adequate performance for this use case. Installed on a compute node it can be configured either as extended main memory or a burst buffer.

## 1.4. Overview of non-volatile memory hardware

A burst buffer is a concept of a fast, non-volatile, intermediate storage layer logically positioned between dynamic random-access memory (DRAM) in compute nodes and a parallel file system. DRAM, which colloquially referred to as RAM, is a type of a fast, volatile memory that serves the role of primary storage. It is widely used in most of the modern devices such as personal computers, mobile devices and server machines. Volatility in the context of memory means that it requires power to maintain stored information. Once the power of volatile memory is down, stored data is quickly lost. In contrast, the non-volatile memory (NVM) is persistent even if not continuously supplied with electric power.

The term NVM refers to a wide variety of memory types. Examples of these include flash memory, read-only memory (ROM), floppy disks, optical discs, hard disk drives (HDD), solid-state drives (SSD), Non-volatile random-access memory (NVRAM) and many others. All of these memory types are characterised by the ability to retain data without applied power. The types of NVM used as burst buffers in supercomputer architectures are SSD and NVRAM. The reason for this is that they offer a much higher performance of data access compared to other persistent memory types.

**Solid State Drive**   SSDs are usually constructed with the use of non-volatile NAND flash memory. Another type of hardware architecture is a DRAM-based SSD, which is a DRAM supplied with an additional source of power such as an internal battery. When external power is lost, the battery provides power, and all data is copied to a persistence backup storage. There also exists a new NVM technology called 3D XPoint announced by Intel and Micron. It could be used both as an ordinary SSD or as NVDIMM (non-volatile dual in-line memory module). SSDs with the 3D Xpoint technology is currently available on the market as Intel Optane SSDs and Micron QuantX X100.

**Non-volatile RAM**   As there is no standardised definition of NVRAM, this term is often ambiguous and interchangeably used with NVM, NVDIMM, DRAM-based SSD and NAND-based SSD. Particularly, ordinary SSDs based on NAND-Flash memory are frequently referred to as NVRAM, for instance in [Dal+17]. NEXTGenIO, an initiative supporting the development of Intel Optane DC Persistent Memory in supercomputing, defines NVRAM as memory "implemented by NVDIMMs, being one kind of NVM, that supports both non-volatile RAM access and persistent storage" [Jac+18]. NVDIMM is a non-volatile random-access memory in the form of a DIMM package, which is a mainstream pluggable form factor for DRAM. Development of NVRAM in the form of NVDIMM raises the potential of using it as a system's main memory.

Historically, there have been numerous different approaches to implementing NVRAM. The modern solutions in implementing NVRAM in hardware include:

- Ferroelectric RAM,

- Magnetoresistive RAM,

- Phase-change RAM,

- Millipede memory,

- FeFET memory.

Out of these solutions, Magnetoresistive RAM (MRAM) and Phase-change RAM (PRAM) are currently in commercial production. An example of PRAM-based solution is the 3D XPoint technology. Intel develops a family of devices based on this technology under the trademark of Intel Optane. The NVRAM devices are offered as Intel Optane Persistent Memory. Everspin is the company specialised in MRAM-based solutions. It produces a variety of MRAM devices including the advanced STT-MRAM technology.

The following list gathers supercomputers equipped with burst buffers together with the information about their NVM devices. The citation in each entry refer to the source of the information.

- Fugaku at the RIKEN Center for Computational Science - about 1.6 TB SSD [RIK20; Don20]

- Summit at the Oak Ridge National Laboratory - 1.6 TB Samsung NVMe SSD [Vaz+18]

- Sierra at the Lawrence Livermore National Laboratory - 1.6 TB Samsung NVMe SSD [Vaz+18]

- Tsubame at the Tokyo Institute of Technology - 2 TB Intel DC P3500 (NVMe, PCI-E 3.0 x4, R2700/W1800) [Tok20]

- Theta at the Argonne National Laboratory - 128 GB Samsung SM961 MZVPV128HEGM SSD [Arg20a]

- Tianhe-2 at the National Supercomputer Center in Guangzhou - 1 TB PCIe SSD [LCC20]

- Trinity at the Los Alamos National Laboratory - 4 TB Intel DC P3608 NAND flash SSD [Los20; Hem+16]

- Cori at the Lawrence Berkeley National Laboratory - 3.2 TB Intel DC P3608 NAND flash SSD (described as NVRAM) [NER20; Bar+16; Bhi+16; Bhi+17]

- Catalyst at the Lawrence Livermore National Laboratory - 800 GB Intel SSD 910 (described as NVRAM) [Law20; Tha+16]

## 1.5. Burst buffers in supercomputer architectures

At the top most level of burst buffer placement classification in supercomputer architectures, there could be distinguished two architectures: node-local and remote shared burst buffers. The node-local burst buffer is a well-established term, which appears in numerous papers [Khe+19; CSB17; Har+16]. The remote shared burst buffers could be further divided depending on their placement in a supercomputer. Khetawat *et al.* [Khe+19] distinguished shared burst buffers between grouped and global. Another terminology is introduced in the Oak Ridge National Laboratory technical report by Harms *et al.* [Har+16]. They classified burst buffer architectures into compute node–local storage, co-located with I/O nodes and the last kind using a separate set of nodes. Glenn K. Lockwood proposed the most granular division on his blog [Loc17], where he distinguished the architectures between compute node attached, I/O node attached, fabric attached and storage fabric attached.

All of the architectures described below, except the hybrid, are presented in Figure 1.2.

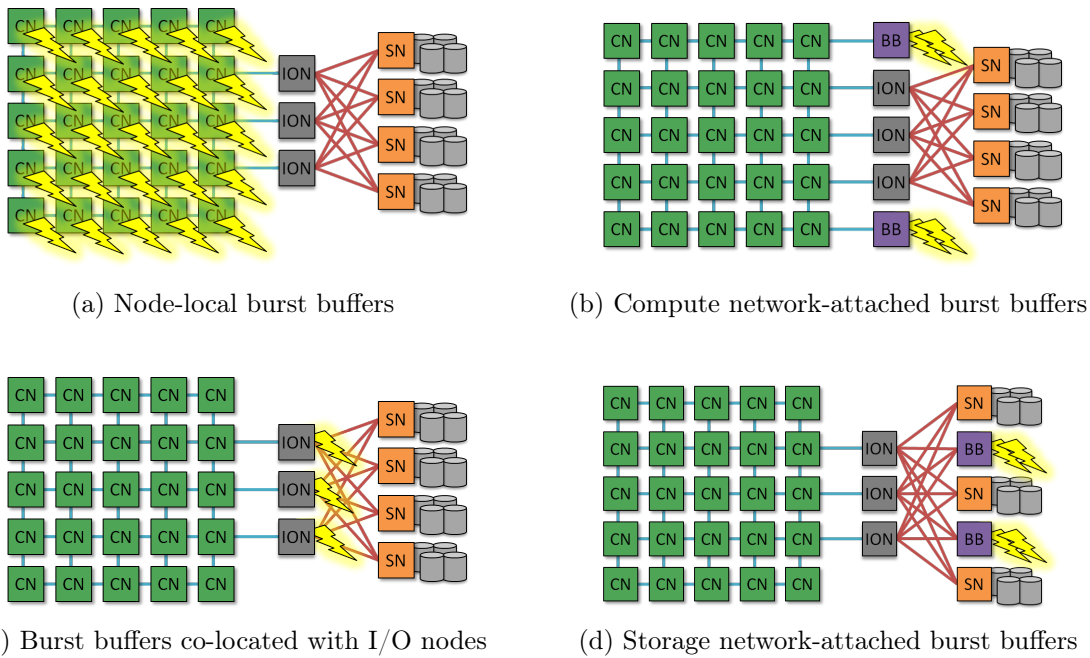

(a) Node-local burst buffers

(b) Compute network-attached burst buffers

(c) Burst buffers co-located with I/O nodes

(d) Storage network-attached burst buffers

Figure 1.2: Illustration of various burst buffer architectures (CN - compute node, ION - I/O node, SN - long-term storage node, BB - dedicated burst buffer node, lightning - burst buffer placement).
Source: `https://glennklockwood.blogspot.com/2017/03/reviewing-state-of-art-of-burst-buffers.html`

**Node-local burst buffers** In the compute node-local burst buffer architecture, burst buffer storage is located on each compute node. The main advantage of this architecture is a performance scaling. The aggregate burst buffer bandwidth scales linearly with the number of compute nodes, which has been confirmed, for example in [Wan+16b; Moo+10]. Another benefit is exclusive and fair access of applications to the storage devices, which provides more consistent and predictable access to storage resulting in lower variation in I/O performance. Further, during application I/O, there is no network traffic, and the large reads or writes do not impact any shared resources such as network or PFS. This architecture is also interface-free, meaning that an operating system can create a local POSIX file system on a burst buffer and applications can use a standard software storage stack.

However, there are several disadvantages of node-local architecture. First of all, there is limited support for shared-file I/O as individual burst buffers are not shared between compute nodes. Consequently, standard file access techniques, such as many readers or writers sharing a single file, called the N-1 model, are difficult to implement. They require an additional I/O middleware. Further, stage-in and stage-out require network traffic to the compute nodes. Another drawback of using this storage model is that the burst buffers are tightly coupled with the compute nodes, which creates a single failure domain. Lastly, any maintenance required by the storage device may demand taking the entire compute node offline.

This architecture was deployed in several systems, including the Summit and Sierra, which are the second and third top supercomputers (as of June 2020) [Str+20]. The more comprehensive list is presented below:

- Summit at the Oak Ridge National Laboratory [Vaz+18]

- Sierra at the Lawrence Livermore National Laboratory [Vaz+18]

- Tsubame at the Tokyo Institute of Technology [Tok20]

- Theta at the Argonne National Laboratory [Arg20a]

**Burst buffers co-located with I/O nodes**  I/O nodes are networking element that route data between compute fabric and backend storage fabric. The I/O nodes could be either commodity routers or specialised forwarding devices.

The positioning of SSDs or NVRAM devices within I/O nodes results in the shared burst buffer architecture, which may readily support N-1 and N-N communication models. Possibly, it is the easiest to realise data staging (stage-in and stage-out) is in this architecture as it might not require any traffic to the compute network. This architecture also provides data resiliency and longer residency times than node-local burst buffers.

There are yet some drawbacks of burst buffers co-located with I/O nodes. Firstly, as storage resources are shared among compute nodes, this might lead to interference among application resulting in unpredictable performance. Secondly, the compute network connections are shared between burst buffer and PFS traffic, which would negatively impact PFS performance at an occurrence of bursty I/O. Furthermore, the storage capacity scalability of this architecture is limited by a fixed number of I/O nodes. The most significant concern of incorporating burst buffers into I/O nodes are maintenance and failure dependency between them.

The architecture of burst buffers co-located with I/O nodes was deployed in the Tianhe-2 system at the National Supercomputer Center in Guangzhou [LCC20].

**Compute network-attached burst buffers**  This design places burst buffers in dedicated nodes within a compute network. Location of these nodes may vary, resulting in different proximity between a compute node and PFS. For instance, a single node of every chassis or cabinet could be dedicated to be a burst buffer. Another approach is to create special storage nodes directly connected with the I/O nodes. This architecture inherits most of the advantages of burst buffers co-located with I/O nodes, including easy data sharing, support for N-1, N-N models, longer residency times and data resiliency. The principal additional advantage of separate dedicated nodes is the transparent maintenance of a supercomputing cluster. SSDs are characterised by a limited number of write operations, which creates a necessity of replacing them over time. Having a separate burst buffer nodes enables to maintain them without downtime of compute or I/O nodes. Conversely, it also provides a possibility of continued use when an I/O node or PFS goes offline.

Similarly to co-location with I/O nodes, compute network-attached burst buffers are burdened with application interference in shared network connection. Furthermore, this architecture brings the highest price cost due to the cost of additional servers and network infrastructure.

Two dedicated hardware and software production systems were created following this burst buffer architecture: Cray DataWarp and Infinite Memory Engine developed by Data Direct Network. The Cray DataWarp was deployed in the Cori system at the Lawrence Berkeley National Laboratory [Bar+16; Bhi+16; Bhi+17], which architecture schema is presented in Figure 1.3 and Trinity at the Los Alamos National Laboratory [Los20; Hem+16].

Currently, the fastest supercomputer in the world according to TOP500 list [Str+20]—Fugaku at the RIKEN Center for Computational Science—utilises a shared burst buffer architecture with SSDs in one of every 16 compute nodes [RIK20].

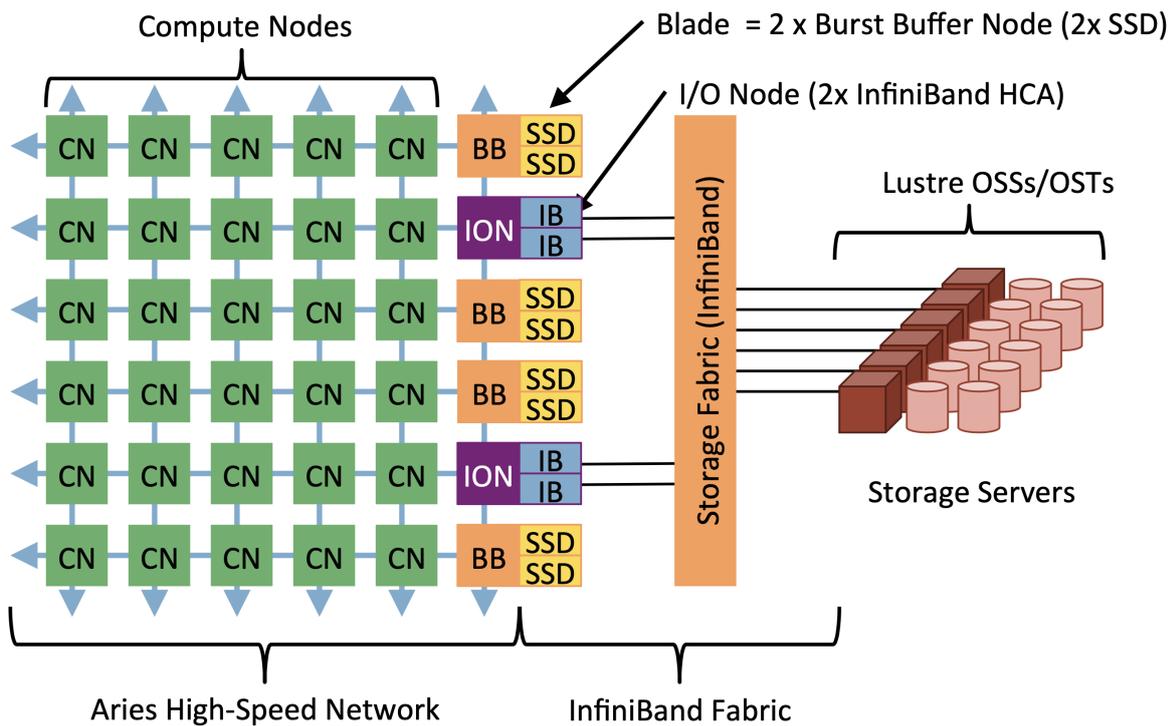

Figure 1.3: The placement of the burst buffer nodes in the Cori System
Source: Accelerating Science with the NERSC Burst Buffer Early User Program [Bhi+16]

**Backend storage network-attached burst buffers** The last of fundamental architectures involves attaching NVM devices to a storage network, either as fast storage on the nodes hosting long-term storage or as a separate set of nodes. This model creates an off-platform burst buffer, which may not provide the highest absolute performance, but may provide significant ease of use benefits via transparent caching. HPC systems with custom compute fabrics are the primary reason for the use of this design as they are usually not amenable to adding third-party burst buffers.

**Hybrid architecture** It is also possible to mix the described architectures creating a multi-level heterogeneous storage model. A particularly interesting variation would be a mixture of node-local burst buffers with NVRAM and dedicated network-attached nodes directly connected to I/O nodes, equipped with SSDs. Such an architecture could handle applications with distinct I/O patterns and offer both high performance, exclusive access to permanent storage as well as support for shared I/O models. Software solutions that transparently move data across multiple storage layers have already been explored in [Wan+18; Tan+18; KDS18].

## 1.6. Resources and Jobs Management Systems

Resource and Jobs Management Systems (RJMS) are complex middlewares, which are the core of platform management and serve several roles. First, they receive job submissions from users and allocate exclusive or non-exclusive access to platform resources (compute nodes, accelerators, memory, burst buffers) for some duration of time, which could be shortly described as management of resource sharing. Second, it provides a framework for starting, executing, and monitoring (usually parallel) jobs on the set of allocated resources. Finally, it arbitrates contention for resources by managing a queue of pending jobs.

RJMSs are therefore the place to take management decisions and to implement management policies and algorithms, hence are often abbreviated as merely—job scheduler or batch scheduler. Despite being commonly called as batch schedulers, they usually support submission of both batch and interactive jobs. RJMSs, besides of functional requirements, are also required to satisfy multiple non-functional requirements such as fault-tolerance, high scalability, fair resource sharing.

The well-known job schedulers in HPC include Slurm [YJG03; Slu20], Moab/TORQUE [Ada20], PBS [Alt20a; Alt20b], and Flux [Flu20].

## 1.7. Scheduling algorithms

As described in Section 1.6, one of the key aspects of RJMSs is queuing and scheduling of jobs given available resource in a platform. For this matter, multiple algorithms were created out of which the most fundamental are FCFS, aggressive and conservative backfilling. We summarise these methods based on the information and observations gathered in [Sri+02a; Sri+02b; Fei05; Car+19]. It is also valuable to distinguish between online and offline scheduling. In online scheduling, new jobs may arrive during the execution of a workload, whereas in offline, all jobs are available for scheduling at the same point of time.

In the existing schedulers, processors are the primary resources that are subject to scheduling. All other resources, such as main memory, are secondary. Batch schedulers were designed to optimise the system utilisation of processors, while the availability of other resources is just a constraint for scheduling, such as the main memory in compute nodes.

In the field of scheduling in HPC systems, jobs are often considered to be parallel, rigid and non-preemptive. The parallelism of jobs means that they run on multiple processors simultaneously, where processors are either compute nodes, CPUs or CPU cores depending on the context. The non-preemptive character states that running jobs cannot be interrupted and interleaved with each other. Rigid jobs are jobs that specify the number of processors they need and run for a particular time using this number of processors. Scheduling of parallel jobs is usually viewed in terms of a 2D chart with time along one axis and the number of processors along the other axis. Each job can be then depicted as a rectangle whose length is the user estimated runtime and width is the required number of processors. This 2D chart is a special kind of a Gantt chart. In particular, it is commonly used to visualise executed jobs in batch job scheduling. Figure 4.1 is an example of the Gantt chart.

Based on the shape of those rectangles, jobs could be classified into four categories. Two basic classes are derived from the runtime of jobs: short and long. Two other classes are orthogonal to runtime and are based on the number of processors requested: narrow (serial and small jobs) and wide (jobs with a large number of processors). With these classes combined, jobs could be categorised into short-narrow, short-wide, long-narrow and long-wide. The distinction between short and long as well as narrow and wide jobs is subject to a concrete problem statement, computing cluster and workload. An example of a specific quantitative definition may be found in Section 2.5.

### 1.7.1. Queuing policies

The job queue, into which jobs arrive, could be a priority queue and implement various priority policies.

The most straightforward scheduling policy is First-Come-First-Served (FCFS), where jobs are executed in the order of their arrivals whenever enough of the resources is available. Although this is a fair policy of resource sharing, this approach suffers from low system utilisation.

Another popular policy is Shortest Jobs First (SJF), a.k.a. Shortest estimated Processing time First (SPF), according to which the jobs are sorted in ascending order based on the estimated running time. There are also the Smallest Requested Resources First (SQF) and Smallest estimated Area First (SAF).

### 1.7.2. Backfilling

Backfilling works by identifying holes between jobs (rectangles) in the 2D chart and moving forward smaller jobs that fit these holes. As described at the beginning of Section 1.7, this 2D chart is a special kind of a Gantt chart. In order to avoid starvation of large jobs, jobs at the front of a waiting queue are assigned reservations of resources with associated start times. Other jobs are not permitted to use the reserved resources.

**Conservative backfilling** In the conservative backfilling, each job is given a reservation when it arrives to the queue, and jobs are allowed to move ahead in the queue as long as they do not cause any queued job to get delayed beyond their reserved start time. In other words, a newly arriving job is given the reservation at the earliest time that will not violate any previously existing guarantees.

The longer the job is, the more difficult it is for it to get a reservation ahead of the previously arrived jobs. Therefore long jobs find it challenging to backfill under conservative backfilling. However, it is beneficial for wide jobs as it guarantees them the start time when they enter the system. In conclusion, conservative backfilling promotes short-wide jobs.

This movement to the head of the queue also becomes possible when at least one of the running jobs finishes prematurely regarding their estimated time (wall time).

**Aggressive backfilling** In the aggressive backfilling, which is also known as EASY backfilling, only one job at the head of the queue is given a reservation. Other jobs are allowed to move ahead of the reserved job, as long as they do not delay that job. The concrete implementation of the canonical EASY backfilling is presented in Section 3.1 by Algorithm 3.

The presence of only one blocking reservation in the schedule helps long jobs to backfill more easily than conservative backfilling. The contrary effect occurs for wide jobs as they do not receive a reservation until they reach the head of the queue. Hence other jobs that entered the queue later or have lower priority may backfill ahead of them if they find enough free processors. EASY backfilling is therefore beneficial for long-narrow jobs.

In practice, for instance, in Slurm, the intermediate approach is applied, where the number of reserved jobs at the head of the queue is a parameter. This parameter is fine-tuned based on characteristics of workloads executed in a given HPC system and a policy imposed by a system administrator.

### 1.7.3. Criteria for scheduling

Job scheduling algorithms should meet specific criteria to be usable in real systems. The most common of them is fairness and starvation-freedom. Freedom from starvation states that all jobs which entered the pending queue should be picked for execution at some point of time. For instance, it is known that the SJF with EASY backfilling might violate this rule [Sri+02b].

The precise definition of fairness is harder to formulate and more ambiguous in the literature. For example, a weak definition of fairness was formulated in [Sri+02b] as follows: "No job is started any later than the earliest time it could have been started under the strictly fair FCFS-NoBackfill schedule.", where FCFS-NoBackfill is an FCFS policy without backfilling.

### 1.7.4. Novel I/O-aware scheduling algorithms

In this section, we describe selected publications on the topic of I/O-aware scheduling, which in various degree contributed to our research.

In [Zho+15], Zhou *et al.* presented an I/O-aware scheduling framework for concurrent I/O requests from parallel applications. It is attempting to solve the issue of I/O contention by queuing I/O requests and specifying the order of their execution. Their approach is based on an integration with high-level RJMS as it holds detailed information on all currently running jobs in a system. It enables the system to keep the network bandwidth saturated and avoid job I/O phase slowdown. They presented an unusual approach which fills the gap between high-level job scheduling and low-level I/O request handling.

Arising from addressing the same issue of I/O contention, Herbein *et al.* created in [Her+16] a completely different solution. Similarly, they extended RJMS by providing it with information about a state of network switches in a system. They exploit this fact by imposing limitations on I/O bandwidth of selected jobs. This way, it is possible to maintain network saturation without I/O congestion.

A high-level methodology was shown by Fan *et al.* in [Fan+19], where a multi-objective optimisation program was defined, which maximises both compute and storage utilisation. At the core of this optimisation technique, a genetic algorithm was used. To limit the computations performed by the job scheduling process, they used window-based scheduling to perform optimisation only for jobs inside the window. Simultaneously, to maintain fairness property, the maximum age of a job inside the window was imposed. We provide an additional description of this publication in Section 2.5.

Yet another noteworthy approach was presented in [LFW19] by Lackner, Fard and Wolf. They faced the issue of I/O throughput between the slow PFS, fast burst buffer layer and main memory in compute nodes. Their research was motivated with the observation that a job waiting in a queue for the availability of burst buffers may finish with higher turnaround time than it would if it was executed without the fast persistent storage. They proposed a solution which automatically chooses whether or not a given job should be started with a requested burst buffer. To evaluate their approach, they used the Batsim simulator (Section 1.10) with the Pybatsim-based scheduler. They modelled the shared burst buffer as a single dedicated node, which the most closely resembles the compute network-attached burst buffer architecture (Section 1.5). The connections between compute nodes, the burst buffer node and PFS node were modelled as separate links, hence there is no I/O contention between application communication and I/O traffic. They performed experiments based on an extended theoretical HPC workload with burst buffer requests derived from main memory requests.

## 1.8. Scheduling metrics

Metrics in job scheduling are used, among others, to compare different scheduling algorithms. As mentioned in Section 1.7, in this dissertation we consider parallel, non-preemptive, rigid jobs. That, however, is not the only model of a computational job. A fine characteristic of computer workloads and systems is presented in [FR98].

Job scheduling metrics could be divided into user metrics (a.k.a. user-level; user-oriented; user-centric; job-level; job-oriented; job-centric) and system metrics (a.k.a. system-level; system-oriented; system-centric). [Cha+99; Sri+02b; EE08; Car+19; Fan+19]. Furthermore, Carastan-Santos *et al.* in [Car+19] claim that there is a distinction between user-centric and job-centric metrics, yet in practise formulating a true user-centric metric would require to capture the overall satisfaction among users, which cannot be simply reproduced in a simulation.

### 1.8.1. Notation

In order to describe the metrics quantitatively, we extend the three-field scheduling notation—$\alpha|\beta|\gamma$—introduced by Graham *et al.* [Gra+79]. Using this notation, the parallel online job scheduling problem could be for instance defined as $P|r_j, size_j|\overline{F}$. For the scheduling with burst buffers problem the $\beta$ is additionally extended with $bb_j$. A descriptive extension of this notation for parallel processing is also presented in [Dro09]. We denote a workload as $J$ and assign an index $j$ to every job contained in $J$. A job is characterised by:

1. Execution time (processing time, running time) $p_j$ - real runtime of a job $j$. Might be unknown to a user.

2. Walltime (duedate, deadline) $d_j$ - user estimated runtime of a job $j$.

3. Requested number of processors $size_j$

4. Requested burst buffer size per processor $bb_j$ - our notation extension

Next, we distinguish specific moments during a lifetime of a given job $j$. The metric names according to the $\alpha|\beta|\gamma$ notation are shown in the brackets.

1. Submit time (release time, ready time) $r_j$ - time when a job $j$ was submitted to a system and added to the queue of pending jobs.

2. Launch time (start time) $s_j$ - time when a job $j$ started its execution and was removed from the queue of pending jobs.

3. Finish time (completion time, end time) $c_j$ - time when a job $j$ finished its execution.

### 1.8.2. User-centric metrics

These type of metrics should represent the quality of scheduling from the perspective of a user. The user could be interested in quantities such as how long was a submitted job waiting to start the execution or what was the ratio of a total time a job spent in a system to the waiting time.

**Waiting time** The waiting time $W_j$ could itself be used as a metric, based on the assumption that the processing time $p_j$ does not depend on the scheduling.

$$W_j = s_j - r_j \tag{1.1}$$

**Turnaround time**  Another metric, the turnaround time $F - J$, which is also called as response time or flow time, denotes the total amount of time job $j$ stayed in the system. The definition of the turnaround and response time sometimes vary in the literature.

$$F_j = c_j - r_j = W_j + p_j \tag{1.2}$$

**Slowdown**  The average turnaround time is a widely accepted metric for online scheduling systems. Nevertheless, it has one significant drawback. As the runtimes of jobs have a considerable variance, it seems that this metric places greater emphasis on long jobs, as opposed to short jobs, which are much more common. A solution to this problem is to use the slowdown metric, also known as stretch, which is defined as the turnaround time normalised by the running time [Fei01].

$$SLD_j = \frac{F_j}{p_j} \tag{1.3}$$

Another way to formulate the definition of slowdown is as the runtime on a loaded system divided by runtime on a dedicated system [FR98].

**Bounded slowdown**  The problem with slowdown is that extremely short jobs with reasonable delays lead to excessive slowdown values. The bounded slowdown avoids this problem thanks to a processing time threshold $\tau$.

$$BSLD_j^\tau = \max\left(\frac{F_j}{\max(p_j, \ \tau)}, \ 1\right) \tag{1.4}$$

There were also proposed other modifications to slowdown such as per-processor slowdown [Fei01], per-processor bounded slowdown [Car+19] or fair-slowdown [Sri+02b].

For each of the above metrics, an average could be taken to characterise the user-centric quality of scheduling in a whole workload. This way are obtained the following averaged metrics:

$$\overline{W} = \frac{1}{|J|} \sum_{j \in J} W_j \tag{1.5}$$

$$\overline{F} = \frac{1}{|J|} \sum_{j \in J} F_j \tag{1.6}$$

$$\overline{SLD} = \frac{1}{|J|} \sum_{j \in J} SLD_j \tag{1.7}$$

$$\overline{BSLD^\tau} = \frac{1}{|J|} \sum_{j \in J} BSLD_j^\tau \tag{1.8}$$

### 1.8.3. System-centric metrics

**Makespan**  Makespan is the time of an execution of a whole workload - a period from the beginning of the simulation the latest completion time among all jobs. It can be thought of as an offline version of turnaround time. It is often the metric of choice for offline scheduling.

$$makespan = \max_{j \in J} c_j \tag{1.9}$$

**System utilisation** System utilisation is the ratio of time when resources were utilised to the total time aggregated for all resources, which means that the idle time of resources is excluded from the numerator. There might be considered many kinds of resources in a system such as processors, network bandwidth or burst buffers. However, the pure term system utilisation usually implicitly refers to the utilisation of computing resources. System utilisation is a highly important metric for offline, batch scheduling. Nevertheless, for online scheduling, the system utilisation is mostly determined by an arrival process and requirements of jobs, rather than a scheduler. For parallel, rigid and non-preemptive jobs, the utilisation could be simply defined with a sum of job rectangles in the Gantt chart, as described in Section 1.7.

$$utilisation = \frac{\sum_{j \in J} size_j \cdot p_j}{M \cdot makespan} \qquad (1.10)$$

where $M$ is the total number of processors available in a system.

For some of the metrics, namely waiting time, turnaround time, slowdown and makespan, the goal is to minimise them. While for utilisation, the goal is to maximise. The storage utilisation may be defined accordingly to the compute utilisation by replacing $size_j$ with $bb_j$.

## 1.9. SimGrid

SimGrid [Cas+14] is a state-of-the-art simulation framework which allows to conduct research of applications in distributed environments. From a technical point of view, it is a library, which exposes interfaces to build custom simulators in C/C++, Python or Java. SimGrid is self-defined as:

> A framework for developing simulators of distributed applications that executed on distributed platforms, which can, in turn, be used to prototype, evaluate and compare relevant platform configurations, system designs, and algorithmic approaches.

SimGrid is the underlying component of Batsim, which is a batch simulator selected to supplement our research (described in Section 1.10). All the technical details of SimGrid are hidden and wrapped inside Batsim, making it transparent for users. The only exposed element of SimGrid is a platform configuration file, which we covered in detail in Section 2.2. The placement of SimGrid in the Batsim architecture is presented in Figure 1.4.

SimGrid has been in active development and maintenance since 1999. According to its website [Sim20] "SimGrid has been used to develop hundreds of simulators", "supported the research in at least 429 articles" and "totaled at least 1873 citations". SimGrid is also considered as the reliable simulation toolkit by the authors of Batsim, as stated in [Poq17, Chapter 3.3], because it is a long-term project that contains deeply validated models. We share the above argumentation and therefore consider SimGrid to be the reliable simulation platform.

## 1.10. Batsim

Built on top of SimGrid, Batsim [Dut+16; Poq17] is a simulator framework of Resources and Jobs Management Systems. In other words, Batsim is a dedicated simulator for analysis of batch schedulers. The Batsim architecture was designed to avoid common drawbacks occurring in job simulators, which are:

- Strong coupling of simulator and algorithm implementation,

- Simulation models restricted to specific phenomena,

- Messy implementation of a simulator created only for a publication.

One of the key aspects that differentiate Batsim from other similar solutions such as GridSim [BM02], Alea [KR10; KSS19] and Simbatch [CG09] is the focus on a separation of decision making logic (scheduling) from RJMS and platform simulation. That is achieved by splitting the whole simulation into two separate operating system processes: the Batsim simulator and a scheduler process, which communicate via a specific network protocol [Bat20a]. As a result, implementations of researched scheduling algorithms could be language-independent from the Batsim implementation. Finally, both Batsim and SimGrid are fully open source and provide extensive online documentation.

The Batsim architecture from the perspective of our research is presented in Figure 1.4.

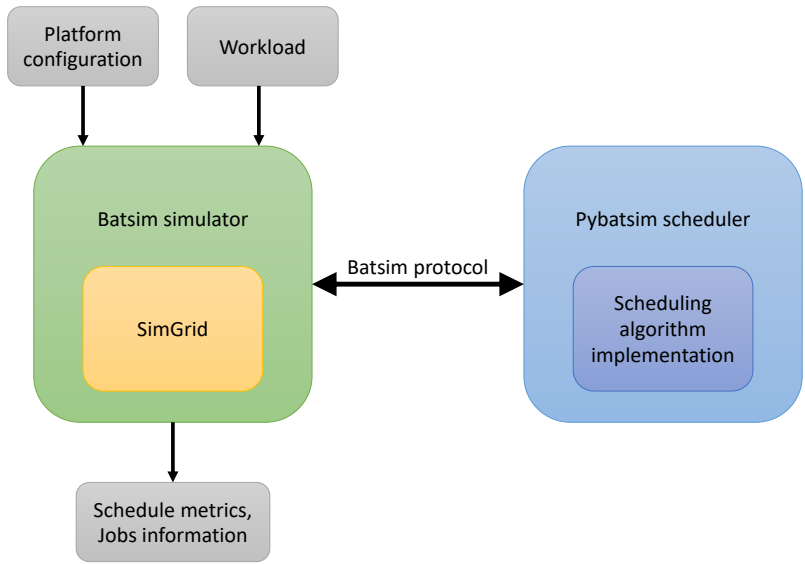

Figure 1.4: Key components of the Batsim architecture from a user perspective

Yet another simulator that could support burst buffer related research is CODES [ARL20; Cop+11; Car15]. This simulator was developed to explore the design of exascale storage architectures and distributed data-intensive science facilities. It is focused on the fine-grained simulation of I/O workloads in high-performance environments. A number of research publications regarding burst buffers have been already carried with CODES [Liu+12; Sny+15; Mub+17].

However, for the research of batch scheduling, we find the higher-level simulation offered by Batsim to be a more suitable approach.

### 1.10.1. Pybatsim

Batsim is provided with four projects in C++, Python, D and Rust implementing the Batsim protocol on the scheduling side. For this dissertation, the Python implementation, named

Pybatsim [Bat20c], was chosen as an origin for implementation of our burst buffers aware scheduler.

Pybatsim exposes low and high-level APIs. The low-level API is merely a wrapper of the Batsim protocol that provides a set of callbacks and functions to interact with the simulator. The high-level API wraps the low-level API and offers numerous functions useful for implementation of basic scheduling algorithms, but at the cost of higher code complexity.

## 1.11. Parallel Workloads Archive

The source of workloads for our performed experiments is the Parallel Workloads Archive [FTK14; Fei20], which is an online archive that contains normalised logs of batch workloads gathered from various supercomputers, clusters and grids. It also includes a selection of parallel workload models derived from those logs.

The logs are stored as text files in the Standard Workload Format (SWF), which is essentially the CSV file format extended with comments in the preamble. A single entry in an SWF file is a description of a job that was executed in a given parallel system. The content and order of the SWF entry is described in detail in [Fei15]. Below, we present a list of the fields which were used in our research together with their original description.

- **Job number**: a counter field, starting from 1.

- **Submit time** in seconds, relative to the start of the log.

- **Runtime** (wallclock) in seconds. "Wait time" and "runtime" are used instead of the equivalent "start time" and "end time" because they are directly attributable to the scheduler and application, and are also suitable for models where only the runtime is relevant.

- **Requested number of processors**.

- **Requested runtime** (or CPU time).

- **Requested memory** (again kilobytes per processor).

The job number and submit time were shifted to 0 on conversion from SWF to the Batsim workload format. In Section 2.5, we investigated a correlation between the requested number of processors and requested memory. Therefore, we created a burst buffer request model using the requested memory field. Lastly, the runtime field and the requested runtime (walltime), together with the created model, were used to generate the Batsim workload, which in detail is described in Section 2.3.

# Chapter 2

# Simulation model

## 2.1. Overview

Based on Batsim, we developed two simulators of parallel jobs with burst buffer requests. The first of them is referred to as Alloc-Only model/simulator/scheduler. It simulates allocations of compute and storage resources, computations and inter-node communication. The second - IO-Aware model - is an extension of the Alloc-Only model, which also simulates I/O network traffic. It is capable of simulating network congestion and I/O contention among running applications. The congestion of the network effectively leads to a slowdown of application runtime. A significant part of a simulation logic responsible for both job scheduling and simulating of burst buffer allocations lies inside a Pybatsim-based scheduler process. The remaining part which simulates computations, communication and I/O traffic is operated be the Batsim simulator process.

## 2.2. Platform model

Describing a simulated supercomputer platform model capable of simulating I/O traffic requires to define many parameters such as the number of compute nodes, network topology, network bandwidth, CPU speed, burst buffer storage model and PFS model. The platform in our simulation is managed by SimGrid, which requires to be provided with a simulated platform configuration. The simplest method to achieve this is to describe a platform using SimGrid XML file format. Batsim inherits and further extends this configuration format.

**Network topology**   SimGrid enables to use of several predefined common cluster topologies, notably:

- Torus

- Fat-Tree

- Dragonfly

We decided to model the dragonfly network topology, which is a relatively novel architecture introduced in [Kim+08]. It was already deployed in Cori [Bhi+16] and Piz Daint [Swi20] supercomputers. The schema of the dragonfly topology is presented in Figure 2.1, where squares represent routers and points represent nodes. Compute nodes in this topology are split into groups that are connected in a full graph (clique). Nodes inside a group could be arranged in various topologies such as a clique or a fat-tree. Figure 2.1 presents a dragonfly cluster network that consists of 9 groups of nodes.

Our model inherits the SimGrid implementation of the dragonfly architecture according to Figure 2.2. In this model compute nodes are attached to routers, which are connected in a clique inside a chassis. Several chassis, again connected in a clique, are then forming a single group.

Listing 1 is the XML configuration of the platform used in our simulations. The tag `cluster` in line 5 defines a cluster in the dragonfly topology with 108 nodes. The tag `topo_parameters` specifies the topology configuration with 3 groups, 4 chassis inside a group, 3 routers per chassis and 3 nodes attached to each router. `topo_parameters` is also explained in Figure 2.2. As shown in this figure, there could be defined several network links between different cluster components. For simplification in our configuration, there is always a single link.

**Compute and storage nodes**  Our cluster model consists of 108 nodes. However, only 96 of them are compute nodes. The other 12 nodes are assigned a storage role and simulate dedicated nodes for burst buffers. To be specific, **a single node in every chassis is dedicated to being a burst buffer node**, which in our cluster model means that there are 8 compute nodes per every burst buffer node. This type of shared burst buffer architecture closely resembles the architecture of Fugaku supercomputer, where one of every 16 compute nodes contains SSDs for burst buffers [RIK20].

Although 96 compute nodes is not a representative number for a modern supercomputing cluster, we decided to choose this number to make the execution traces possible to be analyses in detail during the development and fine-tuning of the simulator parameters. The described burst buffer nodes configuration is not represented in Listing 1 but specified in a Batsim simulator run command.

Another important burst buffer model assumption is no exclusiveness of storage resources for compute nodes. That is every compute node could be allocated with burst buffer located on any storage node in the platform. The performance degradation of placing burst buffer away from a compute node will be taken into account by our IO-Aware simulator. However, when allocating compute and storage resources, a scheduler attempts to find free space for a burst buffer as close as possible to allocated compute nodes according to the cluster topology.

There are also two special nodes in the platform configuration described by tags: `<host id="master" speed="0">` and `<host id="pfs" speed="0">` (lines 12, 18). The master node is unique and is mandatory for running simulation with Batsim. It is not exposed to the scheduler process and cannot be assigned to any job. In our modes, PFS is represented in the platform by a single storage node. There is also a single network link with shared bandwidth from the computing cluster to the pfs node. I/O nodes that usually conduct I/O traffic between a supercomputer and PFS are omitted in this model.

**Network bandwidth**  Two networks could be distinguished in the platform model:

- the main compute network,

- the I/O network.

The main compute network simulates fabrics within the cluster that is between compute nodes and burst buffer storage nodes. We set the bandwidth of a connection to 10 Gbit/s (line 8, `bw="1250MBps"`). The model of burst buffer request size, described in Section 2.5, was collected on the heterogeneous MetaCentrum grid [Cze20]. The computing grid is a collection of multiple clusters with a wide range of interconnects:

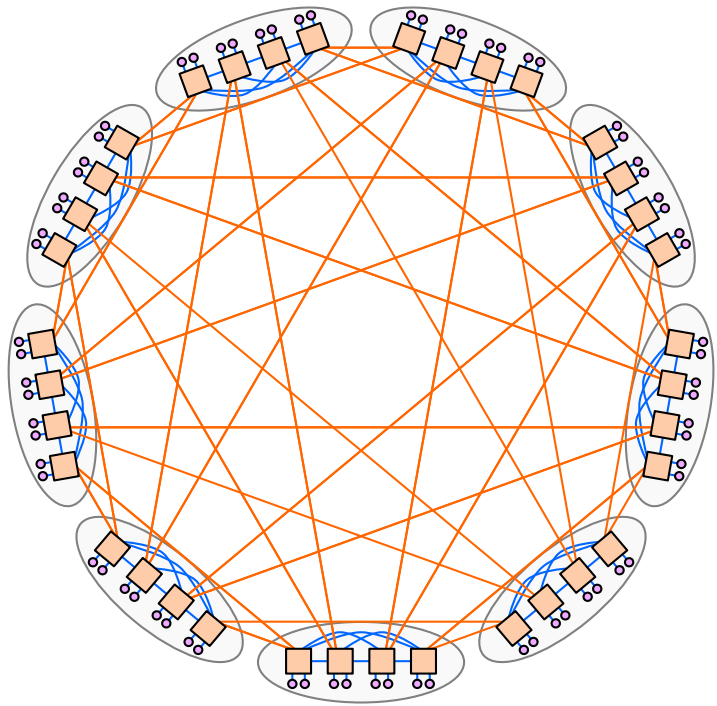

Figure 2.1: Simplified overview of a Dragonfly network topology
Source: `https://commons.wikimedia.org/wiki/File:Dragonfly-topology.svg`

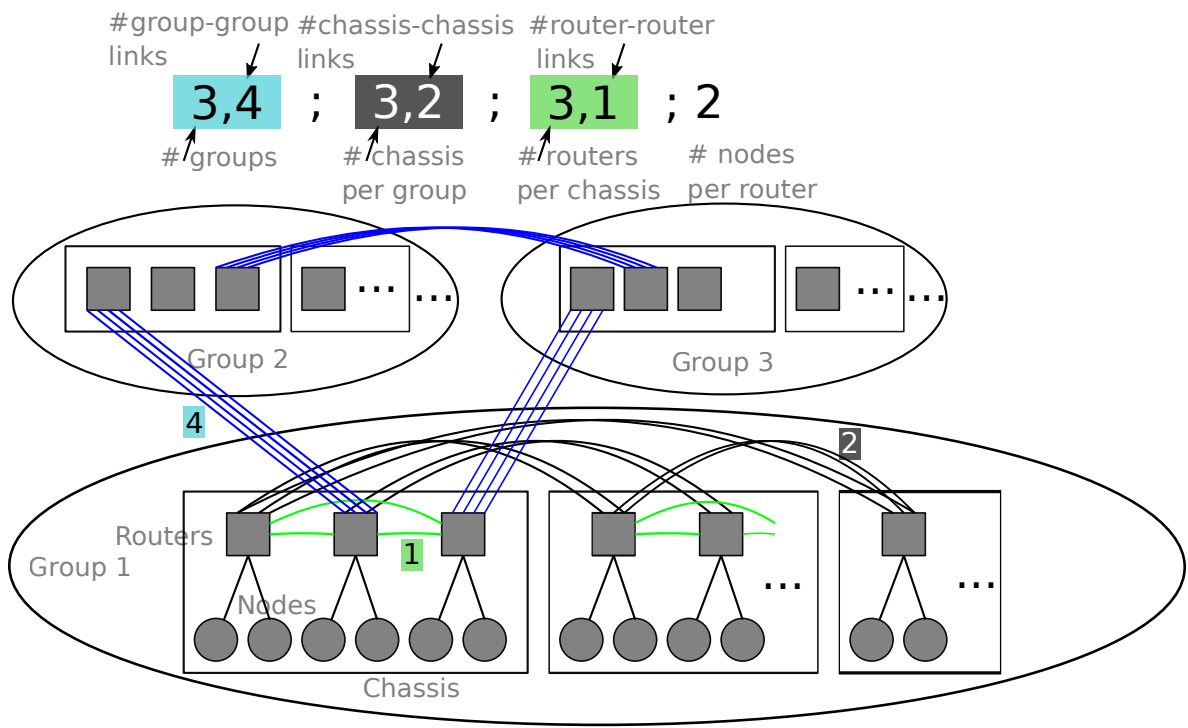

Figure 2.2: Detailed overview of a Dragonfly group definition according to SimGrid
Source: `https://framagit.org/simgrid/simgrid/-/blob/master/examples/platforms/cluster_dragonfly.svg`

```xml
<?xml version='1.0'?>
<!DOCTYPE platform SYSTEM "https://simgrid.org/simgrid.dtd">
<platform version="4.1">
    <zone id="root" routing="Full">
        <cluster id="cluster" topology="DRAGONFLY"
                 topo_parameters="3,1;4,1;3,1;3"
                 prefix="node_" radical="0-107" suffix=""
                 speed="1Gf" bw="1250MBps" lat="50us"
                 loopback_bw="1000MBps" loopback_lat="0"/>

        <zone id="master_zone" routing="None">
            <host id="master" speed="0">
                <prop id="role" value="master"/>
            </host>
        </zone>

        <zone id="pfs_zone" routing="None">
            <host id="pfs" speed="0">
                <prop id="role" value="storage"/>
            </host>
        </zone>

        <link id="link_master" bandwidth="1250MBps" latency="100us"
              sharing_policy="FATPIPE"/>
        <link id="link_pfs" bandwidth="5000MBps" latency="100us"/>

        <zoneRoute src="master_zone" dst="cluster"
                   gw_src="master" gw_dst="node_cluster_router">
            <link_ctn id="link_master"/>
        </zoneRoute>

        <zoneRoute src="pfs_zone" dst="cluster"
                   gw_src="pfs" gw_dst="node_cluster_router">
            <link_ctn id="link_pfs"/>
        </zoneRoute>
    </zone>
</platform>
```

Listing 1: XML configuration of the simulated platform

- Ethernet 1 Gbit/s

- Ethernet 10 Gbit/s

- InfiniBand QDR 8 Gbit/s

- InfiniBand FDR 14 Gbit/s

Furthermore, the number of links in a connection is multiplied from 1 to 4 depending on a concrete cluster, which effectively gives a range of bandwidth from 1 Gbit/s to 56 Gbit/s. The value of 10 Gbit/s lies within this range and is also small enough to make MPI-like communication and I/O traffic contention relevant.

The I/O network is a single link that connects computing cluster to the PFS. When modelling the bandwidth of this link, we wanted to capture a realistic performance of PFS as well as make the effect of I/O traffic contention emphasised enough for network congestion to have a noticeable impact on jobs runtime. The list IO500 [Vir20] provides benchmark results of the most efficient parallel file systems. We investigated the IO500 list and found out that the performance of clusters with 96 nodes varies from 2.8 GiB/s to 9.71 GiB/s, cluster with 100 nodes represented performance from 12.68 GiB/s to 22.77 GiB/s. In general, the list shows an extensive distribution of I/O systems performance. Clusters with the number of nodes from 32 to 128 have benchmark results ranging from 0.26 GiB/s to 368.44 GiB/s. Therefore, we chose the value of 5 GB/s (line 25, `bandwidth="5000MBps"`) for the shared bandwidth of our PFS model. With this value of bandwidth, I/O congestion could be observed when several I/O intensive jobs are simultaneously executed.

Other network parameters, such as latency have lesser salience in our simulation. There were derived from exemplary SimGrid configurations.

**Burst buffer capacity**   In order for the scheduling to be both limited by compute as well as storage resources, we assume that the total burst buffer storage in the platform should be fully utilised when all compute resources are utilised. Based on this assumption, the capacity of a burst buffer node could be estimated. In Section 2.5, we derived a model of burst buffer request size per processor. In the context of our simulation, a processor is equivalent to a compute node. The mean the obtained burst buffer request size distribution is 4.58 GiB. As mentioned in Section 2.2, there are 8 compute nodes per every burst buffer node. Therefore, the expected capacity of a burst buffer node should be equal to:

$$8 \cdot 4.58 \text{ GiB} = 36.64 \text{ GiB} = 39.34 \text{ GB}$$

Finally, we round out this value to 40 GB. Although 40 GB is not a realistic capacity for any modern SSD, this value balances job scheduling well enough between compute resources and storage resources.

However, this estimation has one downside, which comes from another assumption in our model. We assume that burst buffer space for a given compute node cannot be partitioned between burst buffer nodes. This assumption may lead to underutilisation of burst buffer storage when all compute resources are in use.

**Remaining parameters**   The CPU speed is set to 1 Gf (line 8, `speed="1Gf"`). The CPU speed does not impact a simulation as it is only used to convert runtimes from workload log to the number of FLOPS which is an input for the simulator.

Another parameter `loopback_bw="1000MBps"` (line 9) is the bandwidth of a loopback link of a node. This value was left as default from exemplary configurations.

## 2.3. Workload model

To perform experiments on a realistic workload, we decided to transform one of the workloads from the Parallel Workload Achieve into the Batsim input format. We selected the KTH-SP2-1996-2.1-cln log as it was recorded on a cluster with 100 nodes, which is the closest number to the size of our simulated cluster from all the available logs.

This workload log was gathered over 11 months and contains 28475 job entries. The jobs that request more than 96 compute nodes in the original log were simply discarded from our final workload. After transforming there are 28453 jobs left for execution in the simulation.

As we developed two versions of the simulation, there are also two versions of a workload. For both versions, all submission times and walltimes of jobs were preserved according to the original log. Runtimes, however, were adjusted for the IO-Aware simulator as additional IO-phases were added to jobs.

## 2.4. Job model

As described in Section 1.7, we consider only parallel, non-preemptive and rigid job model. For both versions of the simulation, each job in the workload is described by the following parameters:

1. Requested number of compute nodes,

2. Submission time,

3. Walltime,

4. FLOPS per node,

5. Number of bytes per node to send to every other node,

6. Requested size of burst buffer per node.

As described in Section 2.3, the requested number of compute nodes, submission times and walltimes are the same for both simulators and directly derived from the original log.

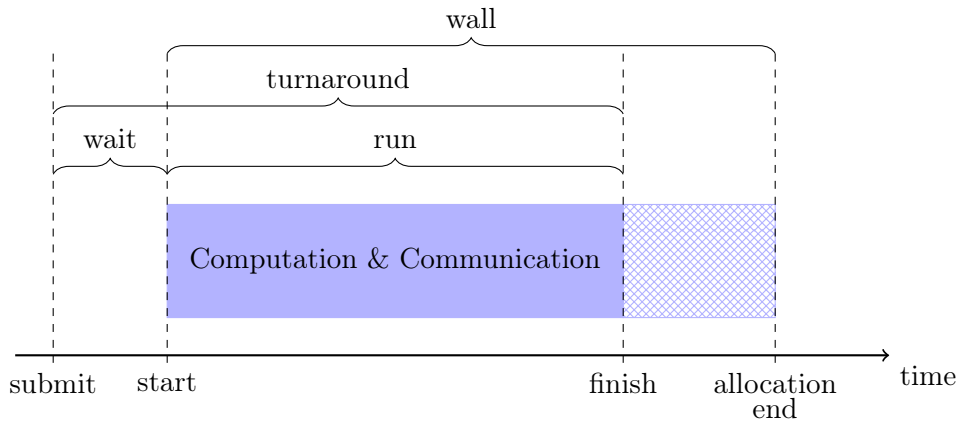

Figure 2.3: Alloc-Only job model

**Alloc-Only job model**  Figure 2.3 presents Alloc-Only job model together with several time metrics. In this model, a job consists of a single phase that simulates both computations and MPI-like communication. The blue rectangle extended with the striped area represents an allocation reserved for a job with the duration equal to a user specified walltime. When a running job reaches the end of an allocation, it is terminated by the system. Naturally, when a job finishes before a walltime, all resources associated with this job are released.

The number of computations per node was generated by multiplying runtime from the original log by CPU speed, which mean that expected job runtimes in the simulation matches runtimes of original jobs. The amount of computations to execute is drawn from an artificial distribution described below. It has been set in such a way that computations dominate the majority of jobs. For simplification of implementation, computations and communication happen in parallel.

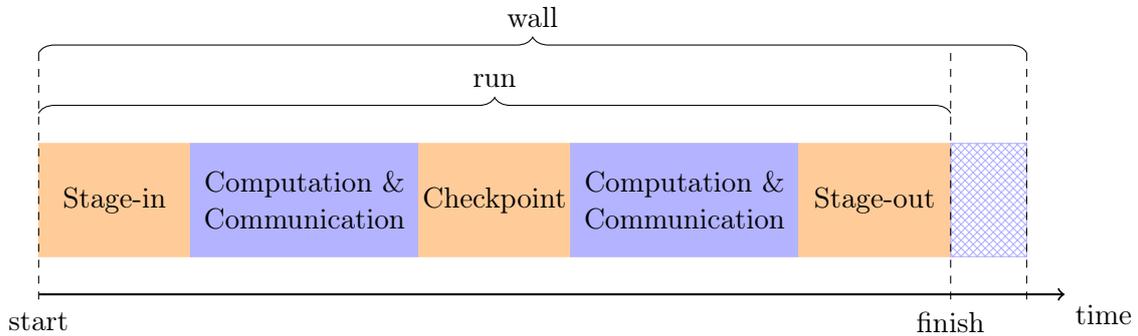

Figure 2.4: IO-Aware job model

**IO-Aware job model**  This model, shown in Figure 2.4, is more complex as it also simulates I/O network traffic. Jobs are divided into a variable number of phases depending on the specified number of computations per node. The number of computation and communication phases varies from 1 to 10 and are all of equally divided for a given job. They are interleaved by I/O phases which simulate checkpointing. We chose checkpointing as a popular application of burst buffers. It is simulated by transferring data from compute nodes to assigned burst buffer nodes, during which other application activities are suspended. After the checkpoint phase is finished, checkpoint data transfer from burst buffers to PFS is triggered, and the next computation phase starts concurrently.

Furthermore, jobs begin and end with data staging phases which simulates input and output data transfers between PFS and assigned burst buffer nodes. The size of the data transfers is equal to the requested burst buffer size.

As jobs in the IO-Aware model were extended with additional I/O phases, the computations phases, derived from original runtimes, had to be shortened accordingly in order for the jobs to avoid walltime exceeding. We adjust the requested total number of computations per node with the following formula:

$$\text{round}(CPU_{speed} \cdot \max(runtime - \kappa \cdot bb/bandwidth, \ 0.05 \cdot runtime))$$

where
$CPU_{speed}$ - processor speed in FLOPS,
$runtime$ - runtime of a job from the original KTH workload log,
$\kappa$ - empirically found factor,

*bb* - requested size of burst buffer per compute node,
*bandwidth* - bandwidth of the main compute network.

We were searching for the value of the $\kappa$ parameter by generating IO-Aware workloads, running a simulation with standard backfilling scheduling and comparing results to those obtained from Alloc-Only simulation. We found that $\kappa = 40$ raises similar results in terms of mean job runtimes, number of timeout jobs and waiting queue characteristics.

We also set an upper bound for the proportion of I/O time to computation time, that is an application could spend up to 95% of the time in I/O phases. Lastly, very short jobs, those with up to 120 seconds runtime, are assigned with a constant minimal burst buffer request size as they are most likely to timeout.

**Network traffic model**   As the amount of network traffic is not included in Standard Workload Format, we could either omit application inter-node communication from simulation or obtain these requirements from other sources. We decided to include inter-node communication in the simulations as we find that interfering between MPI-like traffic and I/O traffic creates a more realistic simulation model. As inter-node communication is not an essential factor in our simulation, we used a normal distribution with $\mu = 100$ MB and $\sigma = 20\%$ to model the amount of network traffic of a job sent between each processor. That is, for a drawn value $\theta$, in each communication phase, each processor sends $\theta$ bytes to every other processor. It is essentially the N-N communication model. The overall application traffic is proportional to the square of the number of processors and grows linearly with the number of communication phases.

## 2.5. Burst buffer request model

As burst buffers are a relatively new technology, there has not been put many studies into them. Notably, the topic of burst buffer aware scheduling has been not well exploited in research. The consequence of the current state of research is a lack of available workload traces and logs that contain data about burst buffers, or models of burst buffer requests. For instance, none of the workloads in the Parallel Workloads Archive contains information about burst buffers. Furthermore, the Standard Workload Format in the latest version 2.2 does not specify a field for burst buffers.

This issue was tackled by Fan *et al.* [Fan+19] for evaluation of their multi-resource scheduling scheme named BBSched, where two different traces and methods were used. The first trace is a four months long SLURM log collected from the Cori system [NER20]. Cori is a supercomputer equipped with burst buffers, so consequently, this log contains real burst buffer requests from users. Unfortunately, this log is not publicly available. The second trace is half-year Darshan log from the Theta system [Arg20b] that provides information about recorded I/O. The Theta system was not deployed at the time with any shared burst buffer. The potential burst buffer requests were estimated by the amount of data moved between PFS and nodes.

Darshan is an HPC I/O characterisation tool, and it is a part of the CODES project (described in Section 1.10). It is a fairly popular data source for I/O workload in HPC research related to storage and burst buffers. Exemplary papers that utilised Darshan I/O logs are [Liu+12; Sny+15].

Another approach to I/O workload modelling was applied in [LFW19]. Lackner *et al.* extended an existing HPC workload model with information about the amount of requested burst buffer storage capacity, stage-in and stage-out data, and the amount of intermediate data written to and read from burst buffers at runtime.

We propose yet another approach to estimate burst buffer requests. *Under the assumption that a burst buffer request size is equal to the main memory request size*, we create a model of burst buffer requests based on the METACENTRUM-2013-3 workload log from the Parallel Workload Achieve. As described in Section 2.3, we use the KTH-SP2-1996-2.1-cln log to generate the Batsim workload as it the best matches our targeted platform. However, it does not contain information on requested memory sizes. Hence, we had to use another log to obtain those information.

From the METACENTRUM-2013-3 log, we extract the data about processors requests, memory requests and execution time of each job. Then we analyse a correlation between processors requests and memory requests. Finally, we perform a fit of existing probability distributions to empirical data and test the quality of obtained models. Such a model can be later used to generate burst buffer requests for jobs in other logs.

**Analysis of a memory request size**  We start with loading the METACENTRUM-2013-3.swf file directly into the Pandas [tea20; McK10] DataFrame, which could be easily achieved with the Evalys library [Bat20b] developed as the Batsim supplementary analysis tool. The Metacentrum log contains 5731099 jobs. We filter out jobs that do not contain information about requested memory (marked as -1) or specify requested memory to 0. After this filtering, there are 5722091 jobs left in the log. This number will be referred to as *all jobs* in the text below. All the remaining jobs have the number of requested processors greater than 0.

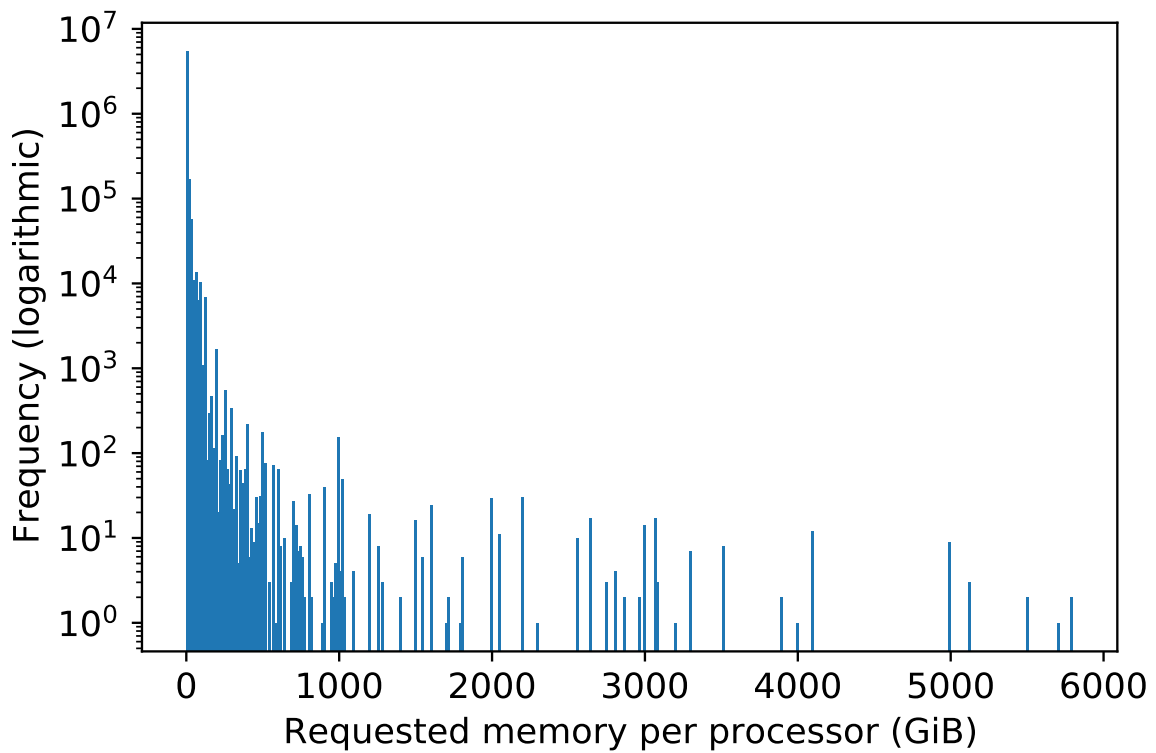

Figure 2.5: Histogram of memory requests of all jobs from the METACENTRUM-2013-3 log.

Figures 2.5 and 2.6 present histograms of memory requests of jobs. In Figure 2.5, we can observe an exponential shape of the empirical distribution with a long tail. Figure 2.6 depicts a histogram with a truncated tail to the value of $2 * 10^7$ kilobytes. The intriguing part to notice there is a significant peak of a count of jobs requesting memory of size around 4 GB. In particular, this peak corresponds precisely to the value of 4 GiB (4294967296 bytes). We calculated that 36.444% (2085360/5722091) of all jobs request 4 GiB of memory per processor. This phenomenon could be easily explained with a hypothesis that 4 GiB was a default size of memory per processor assigned to a job. All jobs which did not have explicitly specified the amount of memory were assigned with the above value.

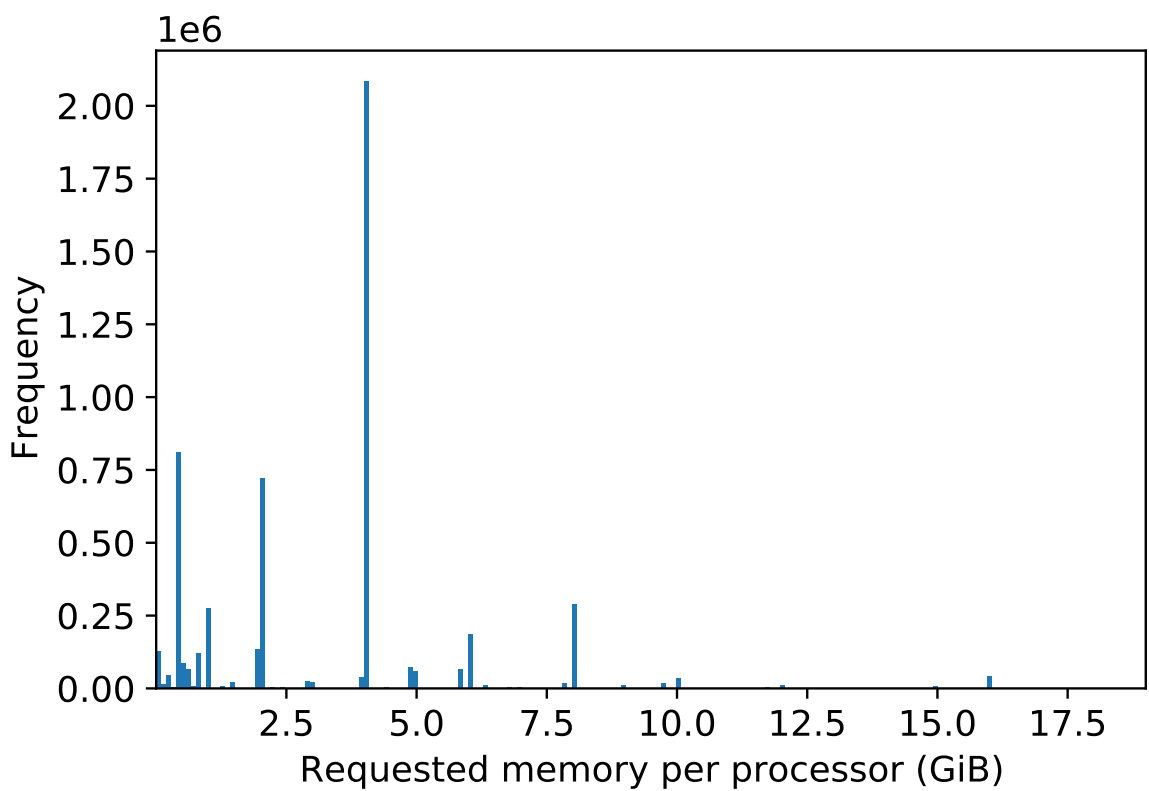

Figure 2.6: Histogram of memory requests from the METACENTRUM-2013-3 log with a truncated tail.

**Cross-correlation between requested memory size and the number of processors**
Following the guides of Dror G. Feitelson outlined in chapter 6. of *Workload Modeling for Computer Systems Performance Evaluation* [Fei15] there often exists several types of correlations between different attributes in computer workloads, namely:

- Temporal locality,

- Spatial locality,

- Cross-correlation among distinct workload attributes,

- Self-similarity,

- Long-range dependence.

In our analysis, we wanted to examine whether there exists a cross-correlation between the requested number of processors and memory per processor. If such a correlation exists, it is appropriate to model a distribution of memory as either a conditional probability distribution or a joint probability distribution with the number of processors.

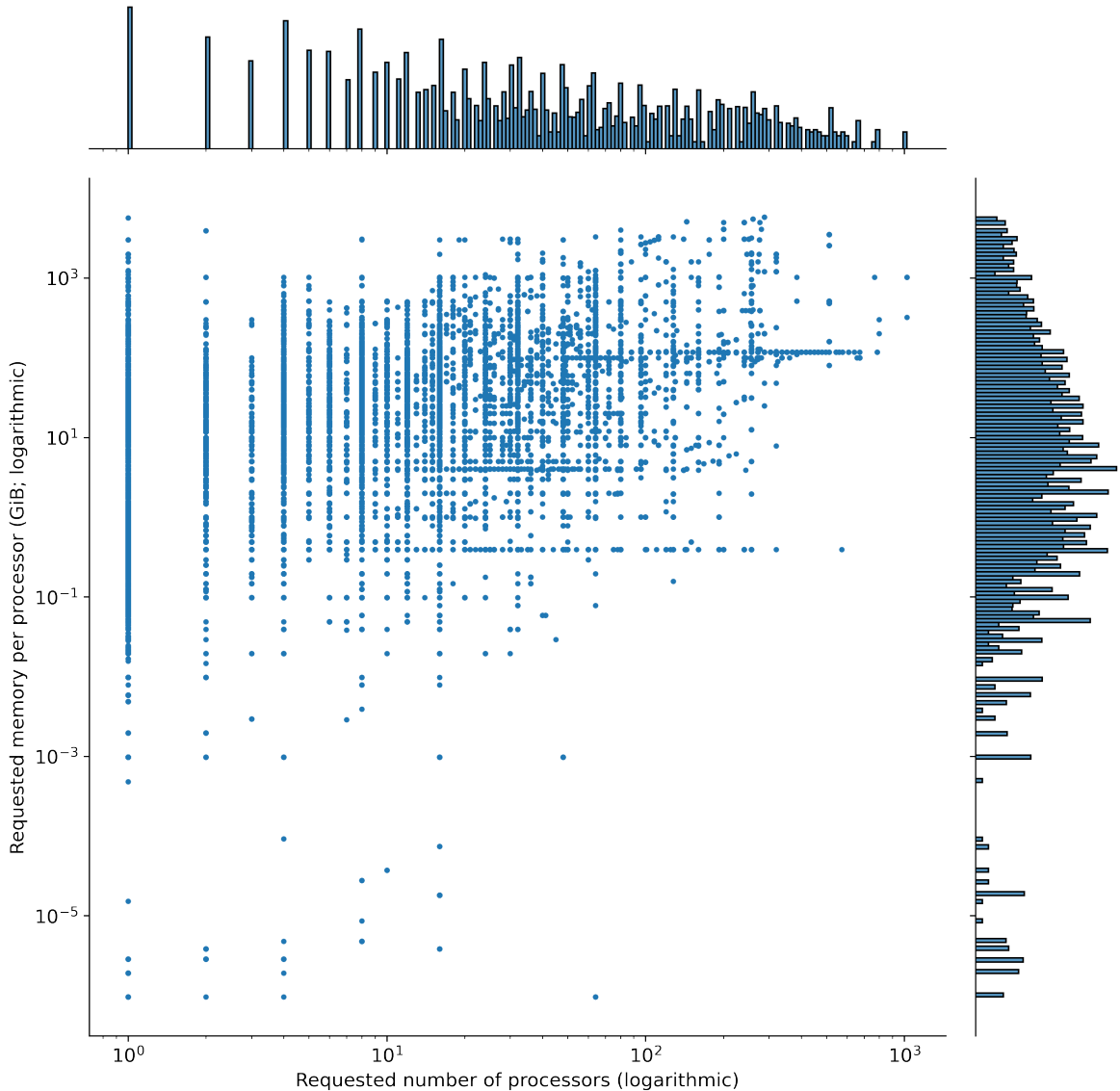

Figure 2.7: Scatterplot of joint distributions of requested number of processors and memory per processor with their related marginal distributions.

Two visual methods of investigating cross-correlations described in [Fei15] are plotting a 2D surface of a joint distribution of attributes and creating a scatterplot with associated marginal distributions as in Figure 2.7. Based on the data points in this scatterplot, it is hard to conclude the existence of a cross-correlation. However, it is visible that the majority of the jobs are concentrated on a small number of requested processors and a small amount of memory. Nonetheless, there are some significant outliers in the data set.

Figure 2.8 presents a heatmap of joint distributions with hexagonal bins and truncated tails of both distributions. Again there is visible a concentration of points at the value of 4

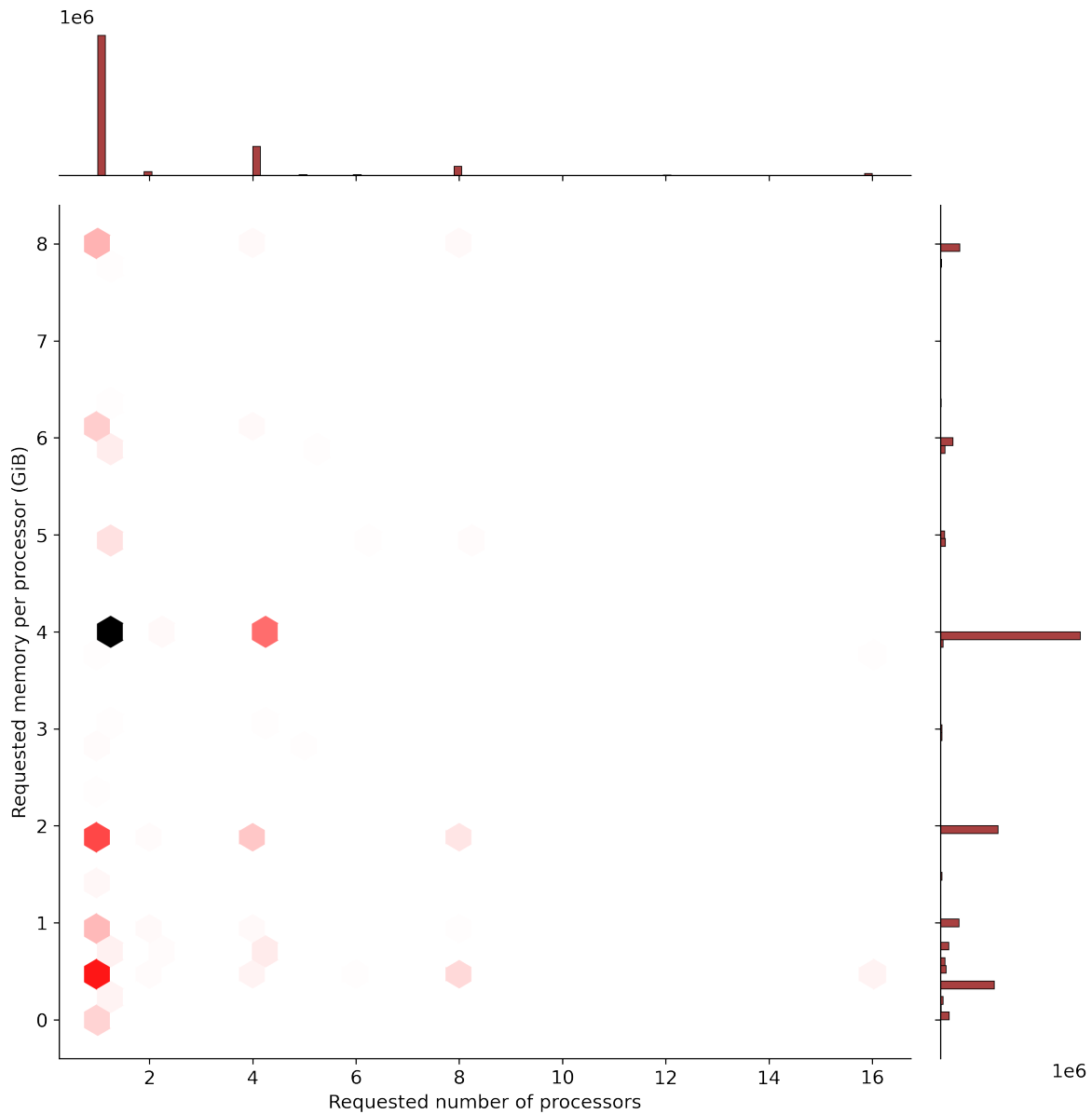

Figure 2.8: Heatmap of joint distributions of the requested number of processors and memory per processor with both tails truncated. The more intense is the colour of a bin, the more frequent it is.

GB of requested memory. Moreover, we can observe the concentration of jobs corresponding to the powers of two of the requested number of processors.

Both of those joint distribution plots were generated with the Seaborn [Was+17] Python library.

As visual methods did not indicate any confident cross-correlation, we computed correlation coefficients. Table 2.1 presents the results of Pearson, Kendall and Spearman correlation coefficients. The Pearson correlation is used to measure the quality of linear correlation in data. Our value of the Pearson coefficient indicates a very weak correlation. The downside of the Pearson coefficient is its sensitivity on outliers, which is a case in this workload. The Kendall and Spearman coefficient are not so sensible to outliers and are capable of detecting non-linear correlations. Our computed values of these coefficients indicate that there is no cross-correlation.

| Pearson | Kendall | Spearman |
|---------|---------|----------|
| 0.345   | 0.073   | 0.088    |

Table 2.1: Cross-correlation coefficients values between the requested number of processors and memory per processor

It may also be possible that cross-correlation does not exist in the whole data set, but only in some range of values. This hypothesis could be investigated by plotting empirical cumulative distribution functions (ECDF) of memory requests conditioned by ranges of numbers of processors. Figure 2.9 shows these distributions divided into ranges by powers of two of the requested number of processors. In this plot, we can observe that ECDFs of serial jobs and small jobs up to the size of 63 processors are relatively similar. The ECDF of jobs of size 64-127 is slightly different. A significant difference is visible in the shape of the ECDF of large jobs with the number of processors equal and above 128. We emphasise these difference in Figure 2.10. The shapes of ECDFs suggests that the probability distribution of requested memory should be modelled taken into account the split into small and large jobs.

In order to finally determine the contribution of large jobs in a whole workload, we calculated the total processors time and memory time taken by large jobs. As we are dealing with rigid jobs, it is sufficient to sum over rectangles representing jobs, where one side is an execution time of a job and the second side it either the number of processors or amount of requested memory by the job. We define large jobs in the METACENTRUM-2013-3 log as those of size 64 or greater. There are only 5936 out of 5722091 of them, which is about 0.1%. They overall contribute to 11% of processor time and 8.4% memory time. Given that this contribution is relatively small, we decided to model the distribution of requested memory independently from the number of processors. We consider the different distribution of large jobs to have a negligible effect on the overall distribution of memory.

**Probability distribution function model**   As we observed in Figure 2.5, the empirical distribution of requested memory values tends to have an exponential characteristic. We want to find a probability distribution that best models the body part as well as the long tail of the empirical distribution. In order to do this, we use the package FITTER [Cok20] to match a sample of 10000 data points to all probability distribution available in the SciPy library [Vir+20] and calculate a sum of squares error. Based on the error values, we selected five fairly popular distributions with different shapes: exponential, log-normal, logistic, half-logistic, beta.

| Distribution | KS statistic average |
|--------------|----------------------|
| Exponential  | $0.2562 \pm 0.0004$  |
| Log-normal   | $0.2200 \pm 0.0009$  |
| Logistic     | $0.2490 \pm 0.0005$  |
| Half-logistic| $0.3047 \pm 0.0029$  |
| Beta         | $0.2247 \pm 0.0082$  |

Table 2.2: Values of Kolmogorov-Smirnov D-statistic for the fitted distributions

We perform 5-fold cross-validation to achieve a reliable fit of distribution and test the goodness of fit. That is, we divide the data set into five parts. Then five times repeat the following procedure: leave out one of the parts as a test set and fit a probability distribution to remaining data points. On the left out part, we perform the Kolmogorov-Smirnov test for the goodness of fit. In Table 2.2 are presented the averaged results of the Kolmogorov-Smirnov test and their standard deviations. The lower is a value of the Kolmogorov-Smirnov D-statistic the better is a fit of a distribution. We rate the quality of obtained models based on the values of the Kolmogorov-Smirnov D-statistic and standard deviations of fitted distribution parameters. These criteria guided us to the conclusion that the log-normal distribution provides the best fit to the empirical data.

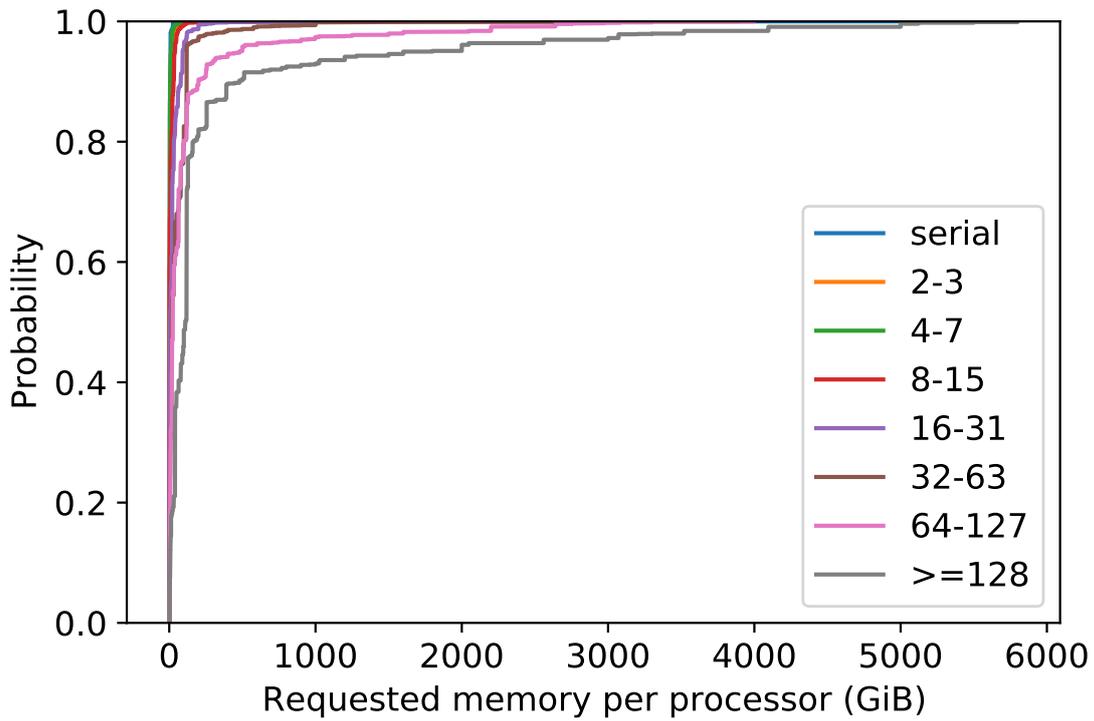

Figure 2.9: Empirical distribution function of requested memory divided into ranges by powers of two of the number of requested processors

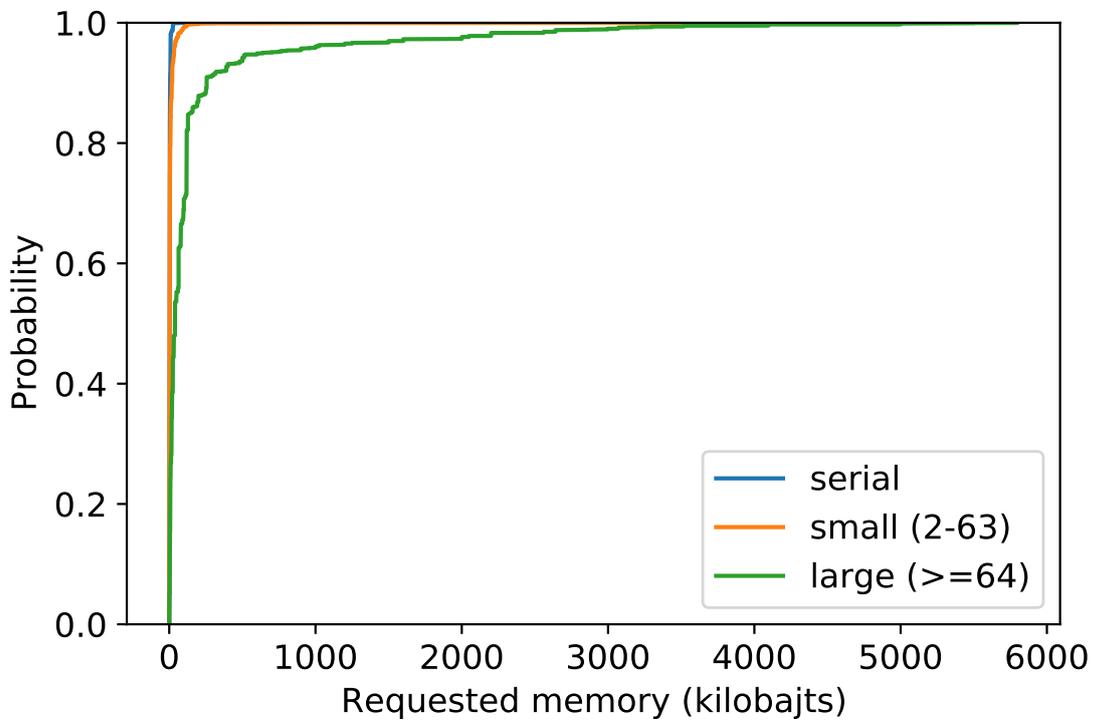

Figure 2.10: Empirical distribution function of requested memory divided into serial, small and large jobs by the number of requested processors

Figures 2.11 and 2.12 show probability distribution functions and cumulative distribution functions of the five selected distributions fitted to the data. We may observe that all selected distributions except the logistic provide natural noise to the peak at 4 GiB in the data.

The general formula for the probability density function of the log-normal distribution is:

$$f(x) = \frac{1}{\sigma x \sqrt{2\pi}} \exp\left(-\frac{\ln((x-\theta)/m)^2}{2\sigma^2}\right); \qquad x > \theta; \; \sigma > 0; \; m > 0$$

where $\sigma$ is the shape parameter, $\theta$ is the location parameter and $m$ is the scale parameter.

As parameters to our distributions, we used averaged values of parameters fitted during the cross-validation. The fitted parameters for the log-normal distribution are in Table 2.3. Selected statistics of this distribution are presented in Table 2.4.

| Shape $\sigma$ | 1.09725 |
| --- | --- |
| Location $\theta$ | -150361 |
| Scale $m$ | 2714115 |

Table 2.3: Parameters of the fitted log-normal distribution.

| Mean | 4804884 |
| --- | --- |
| Standard deviation | 7569200 |
| Median | 2563753 |
| Approximated mode | 665852 |

Table 2.4: Statistics of the fitted log-normal distribution in kilobytes.

The log-normal distribution could be directly programmatically reproduced using the SciPy library with the following lines of code:

```python
from scipy import stats
distribution = stats.lognorm(
    s=1.09725,
    loc=-150361,
    scale=2714115
)
```

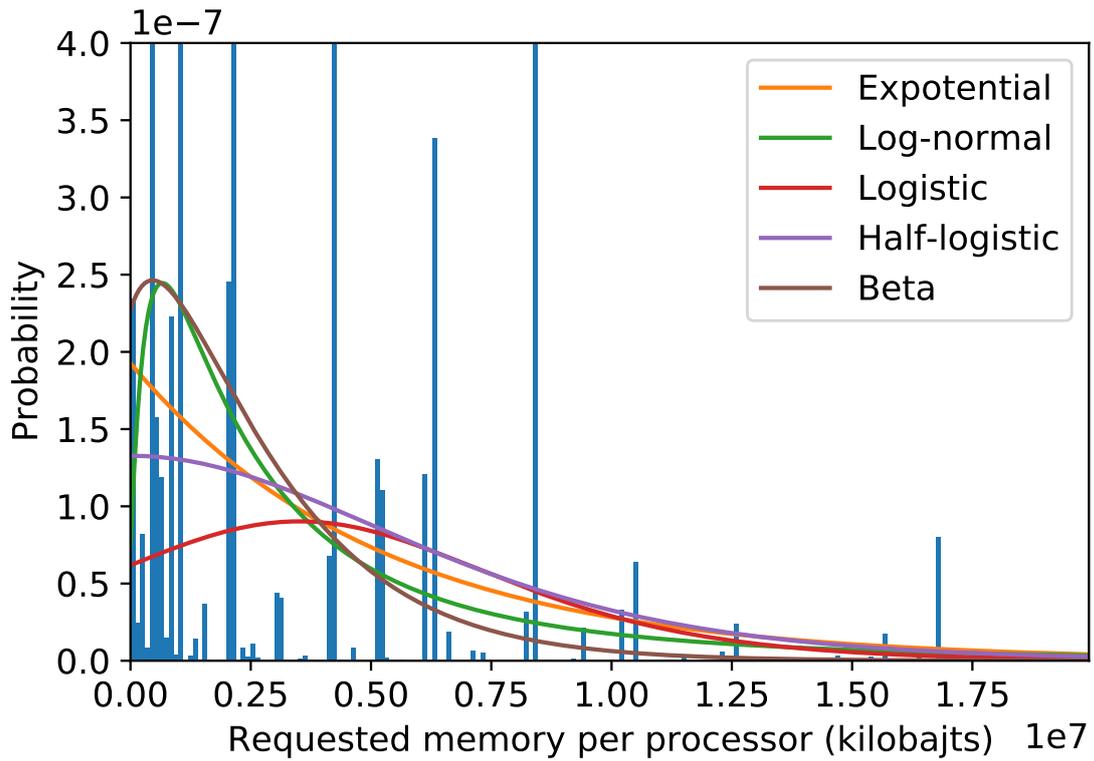

Figure 2.11: Histogram of memory requests and probability density functions of fitted distributions. Both axes of the histogram are truncated.

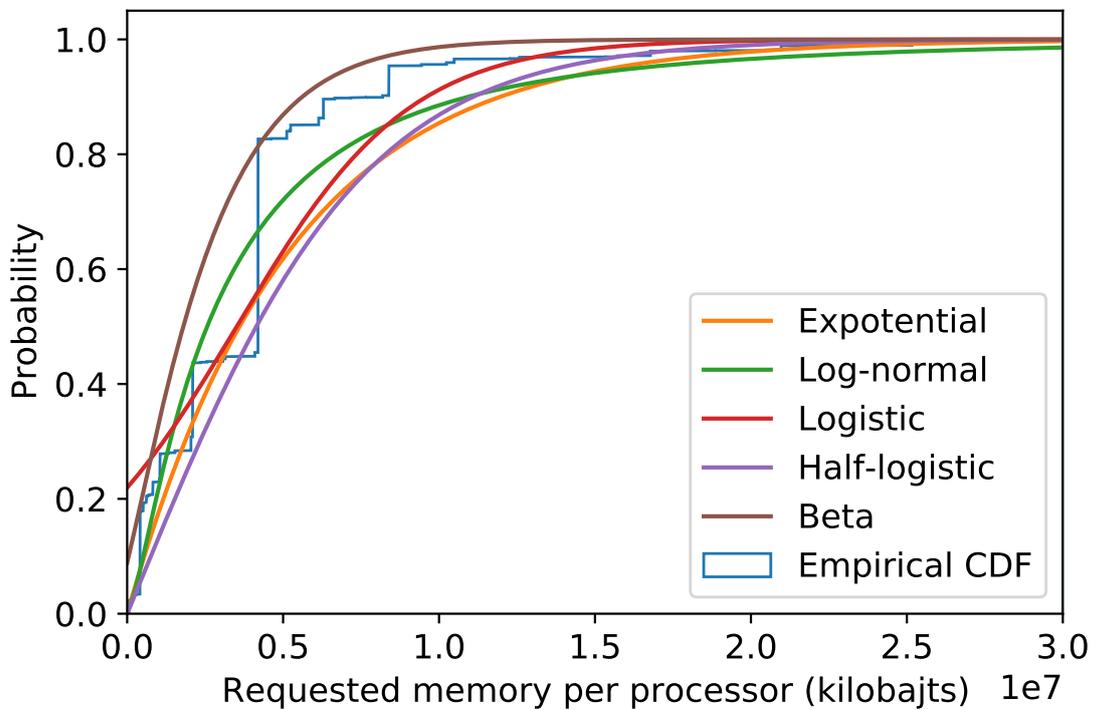

Figure 2.12: Empirical cumulative distribution function of requested memory and cumulative distribution functions of fitted distributions. The X-axis is truncated.

# Chapter 3

# Scheduling algorithms

## 3.1. Burst buffer reservations

The primary motivation for our research was an observation that existing job schedulers, such as Slurm, do not make future reservations of burst buffers during backfilling. Indeed, they only reserve compute resources for jobs at the front of a waiting queue and proceed to backfill remaining jobs in the queue. Available burst buffers are then allocated to those jobs that are expected to start earliest. Our hypothesis states that this procedure may significantly deteriorate the efficiency of scheduling. This effect may occur when jobs with low compute and high storage requirements are backfilled ahead of priority jobs. Priority jobs could be then delayed because of insufficient storage resources available in a system. Furthermore, this scenario could even lead to starvation of priority jobs, which means that backfilling without future burst buffer reservations is not a fair scheduling algorithm.

Slurm also provides a general description of burst buffer management in its documentation (`https://slurm.schedmd.com/burst_buffer.html`).

0. Read configuration information and initial state information

1. After expected start times for pending jobs are established, allocate burst buffers to those jobs expected to start earliest and start stage-in of required files

2. After stage-in has completed, jobs can be allocated compute nodes and begin execution

3. After job has completed execution, begin file stage-out from burst buffer

4. After file stage-out has completed, burst buffer can be released and the job record purged

As the first stage of our research, we analyse the impact of future burst buffer allocations on the overall efficiency of backfilling. To achieve this, we will compare the following scheduling algorithms:

1. First-Come-First-Served (FCFS) without backfilling

2. Greedy FCFS filling

3. FCFS EASY-backfilling without future burst buffer reservations

4. FCFS EASY-backfilling with future burst buffer reservations

---

**Algorithm 1** First-Come-First-Served

---

1: **procedure** FCFS(*queue*)
2:     **for** *job* ∈ *queue* **do**
3:         *procs* ← find available compute resources for *job*
4:         *bb* ← find available storage resources for *job*
5:         **if** *procs* and *bb* were both found **then**
6:             Launch *job* with *procs* and *bb*
7:             Remove *job* from *queue*
8:         **else**
9:             Break

---

**Algorithm 2** Greedy filling

---

1: **procedure** FILLER(*queue*)
2:     **for** *job* ∈ *queue* **do**
3:         *procs* ← find available compute resources for *job*
4:         *bb* ← find available storage resources for *job*
5:         **if** *procs* and *bb* were both found **then**
6:             Launch *job* with *procs* and *bb*
7:             Remove *job* from *queue*

---

**Algorithm 3** Backfilling without future burst buffer reservations

---

1: **procedure** BACKFILL(*queue*, *D*)            ▷ *D* - reservation depth
2:     FCFS(*queue*)
3:     *J* ← pop the first *D* jobs from *queue*
4:     **for** *job* ∈ *J* **do**                   ▷ in FIFO order
5:         Reserve compute resources for *job* in the future
6:     FILLER(*queue*)
7:     Remove reservations for jobs in *J*
8:     Push back *J* at the front of *queue*

---

**Algorithm 4** Backfilling with future burst buffer reservations

---

1: **procedure** BACKFILL(*queue*, *D*)            ▷ *D* - reservation depth
2:     FCFS(*queue*)
3:     *J* ← pop the first *D* jobs from *queue*
4:     **for** *job* ∈ *J* **do**                   ▷ in FIFO order
5:         Reserve compute and <u>storage</u> resources for *job* in the future
6:     FILLER(*queue*)
7:     Remove reservations for jobs in *J*
8:     Push back *J* at the front of *queue*

---

These policies are outlined in algorithm listings 1 to 4. The Greedy Filling policy may be considered as FCFS backfilling without any job reservations. It just attempts to launch jobs whenever enough resources are available. It is important to note that Greedy Filling is not a fair scheduling algorithm. In terms of EASY backfilling, we present a generalised version with a configurable depth of reserved jobs.

None of those elementary algorithms specifies any concrete requirements on resource reservations. Their general approach allows the resource allocation schemes to be implemented differently on each supercomputing platform. It is a significant benefit, because supercomputers often vary widely in components such as network topology or burst buffer architecture, which was outlined in Section 2.2 and Section 1.5 accordingly.

---

**Algorithm 5** Shortest Jobs First Backfilling

1: **procedure** BACKFILL-SJF($queue$, $D$)                                        ▷ $D$ - reservation depth
2:     FCFS($queue$)
3:     $J \leftarrow$ pop the first $D$ jobs from $queue$
4:     **for** $job \in J$ **do**                                                            ▷ in FIFO order
5:         Reserve compute and storage resources for $job$ in the future
6:     $sjf \leftarrow$ sort jobs in $queue$ ascending by walltime
7:     FILLER($sjf$)
8:     Remove reservations for jobs in $J$
9:     Push back $J$ at the front of $queue$

---

In sections 3.2, 3.3, 3.4, we present our algorithms aiming to improve multi-resource scheduling. As a baseline for comparing simulation results, we treat the Shortest Jobs First Backfilling shown in algorithm listing 5. This variant of scheduling has the job queue sorted ascending by walltime for backfilling. However, only the jobs for backfilling are sorted. At first, it attempts to launch as many jobs as possible in First-In-First-Out order. This schema is essential to preserve fairness.

## 3.2. Maximising system utilisation

Canonical scheduling algorithms are focused on providing satisfying performance while preserving fairness. These algorithms are based on greedy heuristics and do not take advantage of optimisation methods. It has been observed that the ordering of jobs in a waiting queue impacts the performance of scheduling [Sri+02b]. Notably, having jobs sorted by walltime for backfilling (Shortest Job First) usually improves results in both mean waiting time and mean slowdown. In general, it is feasible to define an objective function and iterate over permutations of jobs to obtain an optimal resource allocation.

As one of our approaches, we attempt to improve the overall scheduling performance by maximising system utilisation while maintaining load balance in the waiting queue between compute and storage resources. Therefore, we propose algorithm 6, which is based on two concepts:

1. maximise both compute and storage utilisation,

2. ensure a mechanism capable of self-balancing resource load in the waiting queue.

**Optimisation objective**   To meet both goals, our optimisation objective is to lexicography maximise values in a 3-tuple. The order of fields in the tuple differs depending on the queue load. Let $Q$ be a set of jobs in the waiting queue. Let $P$ be a permutation of jobs as in listing 6. It does not include jobs selected for reservations from the front of the queue. Let $\Phi \subset P$ be a set of jobs that could be launched immediately. Let $M$ denote the total number of processors in a platform and $B$ be total capacity of burst buffer nodes. Then our objective tuples, using

---

**Algorithm 6** Maximising system utilisation by job permutation search

1: **procedure** MAXUTIL(*queue*, *D*, *N*, *β*)  ▷ *D* - reservation depth
  ▷ *N* - number of steps
  ▷ *β* - balance factor
2:   $L_{compute} \leftarrow$ calculate compute load in *queue*
3:   $L_{storage} \leftarrow$ calculate storage load in *queue*
4:   **if** $L_{storage} \leq \beta \cdot L_{compute}$ **then**
5:     The objective is to lexicography maximise (compute, storage, mean wait time)
6:   **else**
7:     The objective is to lexicography maximise (storage, compute, mean wait time)

8:   $J \leftarrow$ pop the first *D* jobs from *queue*
9:   FCFS(*J*)
10:   Remove launched jobs from *J*
11:   Reserve resources for jobs in *J* in the future

12:   **if** *queue* is small **then**  ▷ exhaustive search
13:     Iterate over all possible permutations;
14:     $P \leftarrow$ find the jobs permutation with the highest score

15:   **else**  ▷ hill climbing
16:     $P \leftarrow$ find the best permutation among the set of initial candidates
17:     *last* $\leftarrow$ index of the last job in *P* that could be launched
18:     $S_{best} \leftarrow$ score for *P*
19:     *steps* $\leftarrow 0$
20:     **for** *distance* from 1 to $|queue| - 1$ **do**
21:       **for** *index* from 0 to $min(last, |queue| - distance - 1)$ **do**
22:         *steps* $\leftarrow$ *steps* + 1
23:         **if** *steps* > *N* **then** Break both loops
24:         Swap jobs in *P* at positions *index* and *index* + *distance*
25:         $S \leftarrow$ simulate backfilling for *P*
26:         **if** $S > S_{best}$ **then**  ▷ accept new permutation
27:           $S_{best} \leftarrow S$
28:           Update *last*
29:           Break
30:         **else**  ▷ reject new permutation
31:           Swap jobs in *P* at positions *index* and *index* + *distance*

32:   FILLER(*P*)
33:   Remove reservations for jobs in *J*  ▷ will be reacquired in the next iteration
34:   Push back *J* at the front of *queue*

---

notation from Section 1.8.1, are defined as follows.

$$L_{compute} = \frac{1}{M} \sum_{j \in Q} size_j \qquad \leftarrow \text{ compute load in the queue}$$

$$L_{storage} = \frac{1}{B} \sum_{j \in Q} size_j \cdot bb_j \qquad \leftarrow \text{ storage load in the queue}$$

$$(\sum_{j \in \Phi} size_j, \ \sum_{j \in \Phi} size_j \cdot bb_j, \ \frac{1}{|\Phi|} \sum_{j \in \Phi} W_j) \quad \text{for} \quad L_{storage} \leq \beta \cdot L_{compute}$$

$$(\sum_{j \in \Phi} size_j \cdot bb_j, \ \sum_{j \in \Phi} size_j, \ \frac{1}{|\Phi|} \sum_{j \in \Phi} W_j) \quad \text{for} \quad L_{storage} > \beta \cdot L_{compute}$$

where $size_j$ is the requested number of processors and $bb_j$ is the requested size of burst buffer per processor for the job $j$.

In other words, maximising compute resource utilisation is the main priority when demand for compute resources by jobs in the waiting queue is relatively more extensive than demand for storage resources. Analogously, decreasing storage load is the main priority otherwise.

There is a third component of the tuple, which is a mean waiting time of the jobs selected for execution. We added it after observing that various permutations often raise the same resource utilisation, but differ in terms of the waiting time. This change should additionally ensure that, as a side goal, the jobs that spent the most time in the queue are selected earlier.

**Balance factor**   The balance factor $\beta$ is responsible for setting which kind of resource should be favoured when scheduling. For $\beta > 1$, the algorithm is more likely to favour reducing compute load in the waiting queue. In another case, for $\beta < 1$ reduction of storage load is prioritised. Selecting the right value of $\beta$ is a matter of platform and workload characteristics.

**Fairness**   At the beginning of the algorithm in lines 8-11, the first $D$ jobs in the waiting queue are attempted to run immediately or if it is not possible then are reserved with resources in the future. It differs from the canonical backfilling in Algorithm 4, in which jobs were executed in FIFO order as long as possible. Our schema holds fairness property but ensures less restrictive guarantees for it.

At the end of the scheduling algorithm, all reservations are released. Naturally, some running jobs may terminate before achieving walltimes between the consecutive scheduling iterations. The jobs from the $J$ set will receive new reservations, potentially starting earlier, in the succeeding scheduling iteration. It does not impact the fairness, because reservation assignment is performed in the FIFO order. Additionally, the persistent state maintained between the iterations is smaller.

**Exhaustive search**   When the waiting queue is small, we perform an exhaustive search over all possible permutations of jobs. Simulating job execution on a permutation is a relatively fast procedure. Hence we proceed to an exhaustive search for $|queue| \leq 6$, which yields 720 permutations.

**Hill climbing** For longer queues of pending jobs, we switch to the hill-climbing optimisation method. Hill climbing is a simple optimisation technique, in which a transition to a new solution is accepted only in case of a new better solution. We define the transition between permutations as a swap of two jobs. The procedure starts with swapping only adjacent jobs and then gradually moves towards swapping jobs at more and more distant positions. We also ensure that at the beginning, jobs from the front of the queue are selected, as swapping them could have the most significant influence on permutation score. We use *last* variable to track the index of the latest job that was selected for execution in the current solution. Swapping jobs after this index cannot change the result, hence is it dispensable to simulate those permutations. Lastly, we reset the procedure each time after a better solution is found. We run the hill-climbing optimisation for at most $N$ steps (we use 5000 steps in the simulations).

**Candidates for the initial permutation** The initial permutation for optimisation comes from a set of initial candidates. These are permutations with specific properties. Namely, they are sorted according to criteria which makes them potential candidates for global maximum. *We define the set of initial candidates* as jobs sorted:

1. in the order of submission (FCFS),

2. ascending by the requested number of processors,

3. descending by the requested number of processors,

4. ascending by the requested size of burst buffer per processor,

5. descending by the requested size of burst buffer per processor,

6. ascending by the ratio of the requested size of burst buffer per processor to the requested number of processors,

7. descending by the ratio of the requested size of burst buffer per processor to the requested number of processors,

8. ascending by walltime,

9. descending by walltime.

## 3.3. Window-based combinatorial optimisation

A different approach for maximising system utilisation is based on finding an optimal set of jobs, which could be launched immediately. This set should optimise a given objective, which in our case is the same as in Algorithm 6. As the number of all possible combinations is $O(2^n)$, it is vital to limit the search space. One method of doing this is to perform an optimisation on a fixed window of jobs from the front of the waiting queue. A window-based approach with multi-objective optimisation has been already explored by Fan *et al.* in [Fan+19]. They defined two optimisation objectives: compute and burst buffer utilisation. As the optimisation method, they decided to adapt a genetic algorithm which generates a Pareto front. We explore a different custom optimisation technique that can be easily parallelised.

Finding an optimal combination has one major disadvantage compared to a permutation-search–based approach. It requires to incorporate resource allocation into the optimisation method. Searching for optimal permutation is agnostic from resource assignment and hence does not restrict a system to some specific policy nor requires a redesign of an algorithm for a given platform.

---

**Algorithm 7** Window-based combinatorial scheduling

---

1: **procedure** Window($queue$, $N$, $M$, $D$, $\beta$)               ▷ $N$ - max window size
                                                            ▷ $M$ - max age
                                              ▷ $D$ - reservation depth
                                                  ▷ $\beta$ - balance factor

2:     $L_{compute} \leftarrow$ calculate compute load in $queue$

3:     $L_{storage} \leftarrow$ calculate storage load in $queue$

4:     **if** $L_{storage} \leq \beta \cdot L_{compute}$ **then**

5:         The objective is to lexicography maximise (compute, storage, mean wait time)

6:     **else**

7:         The objective is to lexicography maximise (storage, compute, mean wait time)

8:     $W \leftarrow$ pop the first $min(N, |queue|)$ jobs from $queue$

9:     Increase the age of the jobs in $W$

10:     Set the first $D$ jobs in $W$ with age $> M$ as mandatory

11:     Let $C_{best}$ be an empty set                ▷ combination of jobs

12:     $S_{best} \leftarrow (0, 0, 0)$

13:     Let $G_{open}$ be an empty set                ▷ set of combinations

14:     Add a combination made of all jobs in $window$ to $G_{open}$

15:     **while** $G_{open}$ is not empty **do**

16:         Let $G_{unsat}$ be an empty set

17:         **for** $C$ in $G_{open}$ **do**

18:             **if** There are enough resources to run all jobs in $C$ **then**

19:                 $S \leftarrow$ calculate score for $C$

20:                 **if** $S > S_{best}$ **then**

21:                     $S_{best} \leftarrow S$

22:                     $C_{best} \leftarrow K$

23:             **else**

24:                 Add $C$ to $G_{unsat}$

25:         Empty $G_{open}$

26:         **for** $C$ in $G_{unsat}$ **do**

27:             Generate all combinations from $C$ of size $|C| - 1$

28:             Add every combination that contains all mandatory jobs to $G_{open}$

29:     **if** $C_{best}$ is empty **then**

30:         Backfill-sjf($queue$, $D$)

31:     **else**

32:         Launch all jobs in $C_{best}$

33:         Remove jobs in $C_{best}$ from $queue$

34:         Backfill-sjf($queue$, 0)

---

**Fairness**  In our window-based algorithm, presented in Algorithm 7, we use the same optimisation objective as in Section 3.2. However, we even further loosen the fairness guarantees by the introduction of maximum job age. Each job is allowed to spend at most $M$ scheduling runs within the window without being selected to launch. It also applies to the jobs at the front of the queue. Jobs which exceed the age $M$ and are in the first $D$ positions in the queue must be selected for a combination to be accepted.

**Optimisation technique**  We design a custom optimisation technique of searching for an optimal combination within the window. In each step of the loop in line 15, it maintains a set $G_{open}$ of combinations. In the beginning, a single combination made of all jobs inside the window is added. Then, with each step, the size of combinations in $G_{open}$ is decreased by one. If for a given combination $C$, it is known that all jobs in $C$ can be launched immediately, then we compare $C$ to the currently best solution. Any other combination derived from $C$ by taking its subset cannot give a better result. Therefore, we cut this branch of search. If the check for $C$ did not end with a positive result either because the resource allocation is not possible or computation timed out, we generate all combinations based on $C$ of size smaller by one and add them to $G_{open}$.

The whole procedure is computationally intensive. The advantage is that the loop in line 17 could be easily parallelised. A single step of the loop in line 15 has at most $\binom{n}{\lfloor n/2 \rfloor}$ combinations to check. Checking a single combination is a computationally intensive task, hence we propose to assign an $\epsilon$ timeout (1 second in our case) for checking a single combination. Then, for a window of size $n$, $k$ core CPU, checking all combinations should take approximately $\epsilon \cdot \lceil \binom{n}{\lfloor n/2 \rfloor} / k \rceil$ seconds in a single step of the loop in line 15.

**Combination check**  Finding whether there is enough processors for a given combination could be checked with a single *if* statement. Checking for burst buffer allocation may be more complicated depending on a target platform. For this purpose, we decided to use Integer-Linear-Programming. The below linear program finds allocations of burst buffers for jobs in the combination $C$. Let $x_{i,j}$ be an integer variable with the following semantics: $i$-th jobs in the combination $C$ was allocated with $bb_i$ bytes in $j$-th storage node $x_{i,j}$ times. Let $A_j$ denote currently available capacity in $j$-th storage node.

$$0 \leq x_{i,j} \leq M \quad \forall_i \forall_j \tag{3.1}$$

$$\sum_j x_{i,j} = C_i \quad \forall_i \tag{3.2}$$

$$\sum_i bb_i \cdot x_{i,j} \leq A_j \quad \forall_j \tag{3.3}$$

Equation (3.1) bounds the possible values of $x$. In particular, the lower bound is essential. Equation (3.2) ensures that for every job in the window, every processor is assigned with the requested burst buffer size. Equation (3.3) limits burst buffer allocation to currently available space in every storage node.

We utilise the Z3 Theorem Prover [MB08; Mic20] to implement our integer-linear-program.

At the end of the scheduling, we run the Shortest-Job-First backfilling on the whole waiting queue to fill possible gaps of not utilised resources.

## 3.4. Plan-based scheduling

Plan-based scheduling for burst buffers was introduced by Zheng *et al.* in [Zhe+16]. The general idea of plan-based scheduling is to create an execution plan for all jobs in the waiting queue. Naturally, only a few jobs could be launched immediately. Others are reserved with resources at different points in time for the duration of their walltimes. The reservation cannot overlap.

An execution plan is an ordered list of jobs with their scheduled start times and assigned resources. The easiest way to create the execution plan is to iterate over the queue of jobs and for each job find the earliest point in time when sufficient resources are available. The quality of the obtained plan depends on the order of the jobs. It is, therefore, possible to define an optimisation objective and do a search over permutations of the jobs similar to the algorithm outlined in Section 3.2. Zheng *et al.* used simulated annealing to minimise i.a. mean waiting time of all jobs in the plan.

We further investigate and extend the plan-based approach to scheduling with simulated annealing optimisation. First of all, we extend the reservation schema with the reservations for burst buffer requests. This way we obtain awareness of burst buffers in the algorithm. Secondly, we apply several modifications to the simulated annealing. The most important is the introduction of a set of initial permutation candidates used to generate the best and worst initial score. These scores allow us to set optimal initial temperature for the annealing. As as result, our algorithm is capable of finding the optimal permutation and execution plan in only $N \cdot M + |\text{initial set}| = 189$ iterations, compared to 8742 ($\lceil 100 \log_{0.9}(0.0001) \rceil$) iterations in the original publication. The number of iterations has principal importance in online job scheduling, where time available for scheduling is highly limited. The complete list of introduced novelties is following:

1. Set of candidates for the initial permutation to simulated annealing

2. Variable self-adjusting initial temperature for simulated annealing

3. Reservation of resources for jobs at the front of the waiting queue to ensure the fairness property

4. Exhaustive search over all permutations for small sizes of the queue

5. Optimisation objective based on the starting time

6. Generalisation of the waiting time optimisation objective

Algorithm 8 presents the complete plan-based policy.

**Simulated annealing**   Simulated annealing is a probabilistic optimisation method for approximating the global optimum of a given function. It was constituted by Kirkpatrick, Gelatt and Vecchi in [KGV83]. It does not require to compute gradients or Hessians but depend only on the target function values. It is also very flexible in terms of space of solutions, which in our case are permutations. The disadvantage of the canonical simulated annealing is the requirement for the optimised function to evaluate to a single positive number. Hence, simulated annealing is suitable mainly for single-objective optimisation. It is possible to modify simulated annealing schema for multi-objective optimisation, however finding an initial temperature becomes then a significantly more complicated problem. For this reason, we did not use simulated annealing in Section 3.2.

---
**Algorithm 8** Plan-based scheduling
---

1: **procedure** Plan(*queue*, $D$, $r$, $N$, $M$)         ▷ $D$ - reservation depth
        ▷ $r$ - cooling rate
        ▷ $N$ - cooling steps
        ▷ $M$ - constant temperature steps

2:     $J \leftarrow$ pop the first $D$ jobs from *queue*
3:     FCFS($J$)
4:     Remove launched jobs from $J$
5:     Reserve resources for jobs in $J$ in the future

6:     **if** *queue* is small **then**         ▷ exhaustive search
7:         Iterate over all possible permutations;
8:         $P_{best} \leftarrow$ find the jobs permutation with the lowest score

9:     **else**         ▷ simulated annealing
10:         $P_{best} \leftarrow$ find a permutation with the lowest score among the set of initial candidates
11:         $P_{worst} \leftarrow$ find a permutation with the highest score among the set of initial candidates
12:         Create execution plans for $P_{best}$ and $P_{worst}$
13:         $S_{best} \leftarrow$ calculate score for $P_{best}$ based the the execution plan
14:         $S_{worst} \leftarrow$ calculate score for $P_{worst}$ based the the execution plan
15:         **if** $S_{best} \neq S_{worst}$ **then**
16:             $T \leftarrow S_{worst} - S_{best}$         ▷ initial temperature
17:             $P \leftarrow P_{best}$
18:             **for** $N$ iterations **do**
19:                 **for** $M$ iterations **do**
20:                     $P' \leftarrow$ swap two jobs at random positions in $P$
21:                     Create execution plans for $P$ and $P'$
22:                     $S \leftarrow$ calculate score for $P$ based the the execution plan
23:                     $S' \leftarrow$ calculate score for $P'$ based the the execution plan
24:                     **if** $S' < S_{best}$ **then**
25:                         $S_{best} \leftarrow S'$
26:                         $P_{best} \leftarrow P'$
27:                         $S \leftarrow S'$
28:                         $P \leftarrow P'$
29:                     **else if** $random(0,1) < e^{(S-S')/T}$ **then**
30:                         $S \leftarrow S'$
31:              $T \leftarrow r \cdot T$    $P \leftarrow P'$

32:     Filler($P_{best}$)
33:     Remove reservations for jobs in $J$
34:     Push back $J$ at the front of *queue*

---

**Optimisation objective**  We specify two separate optimisation objectives. A concrete plan-based implementation should choose one of them.

1. *Sum of job waiting times raised to power $\alpha > 0$*

2. *Starting time of the latest job*

In both cases, the goal is to minimise the objective.

Let $Q$ denote the job queue. The first objective, for a given plan, could be formally defined as (notation from Section 1.8.1):

$$\sum_{j \in Q} (W_j)^\alpha$$

In [Zhe+16], there were considered two related functions: mean job waiting time ($\alpha = 1$) and mean squared job waiting time ($\alpha = 2$). Our definition is a generalised version of waiting time objective. Indeed, this objective should work for any real $\alpha > 0$. Smaller values, such as $\alpha = 1$, provide much flexibility to the optimisation method. However, as Zheng *et al.* observed, this objective cannot prevent one job from being postponed arbitrarily by many other jobs as long as their total wait time remains the same. In an extreme case, this situation may even lead to the starvation of a job. The higher the value of $\alpha$, the more penalised is the objective function for arbitrary delaying of jobs. It should reduce the number of outliers, but at a possible cost of worse mean waiting time of all jobs. Furthermore, for a given plan minimising a mean is equivalent to minimising a sum. Then we may remove a redundant division. As a consequence, this may require to adjust the initial temperature if it is a fixed value, but as outlined above, we use a variable self-adjusting initial temperature.

Our second objective function is formally defined as:

$$\max_{j \in Q} s_j$$

Initially, we wanted to use a maximum of finish times as a proxy to system utilisation. However, we realised that in case of submission an extraordinarily long-narrow job to the queue, the optimisation method may generate execution plans with significant gaps between scheduled jobs. This result is caused by the fact that the extraordinarily long-narrow job would dominate the finish time of all other jobs.

**Fairness**  As indicated above, plan-based scheduling does not ensure strict fairness criteria such as a reservation depth $D$ in backfilling. Higher values of $\alpha$ prevent extensive delaying of jobs, reduce the number of outliers, but leave lesser flexibility for optimisation. Increasing the value of $\alpha$ parameter may be indeed perceived as a soft mechanism for ensuring fairness. For lower values of $\alpha$, we propose to supplement scheduling with a traditional reservation of resources, which is described in lines 2-5 of algorithm 8.

**Exhaustive search**  Similarly as in Section 3.2, we run the search over all possible permutations of jobs for small queue sizes. Creating an execution plan for all jobs is significantly more computationally intensive operation than a greedy attempt to launch jobs from a permutation. Hence, we limit the exhaustive search to $|queue| \leq 5$, which results in 120 permutations to test.

**Initial permutation and temperature**  We use the same set of candidates for an initial permutation to simulated annealing as defined at the end of Section 3.2. However, in this algorithm, we select both the best and the worst permutations from the set. The scores for these permutations are used to set the initial temperature, which is shown in line 16. We set it as a difference between worst and best score (energy; cost), according to a suggestion from [Ben04]. When the best and worst score obtained from the set of candidates are the same, then we omit the optimisation as in practice, we found it unlikely to yield any improvement.

**Simulated annealing configuration**  Instead of running the simulated annealing until a threshold temperature is achieved, we set a constant number of iterations. Assuming that an optimal initial temperature is found, random transition probabilities should decrease from values near 1 to probabilities of the magnitude 0.001. The constant number of iterations makes scheduling time more predictable and limits the number of iterations with negligible random transition probabilities. We follow the schema of simulated annealing from [Zhe+16] and implement it as two nested loops. The outer loop is responsible for decreasing the temperature exponentially by a cooling rate $r = 0.9$. The inner loop performs iterations with a constant temperature. We set the number of iterations to $N = 30$ and $M = 6$. Additionally, we set a timeout of 20 seconds for the whole simulated annealing.

**Further improvements**  It is also possible to enhance plan-based scheduling schema with a method of predicting actual runtimes of jobs and use them for reservations instead of walltimes. However, we have not found any publications on this approach.

# Chapter 4

# Simulation results

At first, in Section 4.1, we discuss the methodology of performing experiments and evaluating results. In the following six sections, we present the results of numerous experiments.

Section 4.2 is dedicated to a study an influence of resource reservations on the overall efficiency of backfilling. This study is conducted based on both Alloc-Only and IO-Aware models that are described in detail in Section 2.4.

In Sections 4.3, 4.4, 4.5, we seek optimal parameters for the algorithms presented in Sections 3.2, 3.3, 3.4, respectively. This search is based only on the IO-Aware model.

Subsequently, in Section 4.6, we select the most efficient scheduling policies for comparison. We also perform a more in-depth study again based on Alloc-Only and IO-Aware models.

Lastly, Section 4.7 visualises an impact of the *backfilling reservation depth* on the efficiency of selected policies.

## 4.1. Methodology

To capture the whole complexity of differences between investigated scheduling algorithms, we present our results using four metrics (described below) and three main types of plots: mean values, quantiles and tail distributions (scatter plots). The mean value plots are the most popular and easiest to interpret visualisation technique. We use this kind of data representation to enable an easy comparison to other research works. Quantile and tail distribution plots allow us to perform an in-depth analysis of scheduling algorithm behaviour.

**Datasets** From the KTH-SP2-1996-2.1-cln log, described in Section 2.3, we derived four workloads. Two of them are converted full logs for both Alloc-Only and IO-Aware models. The differences between them are covered in Section 2.4.

The other two were created by splitting the KTH-SP2-1996-2.1-cln log into sequential periods, again for both job models. The log was collected over eleven months. To be exact, it was almost 333 days. We divided the log into 16 equal periods, where a single period is approximately three-week-long (20.81 days). The number of jobs within a period varies among instances from 1222 up to 2940. Such divided workloads were then used to verify scheduling algorithms results. They were utilised in Sections 4.2 and 4.6 to obtain figures 4.10-4.13, 4.22-4.25, 4.42-4.45, 4.54-4.57 The distribution of burst buffer requests and other parameters for the split workloads are the same as for the full workloads.

**Metrics** All the results are compared based on four metrics: mean waiting time, mean turnaround time, mean slowdown and mean bounded slowdown. Feitelson in [Fei01] considers three values for the bounded slowdown time threshold $\tau$: 1 second, 1 minute and 10 minutes. He points out that the bounded slowdown with longer time threshold achieves better convergence of confidence intervals. Therefore, we decided to use the longest 10 minutes threshold for the bounded slowdown, yet to avoid short jobs being underrepresented, we also present the standard slowdown. For all the mentioned metrics, time on the Y-axis is given in seconds. Additionally, for Section 4.2, we also present makespan and system utilisation. All of these metrics were defined in Section 1.8.

**Abbreviations**   In all presented charts, we use abbreviations to denote scheduling policies. Every abbreviation starts with the name of a scheduling algorithm and **ends with the *reservation depth* parameter**. In the middle of an abbreviation, there might appear other scheduling policy dependent parameters, for example *maxutil-2-1* or *plan-sum-0*.

**Confidence interval**   All presented bar plots are visualised with confidence intervals. For all figures, we use 95% confidence level. The confidence intervals are generated with a bootstrapping method among all jobs that are included by a given workload or workload part.

**Boxenplot**   Boxenplots are a generalisation of boxplots, which are particularly useful for visualising large datasets. There were introduced as *Letter-value plots* in [HWK17] by H. Hofmann, H. Wickham and K. Kafadar.

Each level in a boxenplot, represent a specific **quantile**. In our case, we use boxenplots of depth 4. That is the boxenplots present the following quantiles:

- first 32-quantile (1/32)
- first hexadecile (1/16)
- first octile (1/8)
- first quartile (1/4)
- median (1/2)
- last quartile (3/4)
- last octile (7/8)
- last hexadecile (15/16)
- last 32-quantile (31/32)

However, the full boxenplots are visible only in figures 4.4-4.7 and 4.16-4.19, where a logarithmic scale was applied. It is due to the fact that we deal with highly skewed data. For other presented boxenplots, we decided to use a linear scale as schedules studied in Sections 4.3-4.7 gave results of the same magnitude. In consequence, we have well represented the higher quantiles but truncated the quantiles below the median.

How to interpret boxenplots? It might be well explained based on Figure 4.34b and schedules: plan-sum-1 (green) and plan-square-1 (red). The plan-sum-1 schedule has lower positioned last quartile, last octile and last hexadecile than plan-square-1, which means that plan-sum-1 has a better distribution of waiting time for 93.75% of jobs with the lowest waiting time. However, for the last quantile, we observe a reversal in this tendency. That is plan-square-1 shows the better distribution for jobs with the lowest waiting time between 93.75% and 96.875%. Therefore, it is not straightforward to distinguish which scheduling policy achieves the best results.

In general, the lower are the levels of a given boxenplot the better is a distribution of a given statistic for a majority of jobs.

**Tail distribution**   In order to visualise outliers and distribution of tail, we use strip plot - a scatter plot where one variable is categorical. For a given statistic, we select a set of jobs with the highest score for each scheduling policy. For instance, in Figure 4.4c, we chose 8000 jobs with the highest waiting times separately for each schedule. This way of representing data has a substantial benefit in the fact that there is no data aggregation applied, which may sometimes reveal surprising results as in Figure 4.4c.

For all tail distribution plots in Section 4.2, we take 8000 data points per schedule. For all other in Sections 4.3-4.7 plots we use only 4000 points.

## 4.2. Impact of burst buffer reservations on backfilling

In this section, we study how resource reservations impact the efficiency of scheduling. We selected four scheduling algorithms presented previously in Section 3.1. In order from the most restrictive to the least restrictive these are:

1. FCFS without backfilling (fcfs, blue, Algorithm 1)

2. FCFS EASY-backfilling without future burst buffer reservations (no-future-1, orange, Algorithm 3)

3. FCFS EASY-backfilling (backfill-1, green, Algorithm 4)

4. FCFS Greedy filling (filler, red, Algorithm 2)

**Relation between makespan and system utilisation**   At first, we would like to emphasise again that in this dissertation, we consider parallel, non-preemptive and rigid job model in an online scheduling environment. These assumptions have a fundamental meaning for makespan and utilisation. In a properly prepared online workload, new job arrivals should be observed until the very end of a simulation. Otherwise, we would observe the accumulation of jobs in the waiting queue. In Figure 4.2, we may see that schedules backfill-1 and filler result in the same makespan, while fcfs and no-future-1 end significantly later. It suggests that **fcfs and no-future-1 were accumulating jobs in the queue until the end of the workload submission was achieved**. A confirmation of this thesis may be found in the compute load plot in Figure 4.8, which shows how many compute nodes are requested by jobs in the waiting queue. Both backfill-1 and filler schedules manage to keep the compute load relatively low over time until the end of the submission is reached. However, for fcfs and no-future-1, the compute load is being accumulated up to the point when it achieves its peak at the end of the jobs submission. After that point, it is gradually unloaded.

System utilisation depends on makespan. All compared experiments are performed on the same workload. Therefore, the numerator in the utilisation Equation (1.10) is constant. The only thing that differs is the denominator, which is the makespan. Hence, both compute and storage utilisation is a reverse function of makespan. This observation may be not valid in the IO-Aware model, where the job running time may differ due to I/O congestion.

In the later Sections 4.3-4.7, we will not present makespan and system utilisation as they do not differ by more than 1% for schedules other than fcfs and no-future-1.

**Job waiting time and starvation**   Waiting time for the Alloc-Only model is presented in Figure 4.4. There are a couple of interesting observations to find. First of all, no-future-1 has almost the same mean waiting time as fcfs, but significantly lower median and substantially higher dispersion (Figure 4.4b). Additionally, we could see the enormous dispersion in the tail of the waiting time distribution. For fcfs jobs with the highest waiting time are concentrated on values near to $10^7$ seconds, but for no-future-1 they are spread from about $0.5 \cdot 10^7$ to $4.3 \cdot 10^7$. For both of these schedules, there is a gap in the plot between the X-axis and scatter column. That is because we present only 8000 highest waiting time entries.

The more intriguing observation is the gap in the tail distribution of no-future-1 schedule starting from about $10^7$ seconds. It suggests that there were no jobs with waiting time in this given range. After closer examination of the trace, we found particularly interesting behaviour of this scheduling algorithm. This gap was caused by a short-wide job, that is a job requesting a large number of processors and even larger burst buffer size. Followingly, this job happened to arrive at the front of the waiting queue and was selected as a priority job for backfilling, which means that it received reservation of processors (without burst buffers) before the backfilling started. Consequently, jobs requiring a small number of processors were backfilled before the priority job. However, together with the priority job, they might require more storage capacity than is available in the platform. In consequence, it leads to the delay of the priority job, which results in the delay of all medium-size and large jobs in the waiting queue. That behaviour might repeat in the subsequent scheduling runs if there is a constant stream of small jobs which could be backfilled before the current priority job. It results in the **starvation** of all the medium and large jobs in the queue. Illustrative examples of the job starvation in the no-future-1 schedule is visible in a Gantt chart in Figure 4.1. Due to the starvation effect part of the simulated cluster remains just idle, which is represented by the large empty gaps in the Gantt chart.

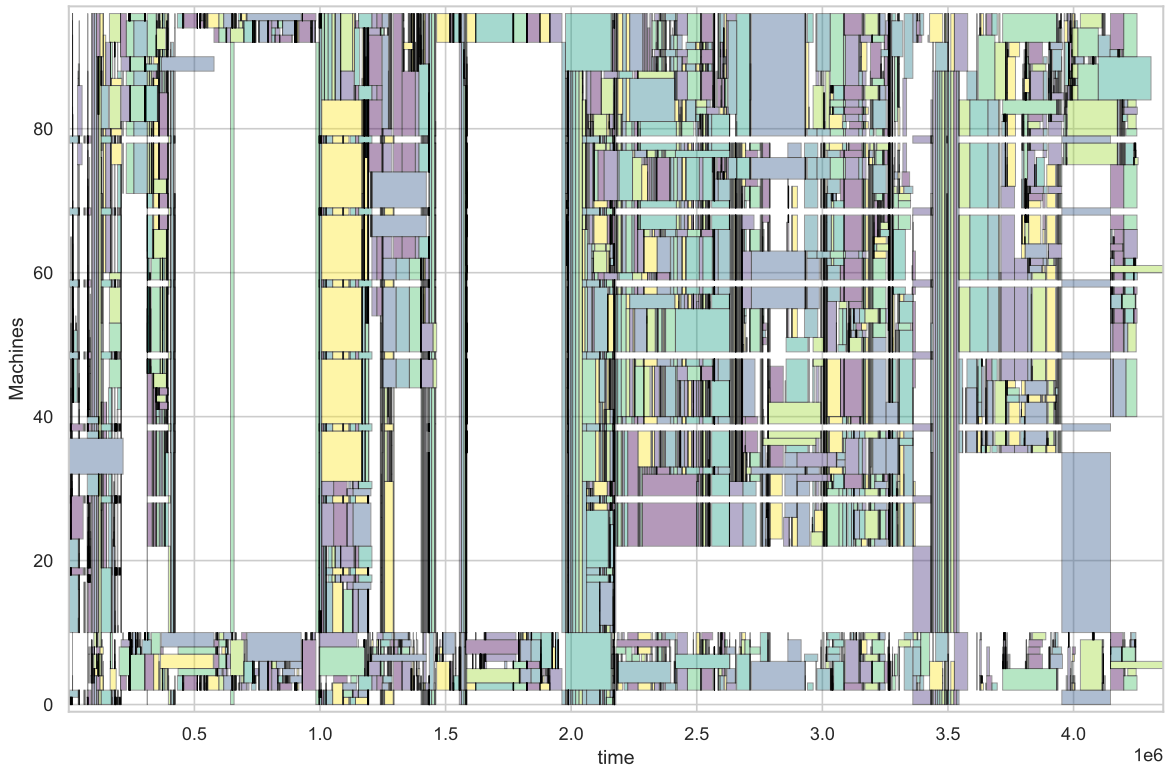

Figure 4.1: Gantt chart of the first 3500 jobs executed by the no-future-1 scheduling policy in the Alloc-Only model (there are 96 compute nodes in the simulated platform).

If we examine Figure 4.16, which is presenting waiting time statistics for the IO-Aware model, we do not find clear evidence for starvation. The mean waiting time of no-future-1 is smaller than fcfs. Plots of quantiles show similar dependencies for both models. There is no noticeable gap in the tail distribution. We also run these experiments for the split workloads. Their results are presented in Figure 4.10 and Figure 4.22, respectively, for Alloc-Only and IO-Aware model. In Figure 4.10, no-future-1 for all the workload parts presents lower mean waiting time than fcfs, but with a varying difference. However, for Figure 4.22, we see one case of a higher mean waiting time (Workload 13). We conclude that the significant starvation effect rarely occurs for no-future-1 schedule, but when it does, it may considerably deteriorate scheduling performance.

In terms of backfill-1 and filler, we can observe that for both models, the first one has a smaller mean waiting time, higher median and similar distribution presented by other quantiles. Tail distribution plots show substantially more outliers for filler, which is expected as it is not a fair scheduling algorithm by definition.

**Turnaround time, slowdown and bounded slowdown**   For turnaround time, we may see similar relations between scheduling algorithms as for waiting time. Naturally, the median has higher values, so all quantiles are distinguishable in the quantiles plots.

In terms of slowdown and bounded slowdown, both backfill-1 and filler have relatively low means and medians close to 1. The most dominate difference can be noticed for fcfs in the tail distribution. For slowdown, there are multiple outlying values, which do not occur for bounded slowdown. It suggests that fcfs could significantly delay short jobs. Nevertheless, no-future-1 have extensively more outliers for both slowdown and bounded slowdown than fcfs. Those conclusions apply for both models.

**Waiting queue load**   Compute load and storage load plots, shown in Figure 4.8, Figure 4.9 Figure 4.20 and Figure 4.21, present the summarised number of processors and burst buffer size requested by all jobs that are in the waiting queue at a given moment. The meaning of lines on the plots is following:

- Orange - total amount of resources available in the platform

- Blue - summarised resource requirements from jobs in the queue

- Dotted green - mean resource requirement (may differ from other plots due to number rounding)

- Red vertical lines at the bottom - queue reset events (situation when the the waiting queue was empty)

We may observe, by an increase in the compute load, that for backfill-1 and filler schedules, the last job submission happened just before the simulation end. Therefore, we conclude that our workloads were correctly calibrated for online scheduling.

Compute and storage load plots are visually similar for corresponding schedules. It may suggest that compute load and storage load are highly correlated in the workloads.

**Conclusion**   Our results indicate that **backfilling without future reservations of burst buffers (no-future-1) may significantly deteriorate the efficiency of job scheduling** in terms of waiting time and slowdown. In extreme cases, **backfilling without future reservations of burst buffers may result in even higher mean waiting time than first-come-first-served scheduling without any backfilling**. Furthermore, **backfilling without future reservations of burst buffers is not a fair sharing policy** as it effects with an enormous number of arbitrarily delayed jobs. Its tail distribution of bounded slowdown shows a very high number of outliers.

### 4.2.1. Alloc-Only model

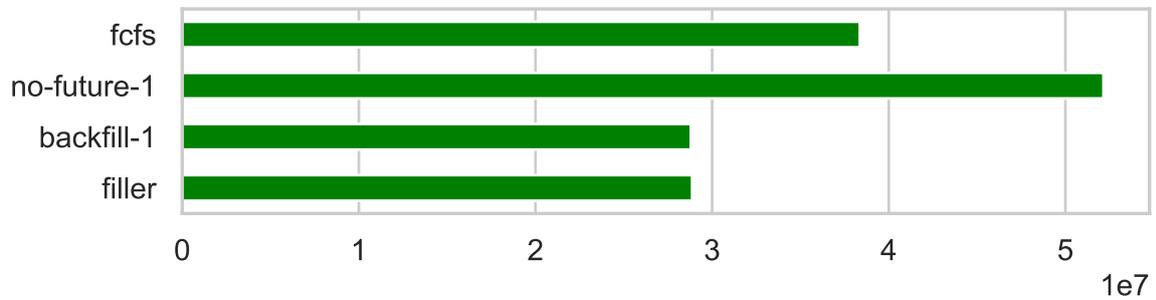

Figure 4.2: Makespan

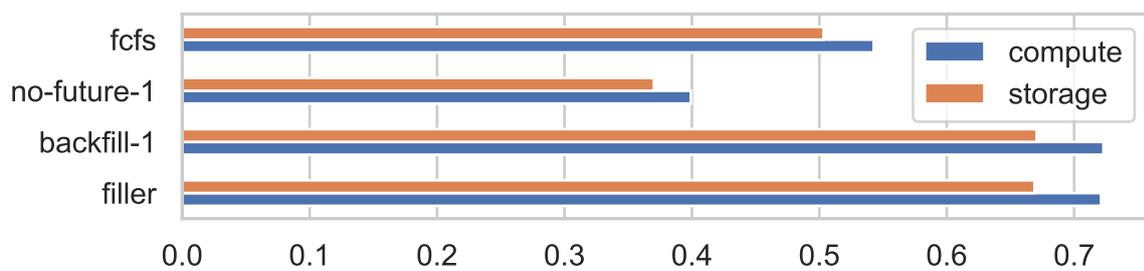

Figure 4.3: System utilisation

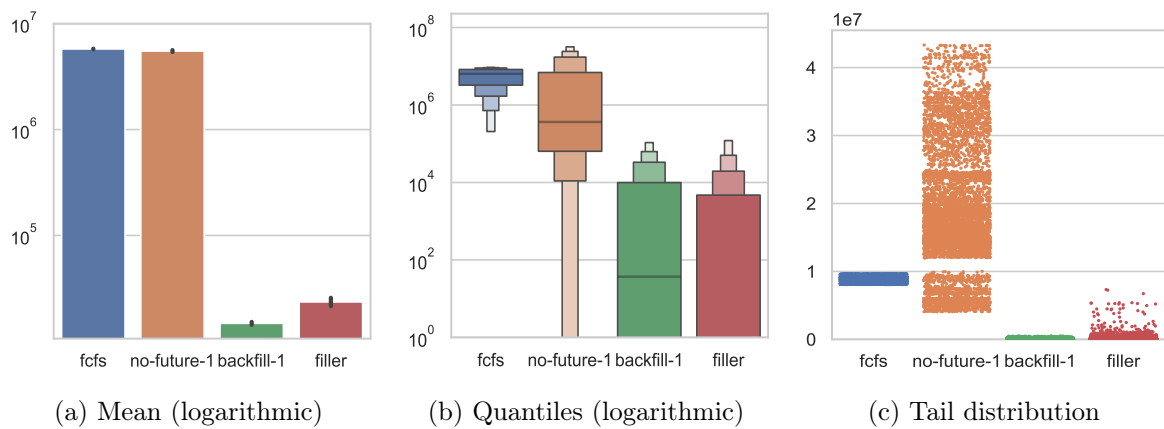

(a) Mean (logarithmic)   (b) Quantiles (logarithmic)   (c) Tail distribution

Figure 4.4: Waiting time

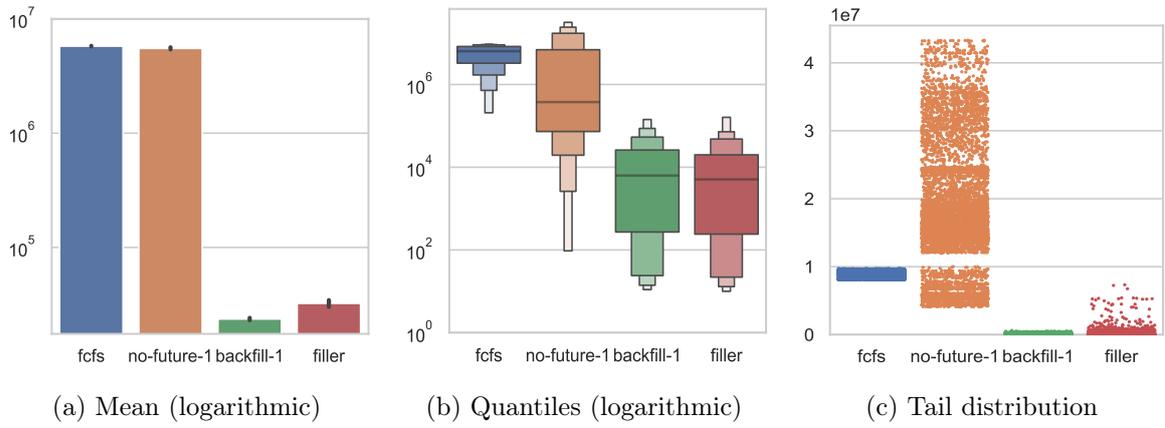

(a) Mean (logarithmic)   (b) Quantiles (logarithmic)   (c) Tail distribution

Figure 4.5: Turnaround time

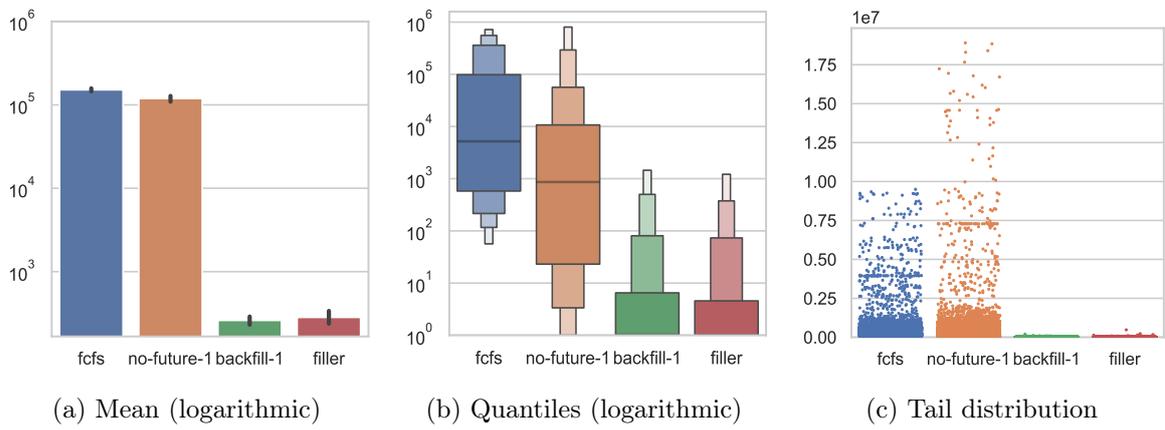

(a) Mean (logarithmic)   (b) Quantiles (logarithmic)   (c) Tail distribution

Figure 4.6: Slowdown

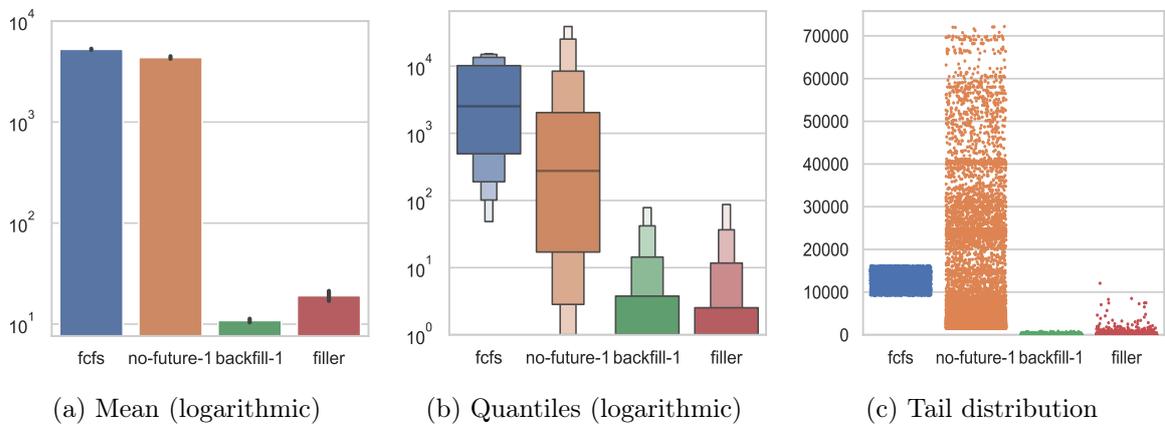

(a) Mean (logarithmic)   (b) Quantiles (logarithmic)   (c) Tail distribution

Figure 4.7: Bounded slowdown

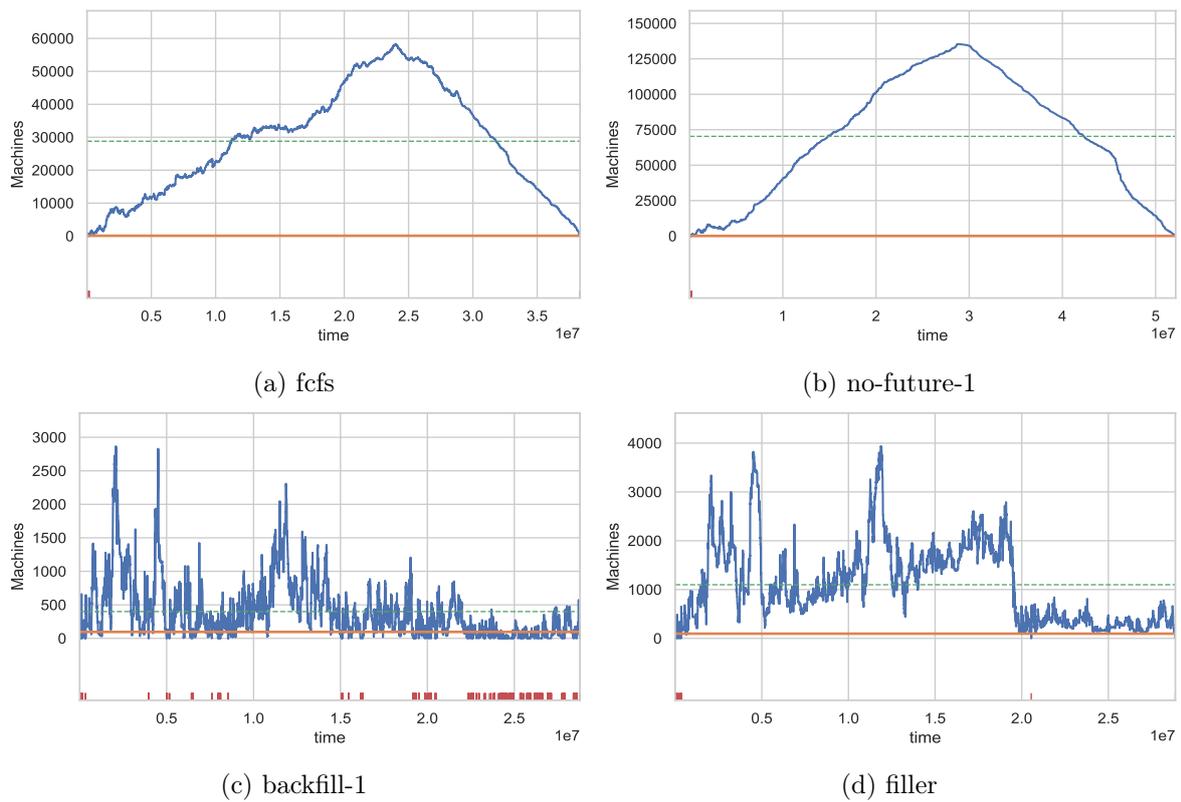

(a) fcfs  (b) no-future-1

(c) backfill-1  (d) filler

Figure 4.8: Compute load (different axes ranges)

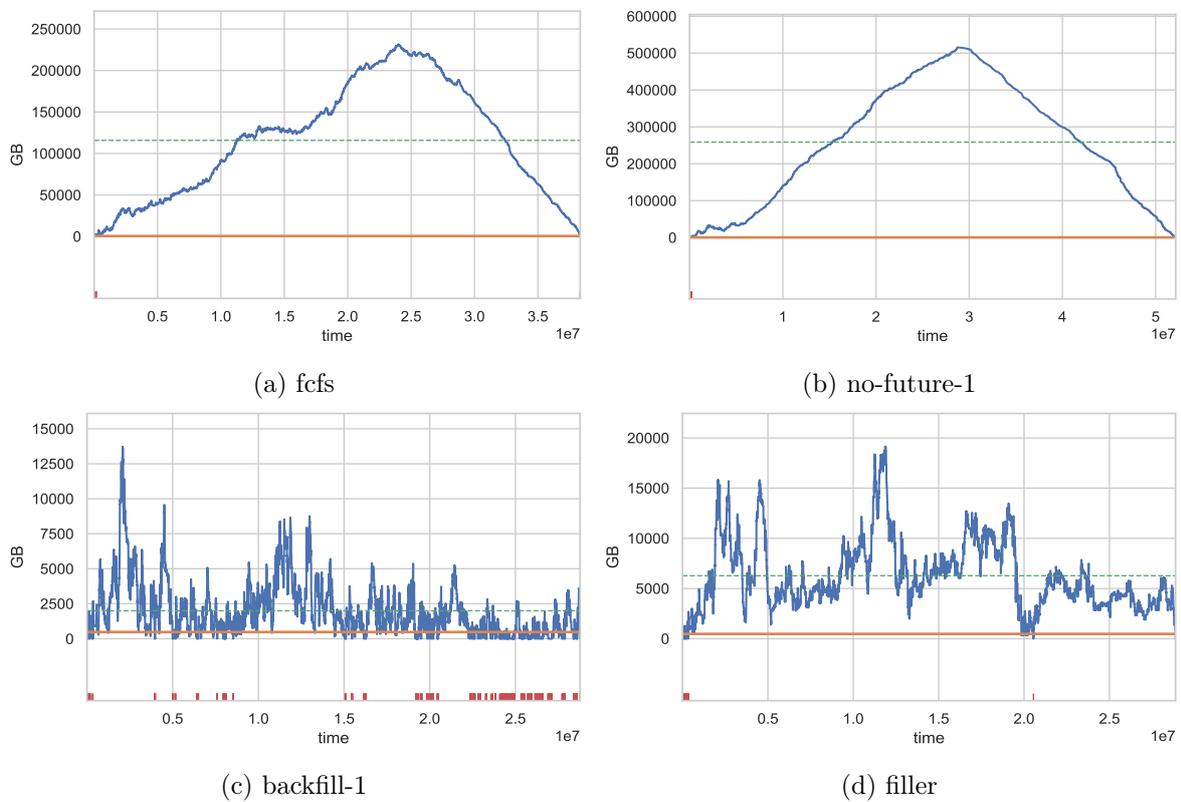

(a) fcfs  (b) no-future-1

(c) backfill-1  (d) filler

Figure 4.9: Storage load (different axes ranges)

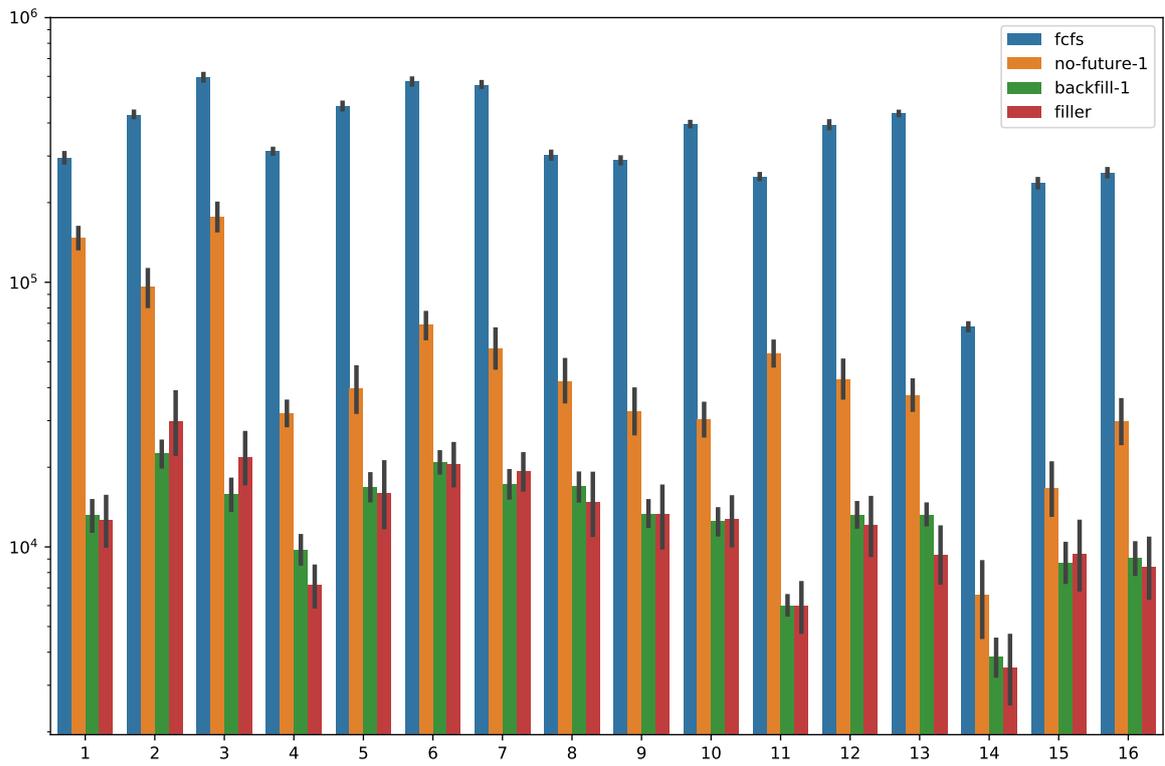

Figure 4.10: Mean waiting time of the split workload

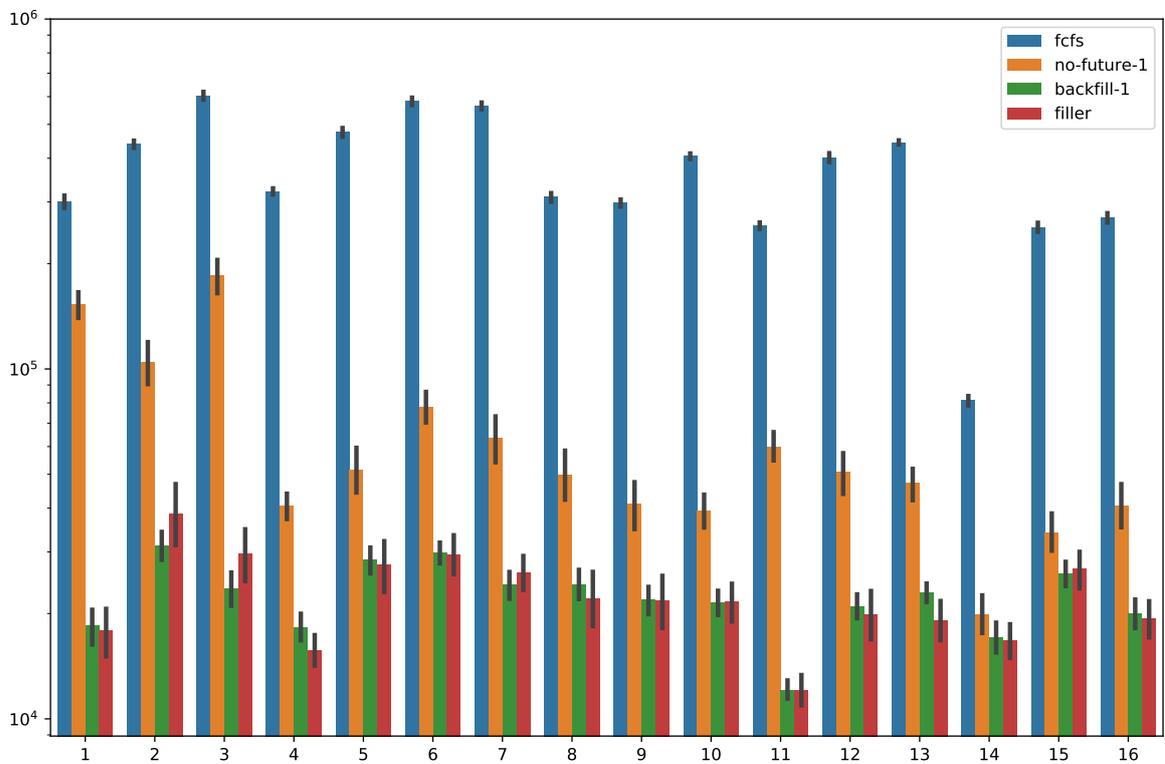

Figure 4.11: Mean turnaround time of the split workload

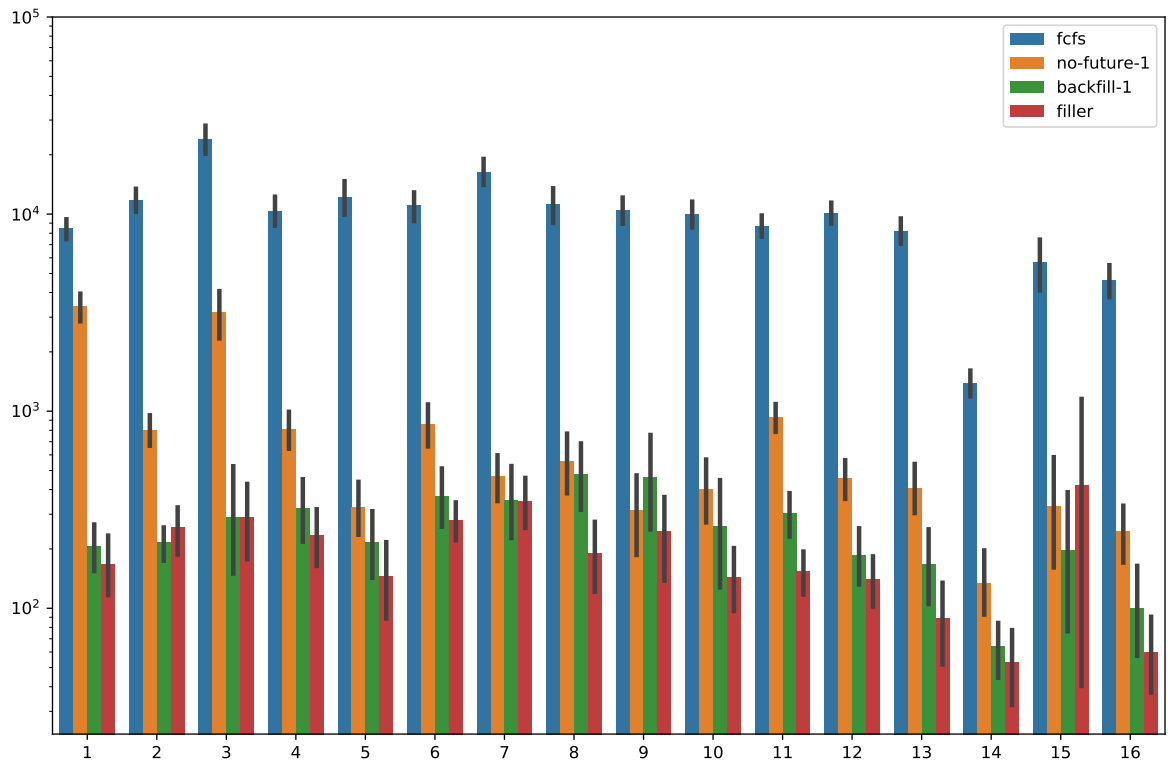

Figure 4.12: Mean slowdown of the split workload

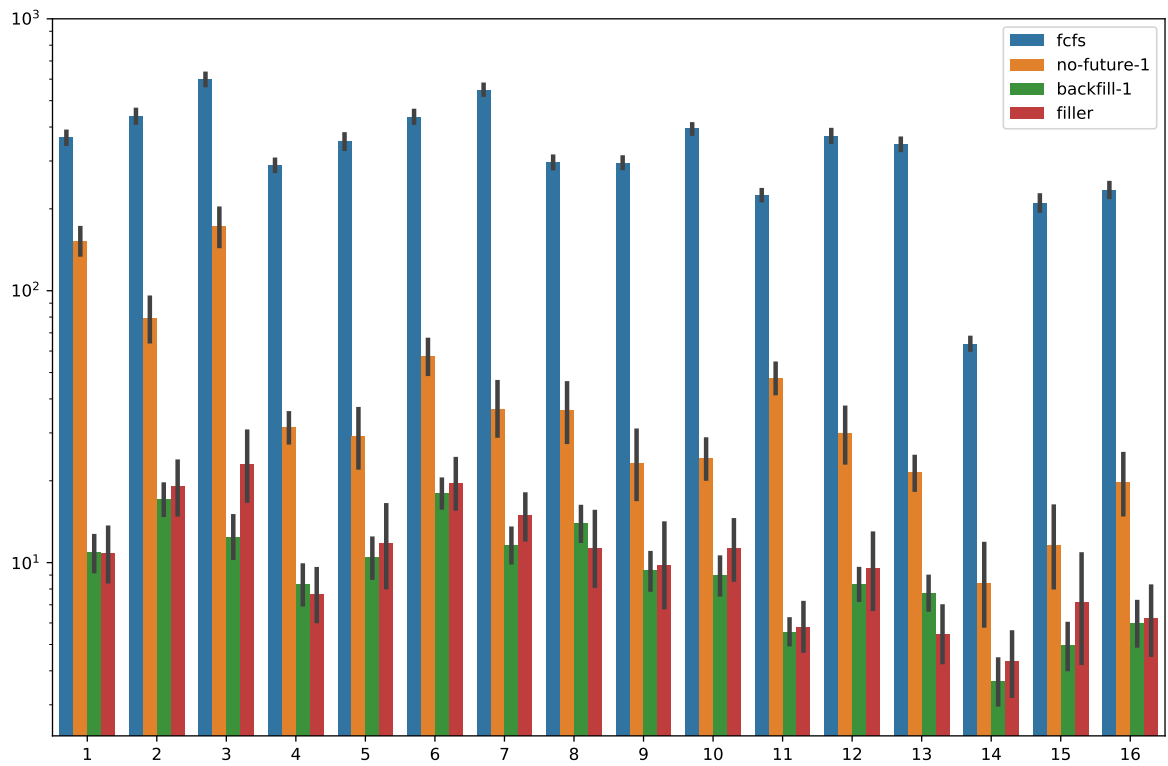

Figure 4.13: Mean bounded slowdown of the split workload

### 4.2.2. IO-Aware model

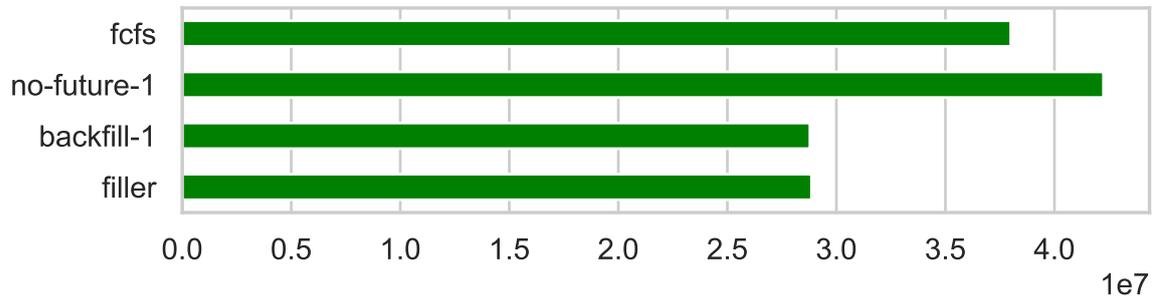

Figure 4.14: Makespan

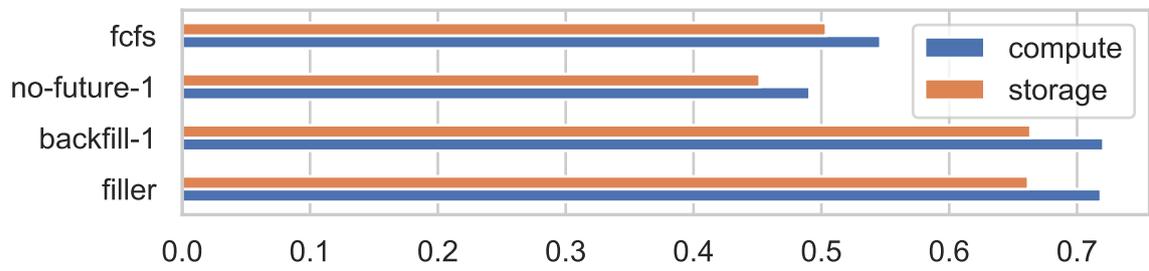

Figure 4.15: System utilisation

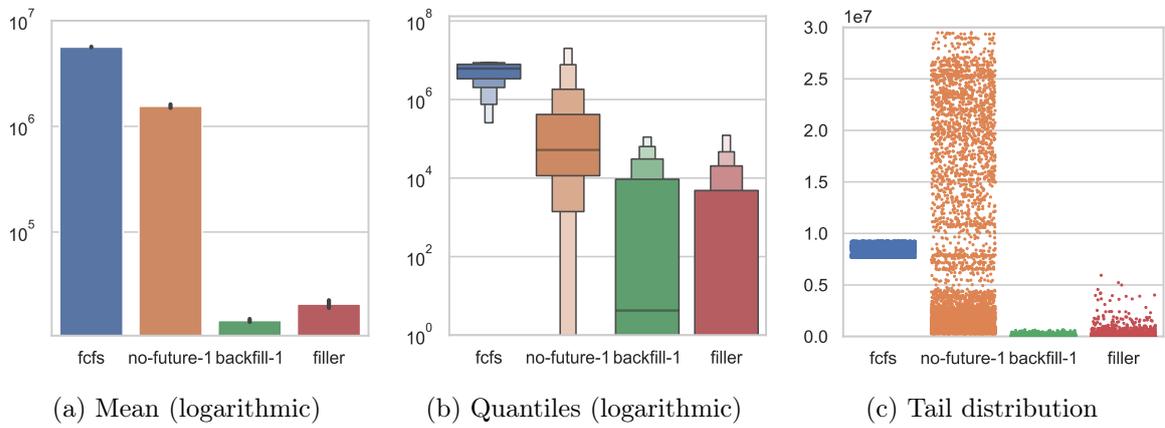

(a) Mean (logarithmic)  (b) Quantiles (logarithmic)  (c) Tail distribution

Figure 4.16: Waiting time

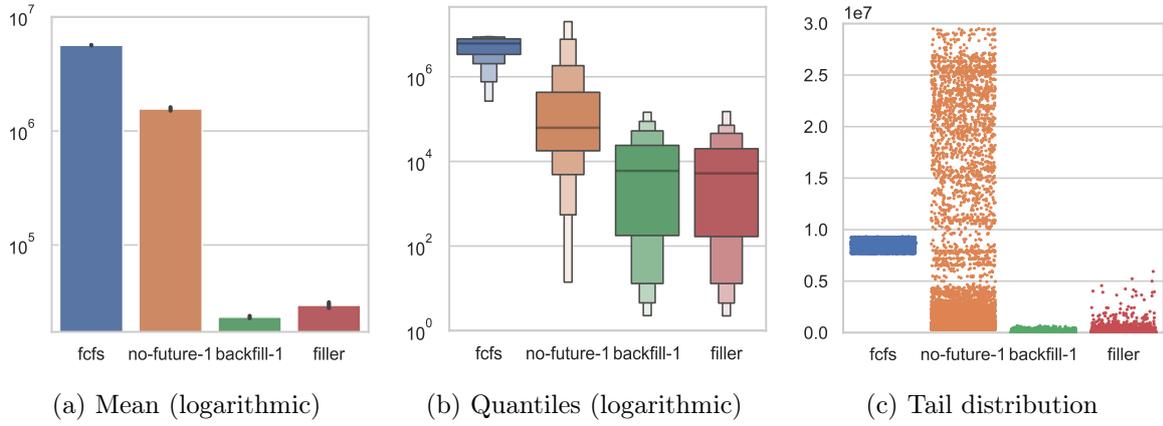

(a) Mean (logarithmic)　　(b) Quantiles (logarithmic)　　(c) Tail distribution

Figure 4.17: Turnaround time

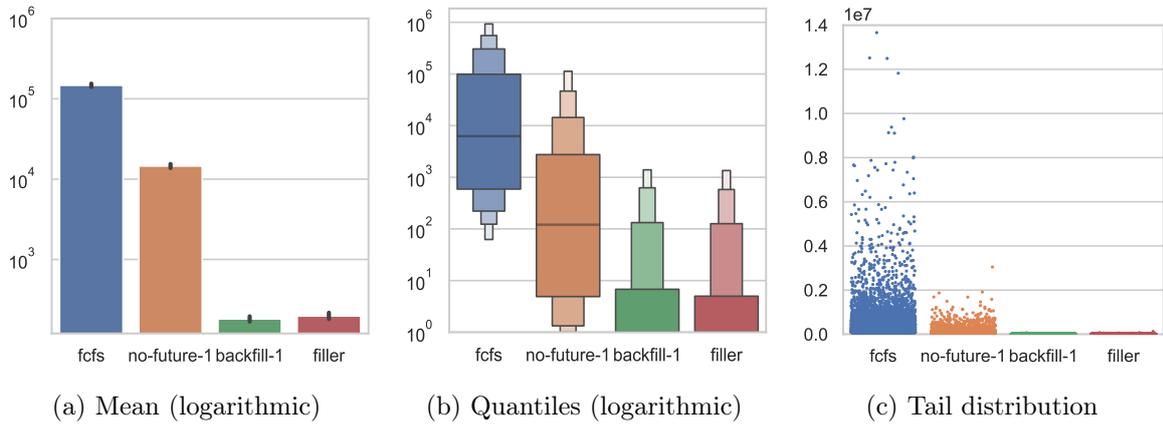

(a) Mean (logarithmic)　　(b) Quantiles (logarithmic)　　(c) Tail distribution

Figure 4.18: Slowdown

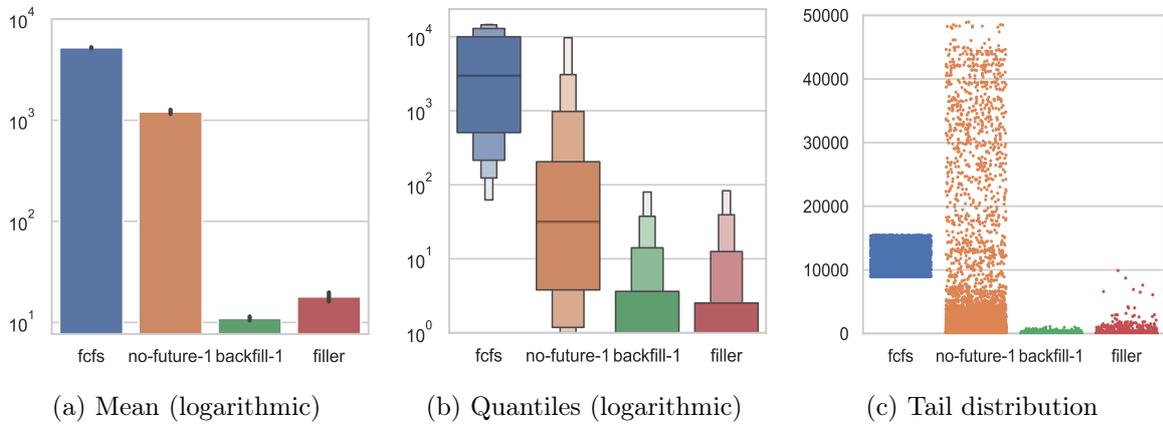

(a) Mean (logarithmic)　　(b) Quantiles (logarithmic)　　(c) Tail distribution

Figure 4.19: Bounded slowdown

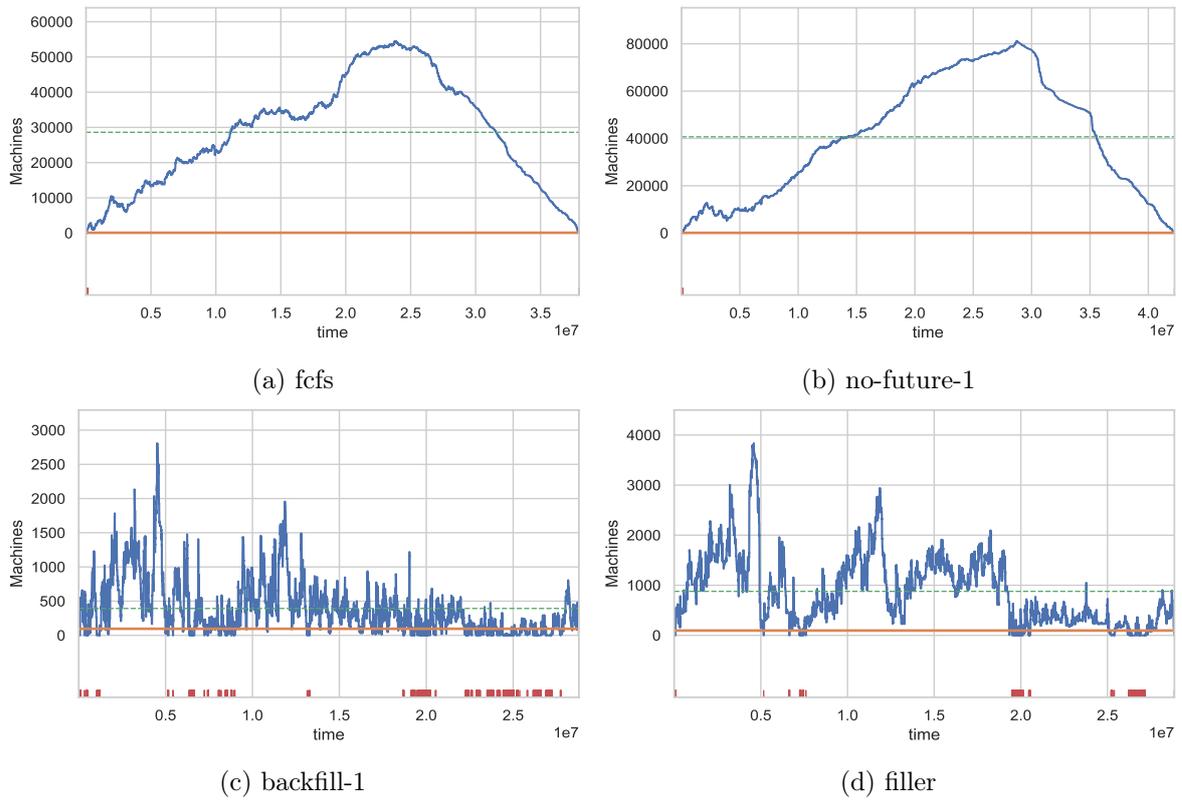

(a) fcfs

(b) no-future-1

(c) backfill-1

(d) filler

Figure 4.20: Compute load (different axes ranges)

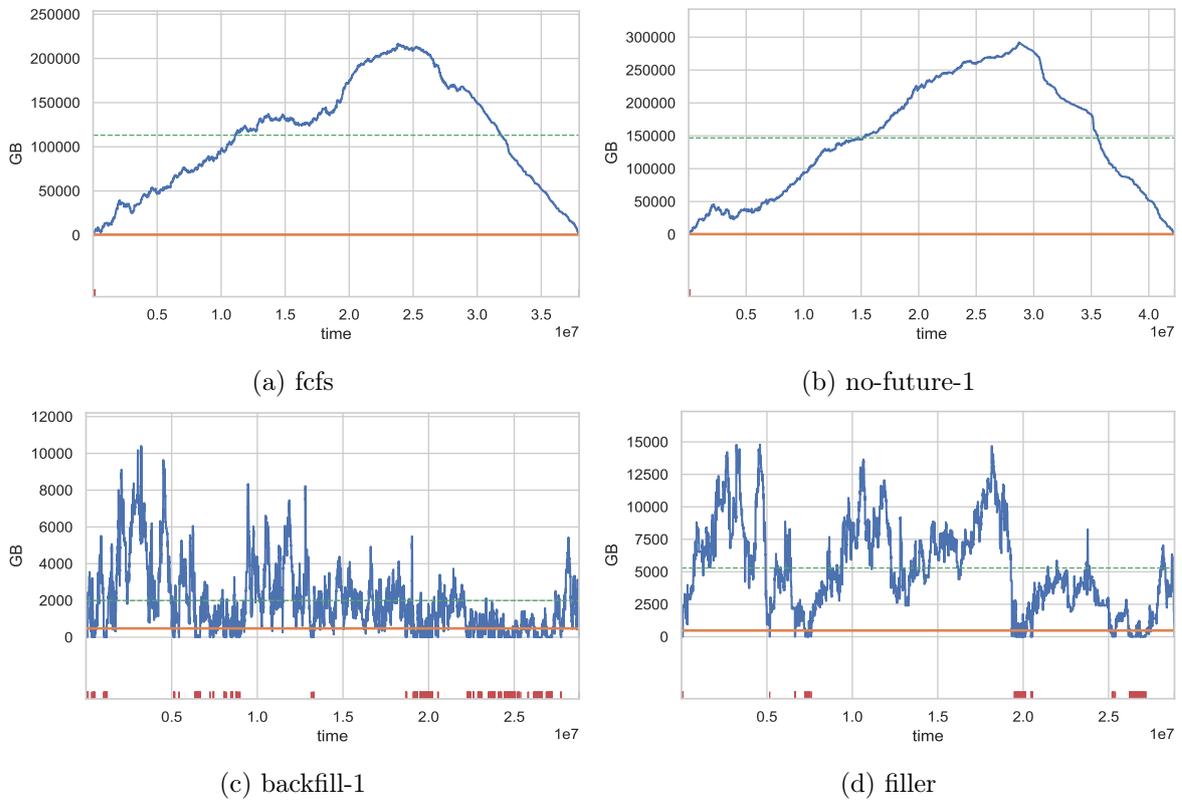

(a) fcfs

(b) no-future-1

(c) backfill-1

(d) filler

Figure 4.21: Storage load (different axes ranges)

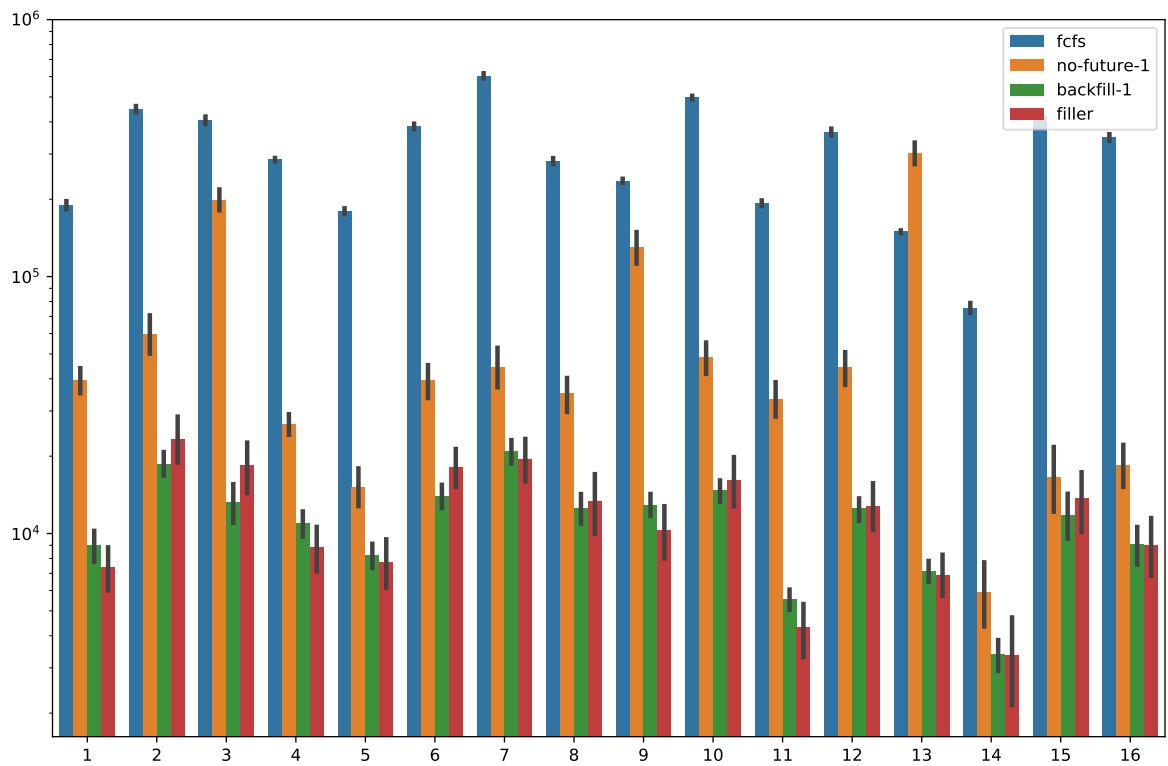

Figure 4.22: Mean waiting time of the split workload

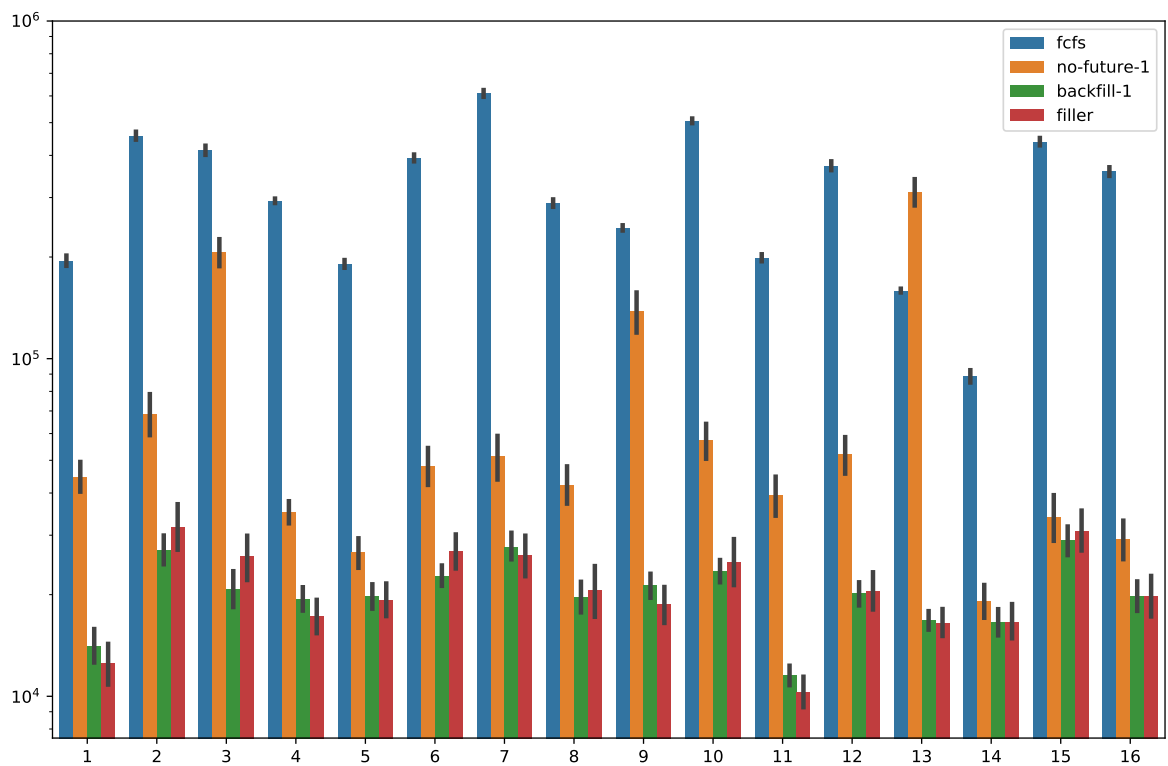

Figure 4.23: Mean turnaround time of the split workload

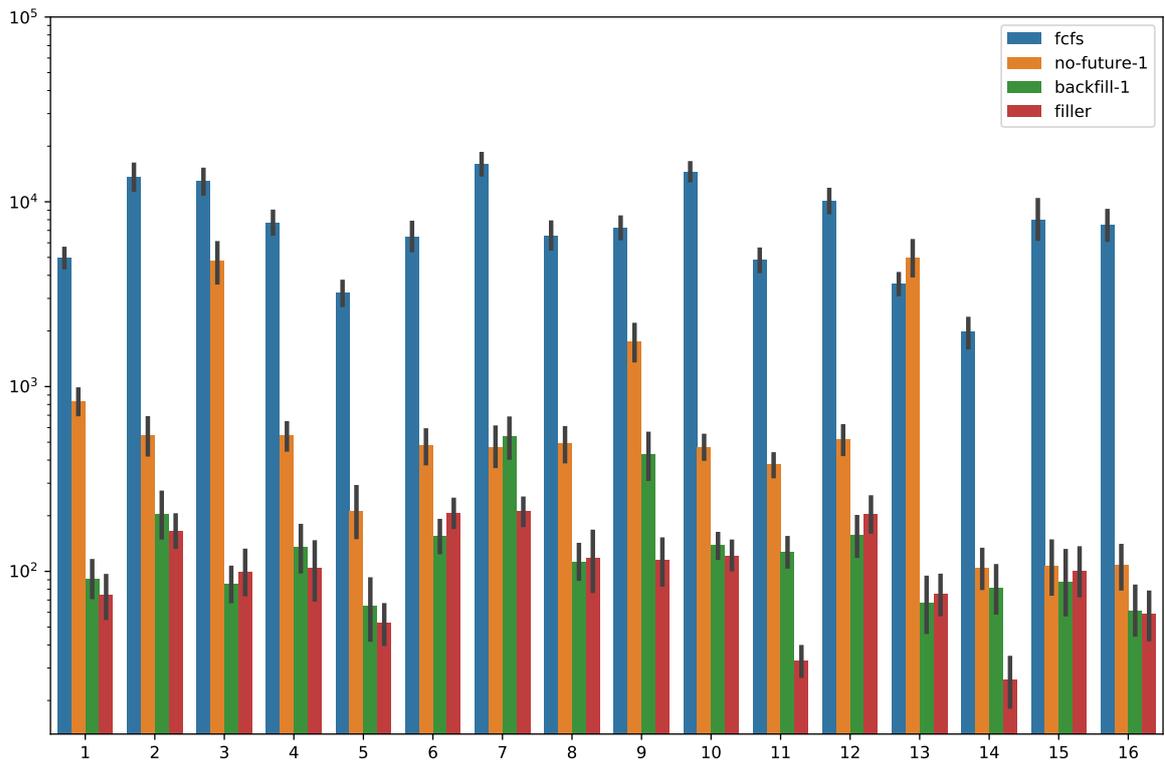

Figure 4.24: Mean slowdown of the split workload

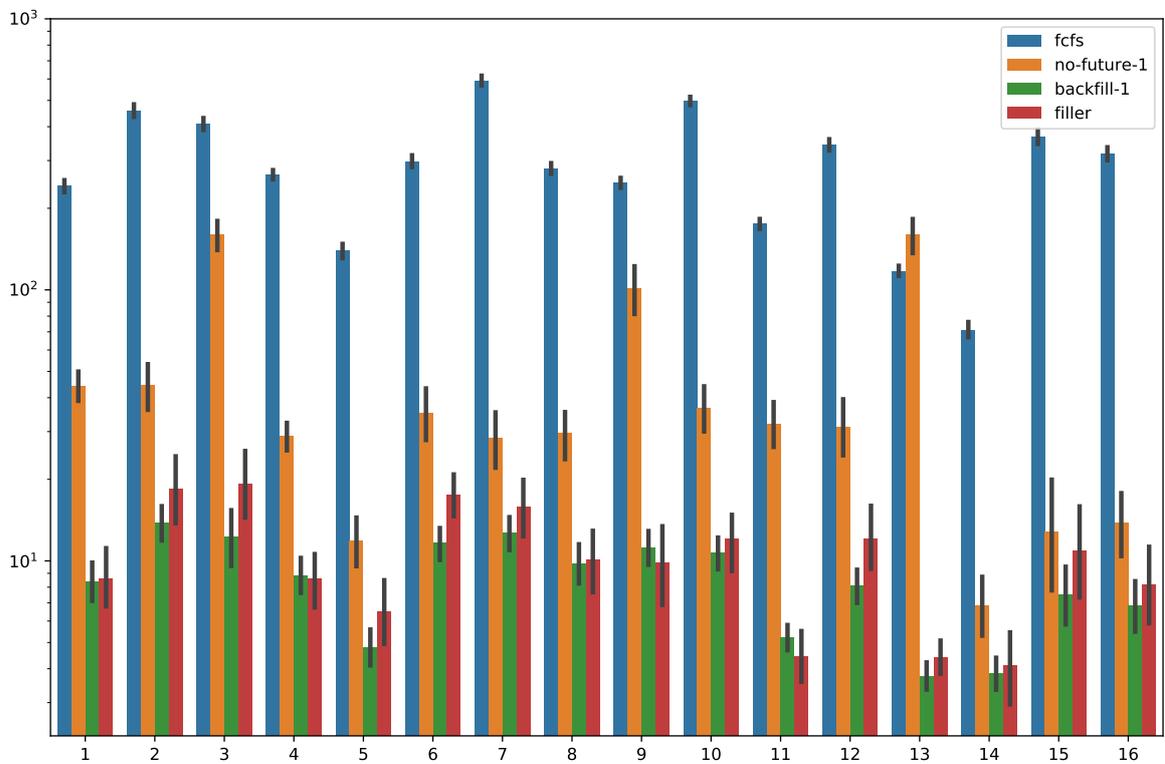

Figure 4.25: Mean bounded slowdown of the split workload

## 4.3. System utilisation maximising scheduling

We dedicate this section for finding the optimal *balance factor* $\beta$ for Maxutil algorithm shown in listing 6. As described in Section 3.2, the value of $\beta$ determines how likely is the algorithm to reduce either compute or storage load. We test four different values of $\beta$: 0.5 (maxutil-0.5-1, green), 1 (maxutil-1-1, red), 1.5 (maxutil-1.5-1, purple), 2 (maxutil-2-1, brown). We compare these schedules with two canonical scheduling algorithms: First-Come-First-Served backfilling (backfill-1, blue) and Shortest-Job-First backfilling (backfill-sjf-1, orange). For all experiments we set the *reservation depth D* to 1, which ensures the fairness property. The *maximum number of steps for hill climbing N* is 5000 in each case, as stated in Section 3.2.

From Figure 4.26, we see that all scheduling policies except backfill-1 have similar mean waiting time. Mean waiting time of backfill-1 is slightly higher than the others. The quantile plot also presents similar scores for all schedules, except for backfill-sjf-1, for which the last 32-quantile is higher than other. In terms of tail distribution, maxutil-1.5-1 and maxutil-2-1 have more significantly outlying waiting times than others. However, all the above differences are relatively small, so we find that waiting time does not indicate the optimal value for $\beta$.

Plots for turnaround time in Figure 4.27 presents the same relations as the waiting time.

For slowdown, backfill-sjf-1 shows significant improvement in all compared statistics.This improvement, however, does not transit to bounded slowdown. It shows only a slight decrease in mean in favour of the backfill-sjf-1. The schedule backfill-1 achieves higher mean and quantile statistics than the other schedules. For all values of $\beta$, maxutil schedules present negligible differences in bounded slowdown based statistics.

For the IO-Aware model with the full workload, we did not found any vital differences in scheduling results for the Maxutil algorithms depending on the *balance factor* $\beta$. Therefore, for comparing Maxutil with other algorithms in the following sections, we use the default value of $\beta = 1$.

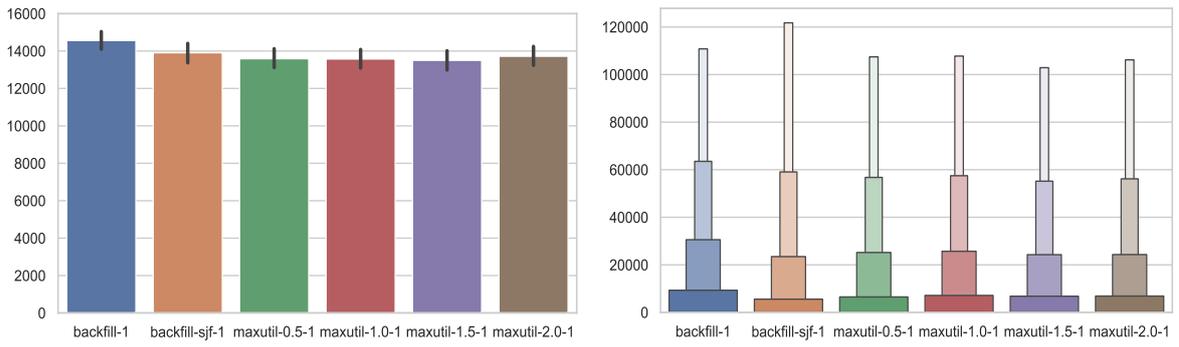

(a) Mean          (b) Quantiles

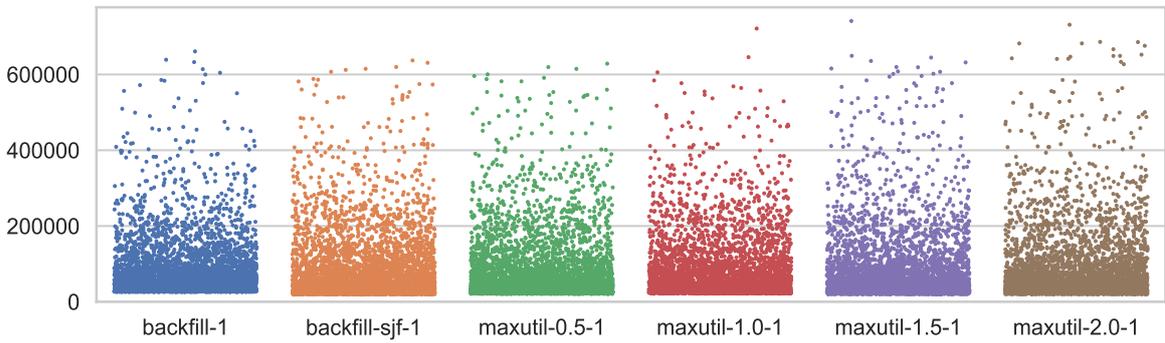

(c) Tail distribution

Figure 4.26: Waiting time

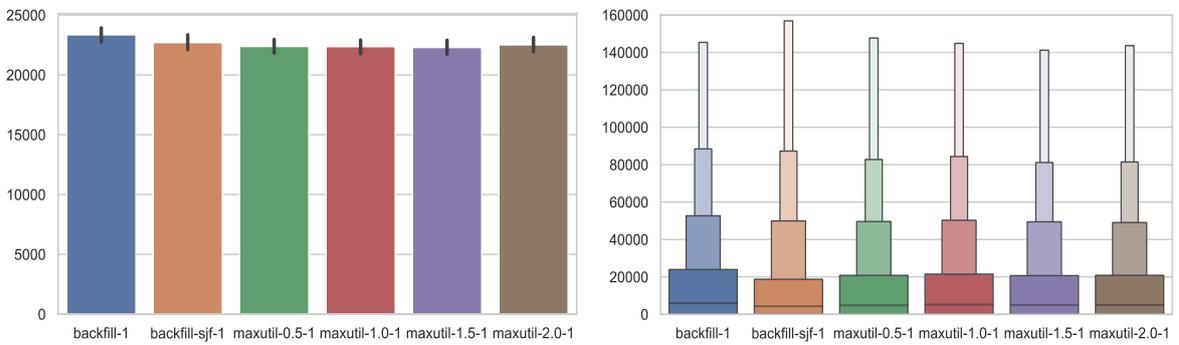

(a) Mean          (b) Quantiles

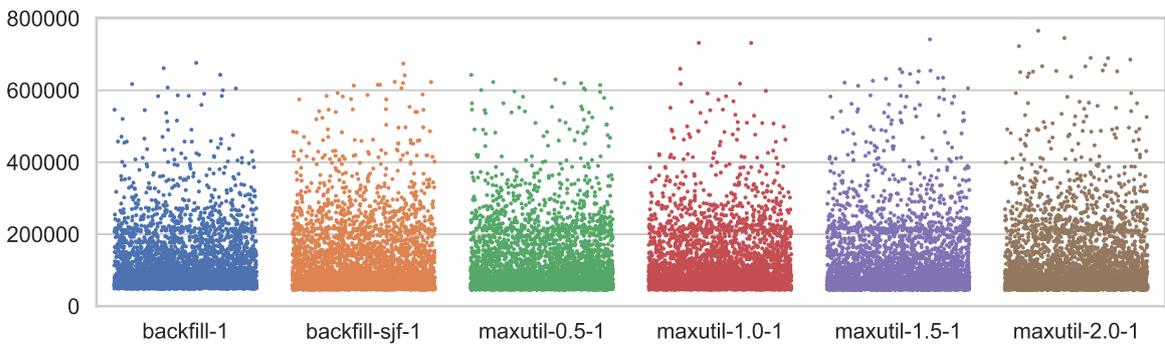

(c) Tail distribution

Figure 4.27: Turnaround time

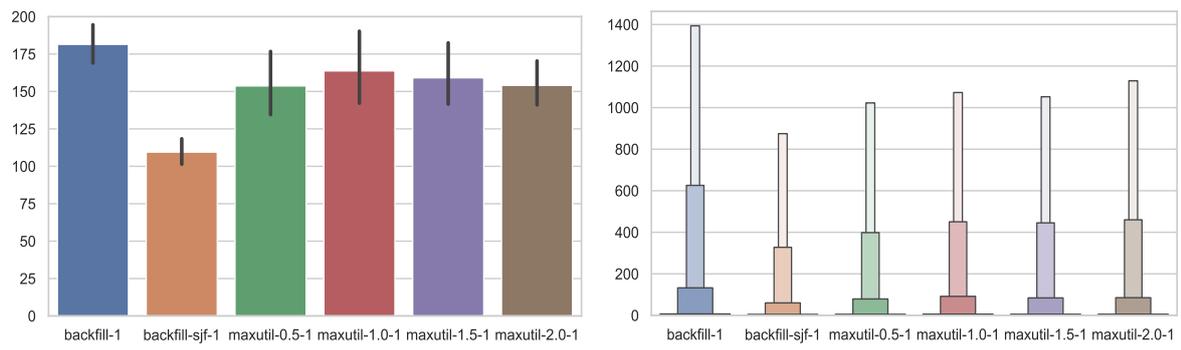

(a) Mean | (b) Quantiles

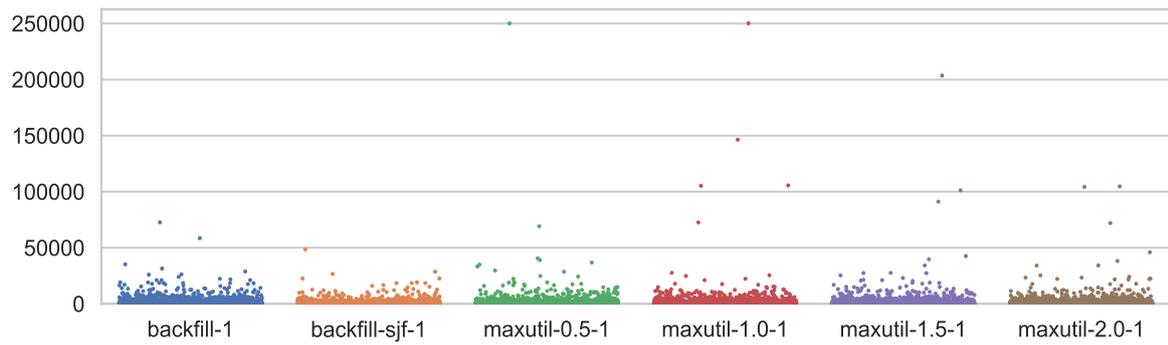

(c) Tail distribution

Figure 4.28: Slowdown

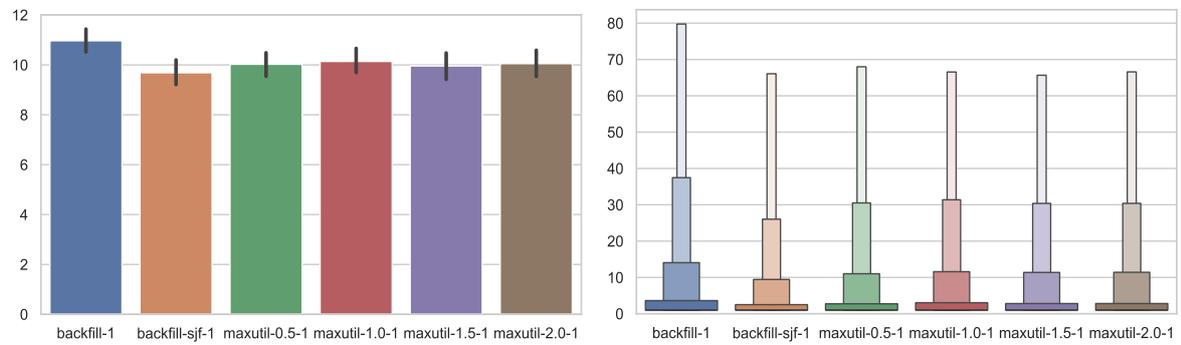

(a) Mean | (b) Quantiles

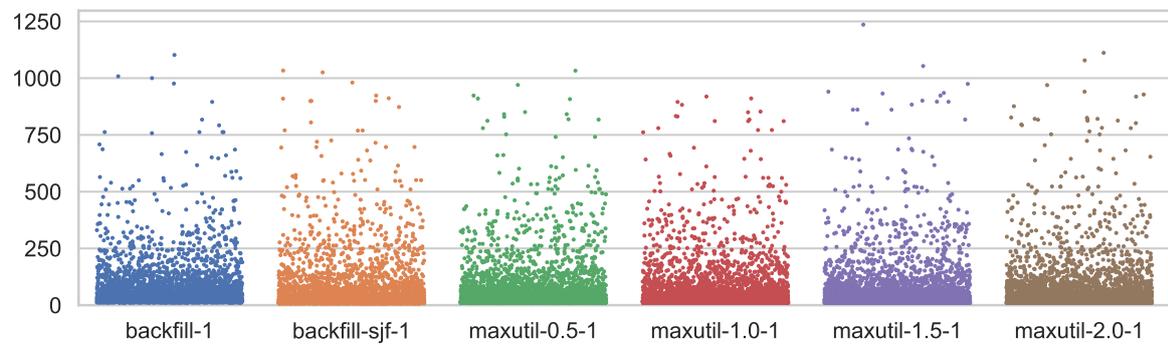

(c) Tail distribution

Figure 4.29: Bounded slowdown

## 4.4. Window-based combinatorial scheduling

The main parameter in Algorithm 7 is the *max window size $N$*. A different approach to window-based scheduling was explored in [Fan+19]. Experiments there were performed for window size ranging from 1 to 20. We were, unfortunately, unable to test our approach for the window size greater than 10 due to a multi-threading issue in one of our software dependencies - Z3 Solver. Therefore, we only test the influence of resource reservation on the algorithm for a fixed $N = 10$. We compare the following scheduling policies:

- FCFS backfilling (backfill-1, blue)

- SJF backfilling (backfill-sjf-1)

- Window-based combinatorial scheduling with a reservation of the first job in the queue (window-1, green)

- Window-based combinatorial scheduling without resource reservations (window-0, red)

In terms of waiting time and turnaround time, window-1 presents almost identical statistics as backfill-sjf-1. Whereas, window-0 show better results in quantiles, but much higher mean. This schedule is less restrictive than window-1 as it does not implement any fairness mechanism. The value of the mean could be easily explained by the tail distribution, where we see many outliers for window-0, which may evince starvation of jobs.

For bounded slowdown, there can be distinguished ordering between the algorithms. The best results are achieved by backfill-sjf-1. Then window-1 shows better results than backfill-1. While window-0 has significantly worse mean than all other algorithms.

For the following sections, we select window-1 for comparison as window-0 does not provide fair sharing.

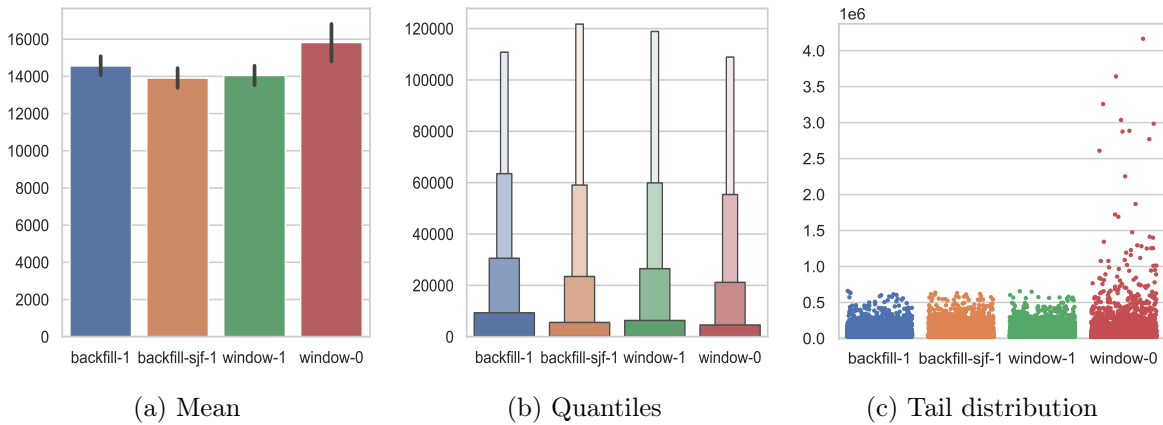

(a) Mean      (b) Quantiles      (c) Tail distribution

Figure 4.30: Waiting time

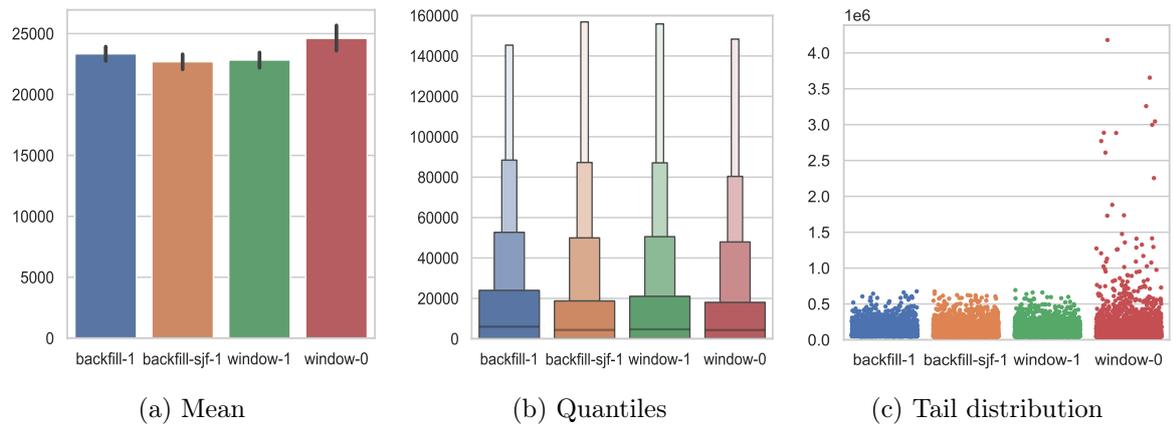

(a) Mean     (b) Quantiles     (c) Tail distribution

Figure 4.31: Turnaround time

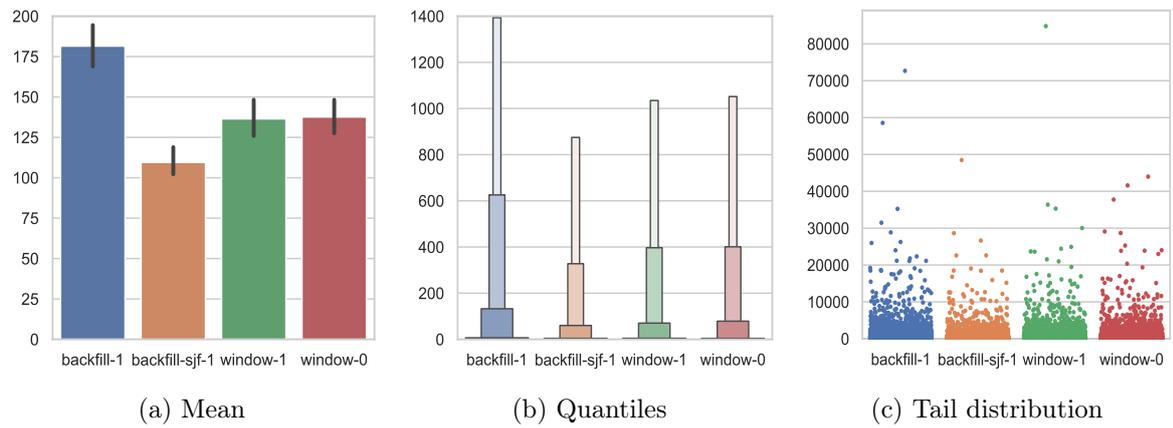

(a) Mean     (b) Quantiles     (c) Tail distribution

Figure 4.32: Slowdown

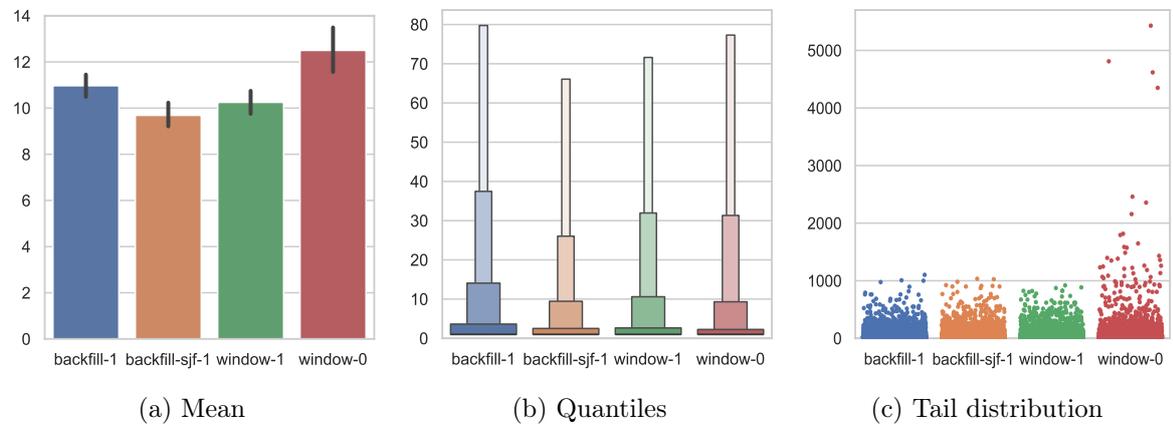

(a) Mean     (b) Quantiles     (c) Tail distribution

Figure 4.33: Bounded slowdown

## 4.5. Plan-based scheduling

In this section, we study multiple variants of plan-based scheduling introduced in Section 3.4. Our goal is to test how resource reservations and optimisation objectives influence the overall efficiency of scheduling. As mentioned in Section 4.1, we present results of simulations for only IO-Aware model. We compare the set of the following policies:

- FCFS EASY-backfilling (backfill-1, blue)

- SJF EASY-backfilling (backfill-sjf-1, orange)

- Plan-based (plan-sum-1, green)
  - reservation of resources for the first job ($D = 1$)
  - minimise the sum of waiting time ($\alpha = 1$)

- Plan-based (plan-square-1, red)
  - reservation of resources for the first job ($D = 1$)
  - minimise the sum of squared waiting time ($\alpha = 2$)

- Plan-based (plan-start-1, purple)
  - reservation of resources for the first job ($D = 1$)
  - minimise starting time of the latest job

- Plan-based (plan-sum-0, brown)
  - no reservations ($D = 0$)
  - minimise the sum of waiting time ($\alpha = 1$)

- Plan-based (plan-square-0, pink)
  - no reservations ($D = 0$)
  - minimise the sum of squared waiting time ($\alpha = 2$)

- Plan-based (plan-cube-0, grey)
  - no reservations ($D = 0$)
  - minimise the sum of cubed waiting time ($\alpha = 3$)

- Plan-based (plan-start-0, gold)
  - no reservations ($D = 0$)
  - minimise starting time of the latest job

**Waiting time and turnaround time**   All scheduling algorithms with resource reservations achieve comparable results in mean waiting time. They, however, differ in ordered statistics. Plan-sum-1, plan-square-1 and plan-start-1 are visibly better in the last and last but one quantile than backfill-sjf-1. Further relations between them are hard to distinguish as plan-square-1 has better distribution for the last quantile but worse for others compared to plan-sum-1 and plan-start-1. The tail distribution shows no outstanding results for all of these schedules. The especially interesting observation is that plan-sum-1 and plan-start-1 demonstrate an almost identical outcome. For turnaround time, all these relations remain accurate.

The next subset of scheduling policies with similar results is plan-sum-0 and plan-start-0. As we indicated in Section 3.4, they do not implement any fair sharing mechanism. We can see the confirmation of this fact in Figure 4.34c, where that both expose a significant number of outliers in waiting time statistic. These outliers explain very high mean waiting time achieved by both schedules. For distribution of quantiles, plan-sum-0 and plan-start-0 show better results than policies with reservation but are only comparable with plan-square-0 and plan-cube-0. To summarise, plan-sum-0 and plan-start-0 are not practical candidates for implementation in a real RJMS.

Lastly, plan-square-0 and plan-cube-0 are present considerably lower mean waiting and turnaround time than other algorithms. Compared to scheduling policies with resource reservations, plan-square-0 and plan-cube-0 have a few outstanding values. However it is relatively negligible when compared to the distribution of plan-sum-0 and plan-start-0. Plan-square-0 achieves slightly better results in mean and quantiles than plan-cube-0.

**Slowdown and bounded slowdown**   For mean slowdown, the best results are presented by backfill-sjf-1, plan-sum-0, plan-square-0, plan-cube-0 and plan-start-0. However, it does not hold for the bounded slowdown, for which plan-square-0 is better than all other algorithms. Plan-cube-0 follows its score. The value of mean is not reflected by the quantiles plots (Figure 4.36b, Figure 4.37b), where the best distributions are presented by plan-sum-0 and plan-start-0 for both slowdown and bounded slowdown. The tail distribution of slowdown shows similar results for all algorithms. For bounded slowdown, tail distribution presents a few highly outstanding values only for plan-sum-0 and plan-start-0, yet there are considerably fewer outliers that for waiting time.

**Conclusion**   Plan-based scheduling algorithms with resource reservations do not demonstrate better results than the canonical scheduling algorithms with burst buffer reservations. Plan-sum-0 and plan-start-0 proved that they lack a fairness mechanism and consequently can arbitrarily delay jobs in the waiting queue. Similarly, plan-square-0 and plan-cube-0 do not implement any strict mechanism for ensuring fairness. However, in the experiments, they did not indicate any relevantly outlying values in waiting time. Therefore, we consider plan-square-0 and plan-cube-0 as the best scheduling algorithms discussed in this section as they showed significant improvement in the mean and remarkable distribution for all statistics. Furthermore, plan-square-0 was slightly better than plan-cube-0 in all measurements.

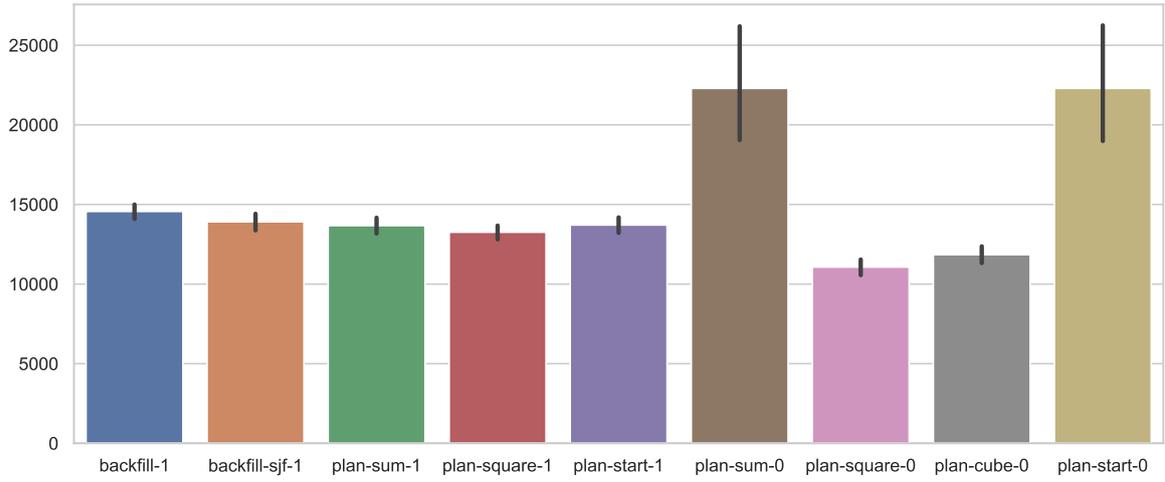

(a) Mean

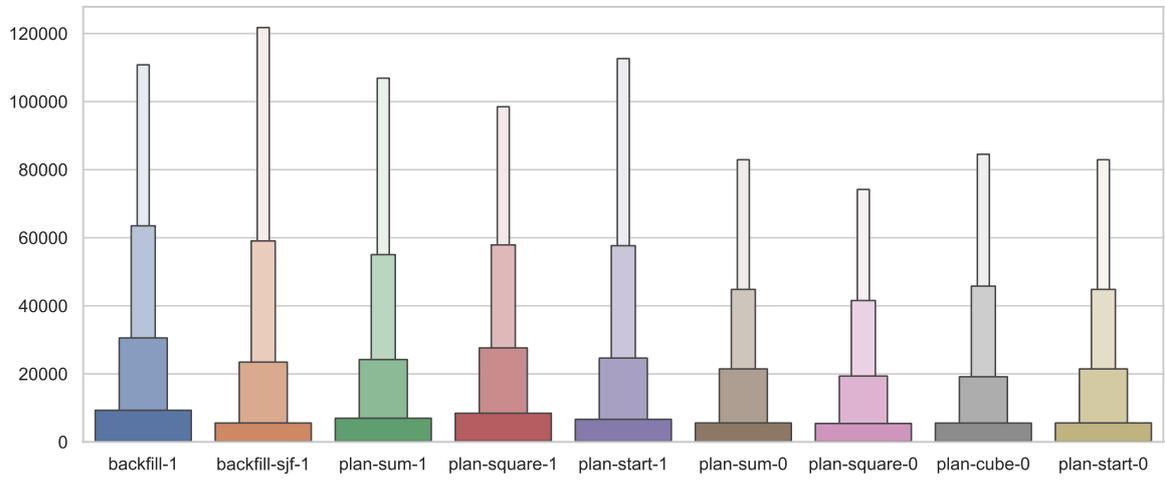

(b) Quantiles

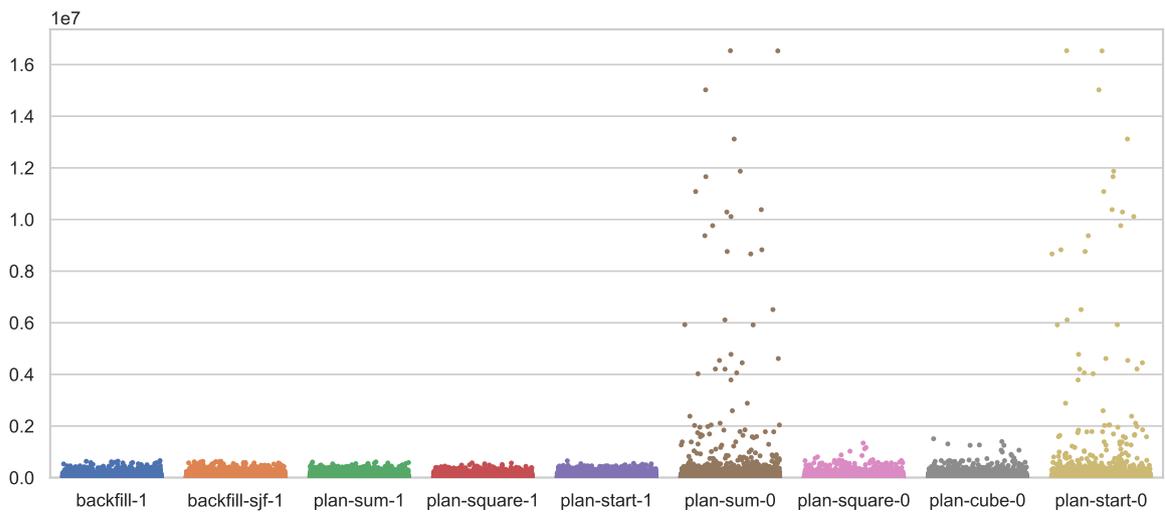

(c) Tail distribution

Figure 4.34: Waiting time

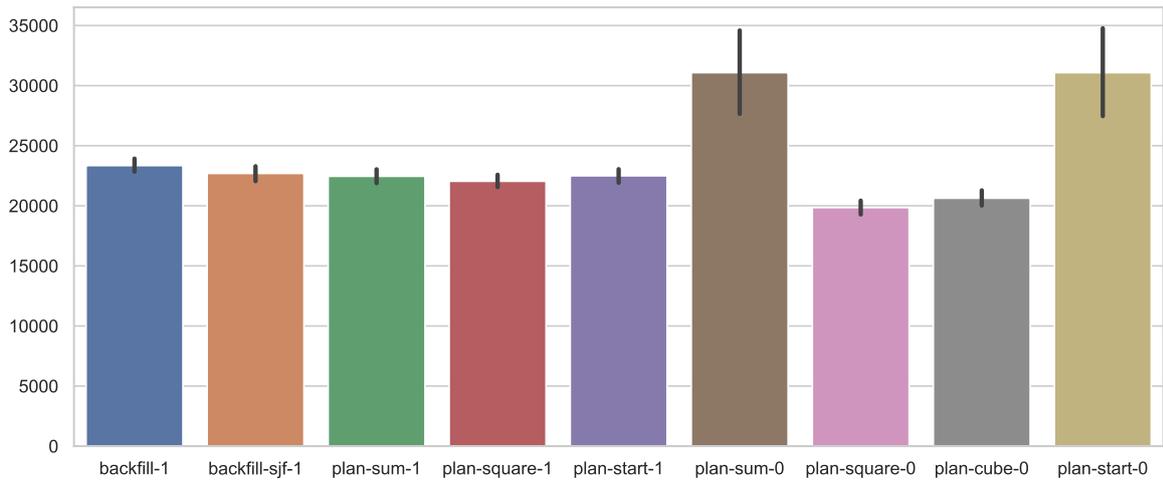

(a) Mean

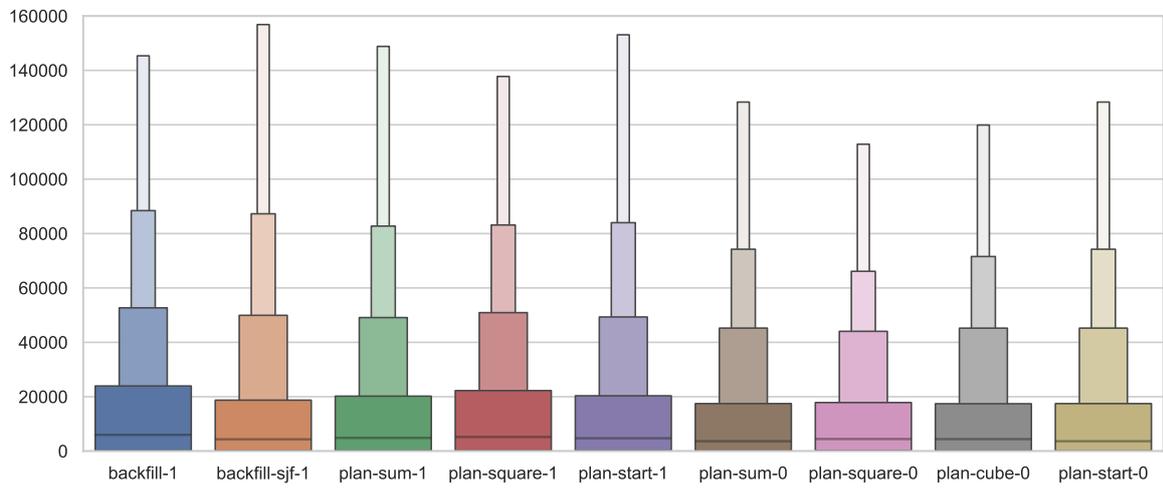

(b) Quantiles

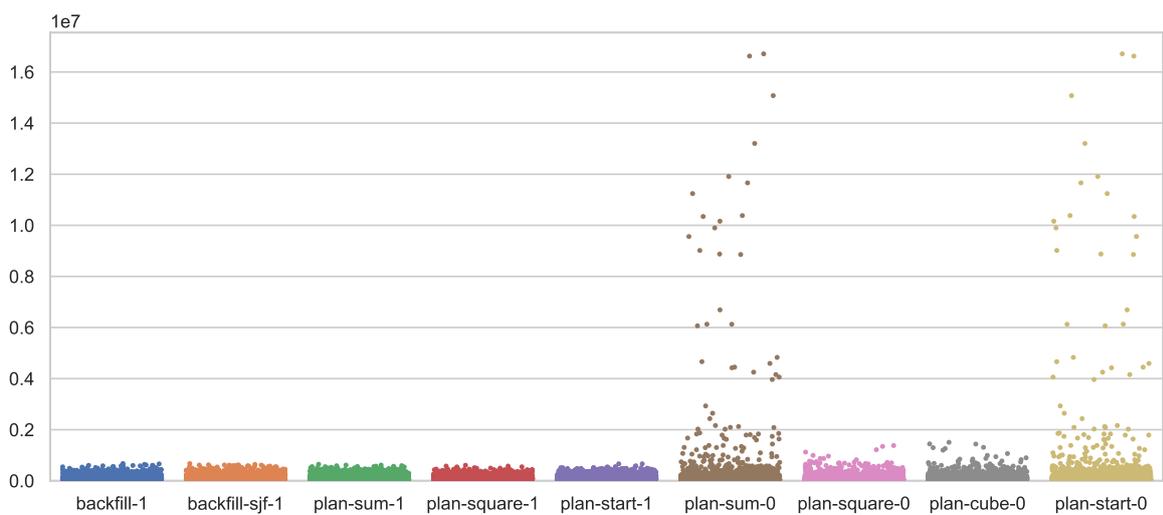

(c) Tail distribution

Figure 4.35: Turnaround time

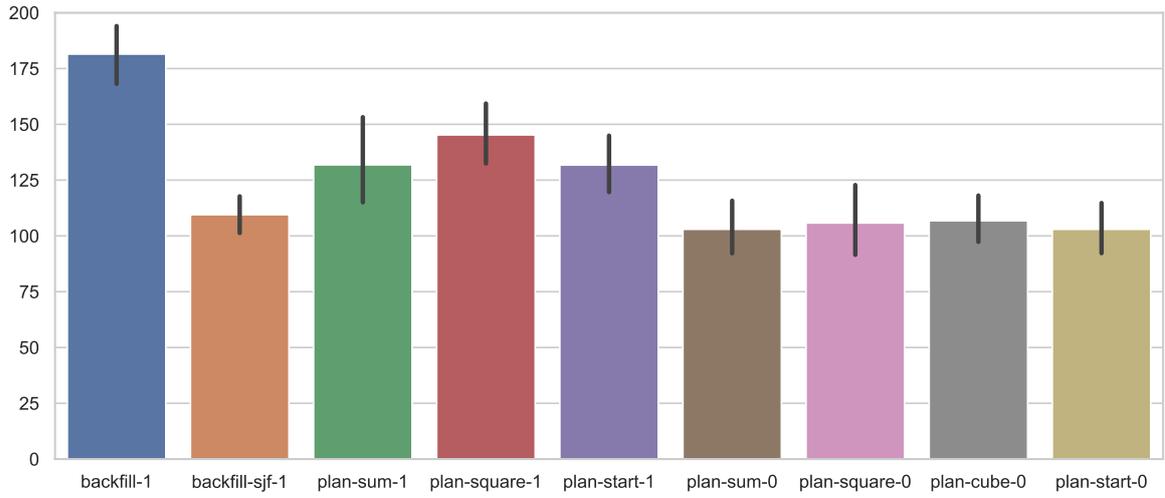

(a) Mean

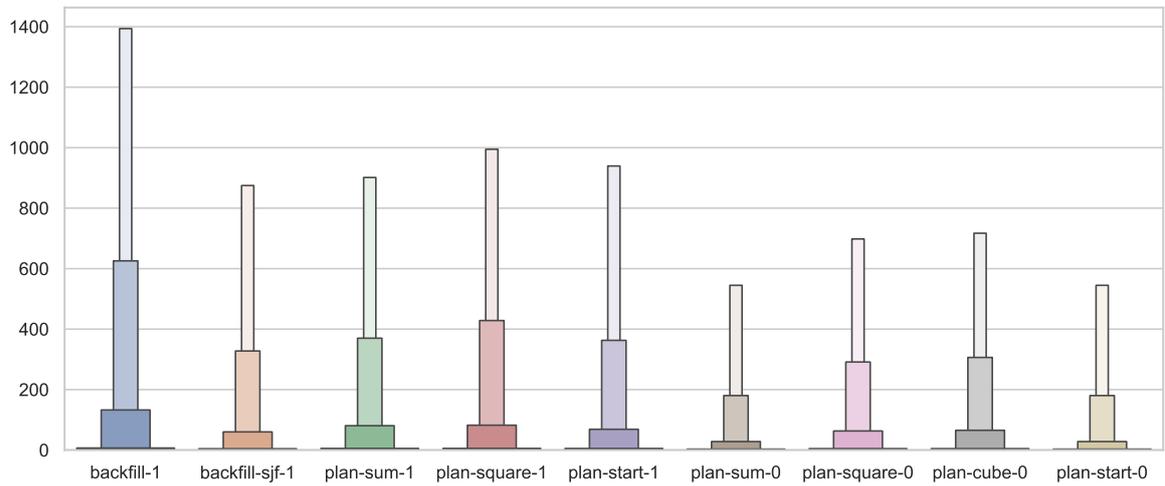

(b) Quantiles

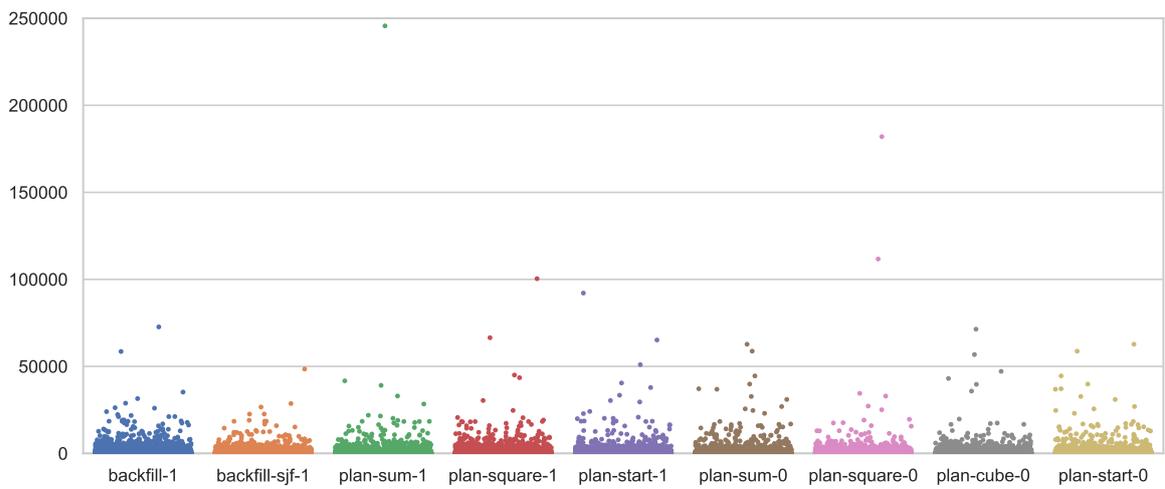

(c) Tail distribution

Figure 4.36: Slowdown

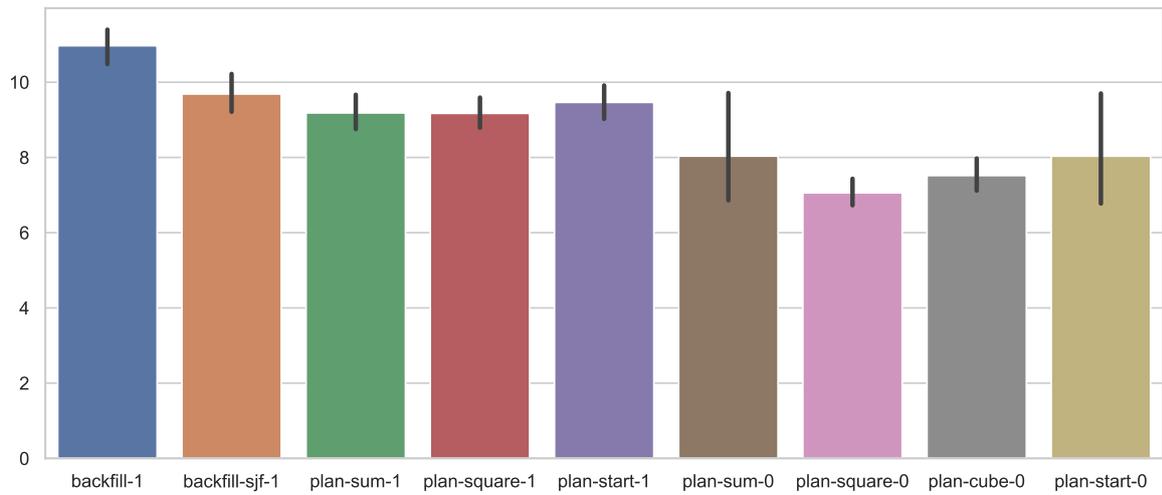

(a) Mean

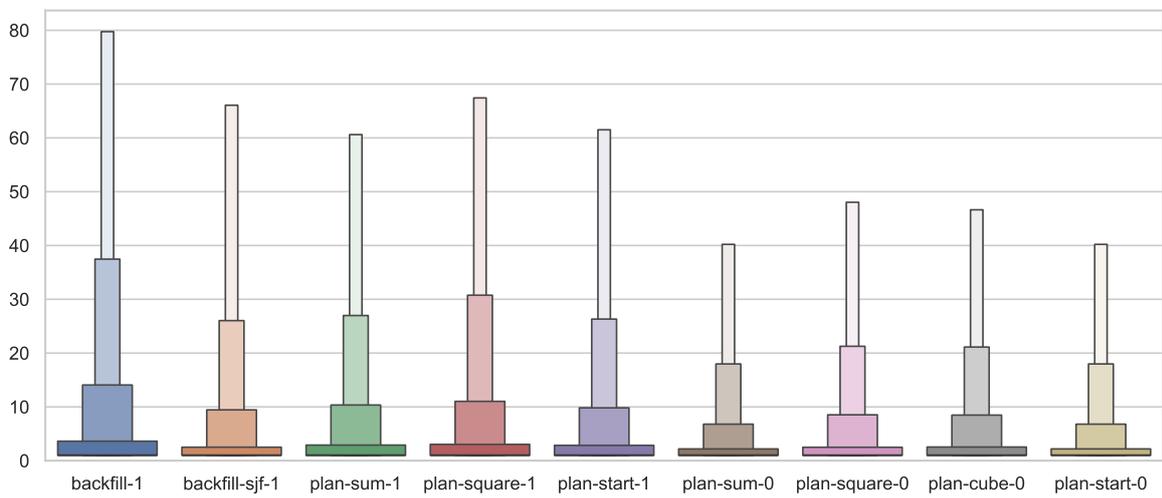

(b) Quantiles

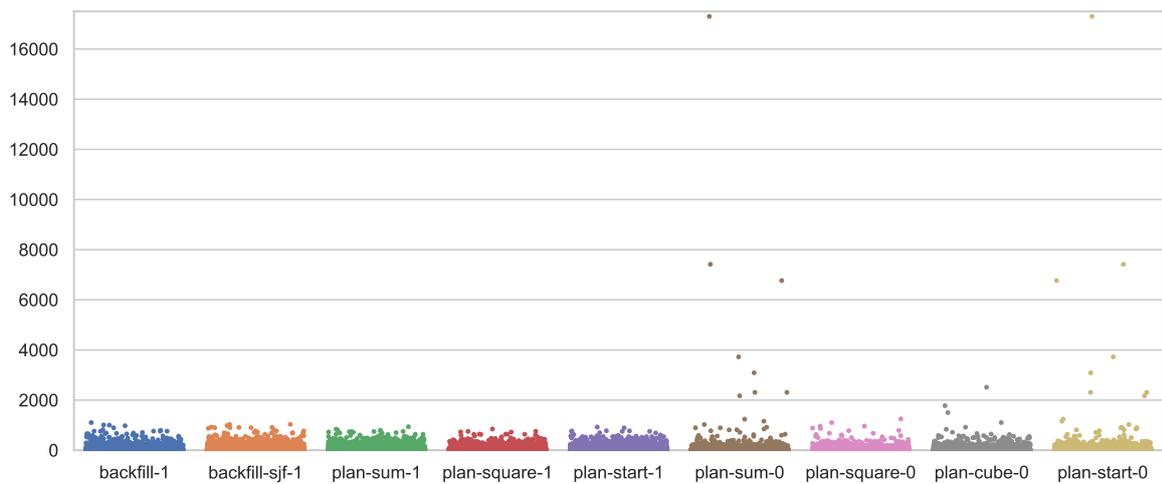

(c) Tail distribution

Figure 4.37: Bounded slowdown

## 4.6. Comparison of the best scheduling algorithms

From previous sections, we select scheduling policies that showed the best results for each algorithm. We compare them based on the in-depth analysis performed for both Alloc-Only and IO-Aware model. For each selected algorithm, we present the results for full workloads and split workloads. The selected algorithms are:

- FCFS EASY-backfilling (backfill-1, blue)

- SJF EASY-backfilling (backfill-sjf-1, orange)

- Maxutil with *balance factor* $\beta = 1$ (maxutil-1.0-1, green)

- Window-based combinatorial scheduling with a reservation of the first job (window-1, red)

- Plan-based (plan-square-0, purple)

    - no reservations ($D = 0$)
    - minimise the sum of squared waiting time ($\alpha = 2$)

- Plan-based (plan-cube-0, brown)

    - no reservations ($D = 0$)
    - minimise the sum of cubed waiting time ($\alpha = 3$)

**Alloc-Only model**  In Sections 4.3, 4.4, 4.5, we compared algorithms based on only the IO-Aware model. We supplement those results with experiments performed for the Alloc-Only model. Comparison of the results for both models provides us with information on the influence of I/O contentions and I/O congestion effects on burst buffer aware job scheduling.

The first outstanding observation is that backfill-sjf-1 for the full workload demonstrates worse results for mean and distribution of waiting time than backfill-1 (Figure 4.38), which was not a case in the IO-Aware model. However, this relation is not reflected in Figure 4.42, presenting the mean waiting time for the split workload. For most of the parts, backfill-1 achieves worse results than backfill-sjf-1. Therefore, we find it probable that backfill-sjf-1 for the full workload was affected by unfavourable interlace of jobs.

Focusing on scheduling algorithms which optimise resource utilisation, we observe that maxutil-1.0-1 and window-1 achieve almost identical results in mean and distribution of waiting and turnaround time. For slowdown and bounded slowdown, window-1 tend to present a better outcome for the full workload. In terms of the split workload, we do not see a strict relation between those schedules. Moreover, the confidence intervals indicate a considerable uncertainty of the results. For Figure 4.45, the uncertainty level is so great that it is at the same order of magnitude as the mean slowdown for each part. Consequently, we find the results of the mean slowdown as unreliable. In general, backfill-1, backfill-sjf-1, maxutil-1.0-1, window-1 show comparable results. For a given statistic, there are visible differences between them for the full workload. In plots 4.42-4.45, it is, however, visible that their results interleave for various workload parts.

In Section 4.5, we stated that plan-square-0 is the best plan-based scheduling algorithm based on the analysis of the full workload in the IO-Aware model. Plan-cube-0 directly followed its results. We observe the same relations for the Alloc-Only model. From summary statistics plots in figures 4.38-4.41, it is clear that plan-square-0 and plan-cube-0 dominate other schedules. Furthermore, for the slowdown and bounded slowdown, all algorithms present similar tail distributions. The only statistic for which plan-square-0 and plan-cube-0 show a deterioration is the tail distribution of waiting and turnaround time. However, they present only a few jobs that have higher values than outliers of other schedules.

**IO-Aware model**  For the canonical scheduling algorithms backfill-1 and backfill-sjf-1, the results for the IO-Aware model are discussed in Section 4.3. The plan-based algorithms plan-square-0 and plan-cube-0 are compared in Section 4.5. For waiting time and turnaround time, we generally observe similar relations between algorithms as they are in the Alloc-Only model. That is backfill-1, backfill-sjf-1, maxutil-1.0-1 and window-1 have comparable performance. Similarly, plan-square-0 and plan-cube-0 present a much better performance in summary statistics but worse results in the tail distributions. The results for the bounded slowdown are similar for IO-Aware and Alloc-Only models. The schedules backfill-sjf-1, maxutil-1.0-1 and window-1 are comparable, backfill-1 is slightly worse, and plan-based scheduling policies strictly dominate others. For slowdown, as we already noticed backfill-sjf-1 indicates excellent results that are comparable with plan-square-0 and plan-cube-0. Window-1 present considerably worse results and is followed by maxutil-1.0-1.

**Normalised mean plots**  The bar plots presenting values of mean enables to determine whether or not one scheduling algorithm dominate another for a given statistic. However, they do not directly inform if one algorithm is on average better than another. For this reason, we created figures 4.46-4.49 and 4.58-4.61. They present how many times a given scheduling algorithm is worse than backfill-sjf-1. This plot for a given statistic was created in the following way. At first, we calculate the mean for each algorithm and each workload part. Then we normalise them by the mean of backfill-sjf-1, that is for each algorithm and each workload part we divide the mean for this part by the mean of the corresponding part from backfill-sjf-1. Lastly, we create a box plot for each algorithm based on the obtained 16 values. The scattered black dots on those plots represent the individual mean values for all the workload parts. In conclusion, we can easily compare algorithms for the split workload using a double aggregation of data: mean followed by ordered statistics.

For instance, by looking at Figure 4.46, we may conclude that backfill-1 is about 1.1 times worse in waiting time than backfill-sjf-1. Similarly, plan-cube-0 achieves about 0.7 times lower results in waiting time than backfill-sjf-1.

By analysing plots 4.46-4.49, we conclude that for the Alloc-Only model backfill-sjf-1, maxutil-1.0-1 and window-1 are comparable scheduling algorithms in terms of all metrics. Window-1 tends to present slightly better results. In contrast, backfill-1 generally shows worse results. The results for plan-based scheduling are far better than other algorithms in all metrics. Almost the same conclusion applies to the IO-Aware model presented in figures 4.58-4.61. In this case, the performance of backfill-1 is similar to backfill-sjf-1. Plan-based schedules again indicate dominance over other algorithms.

### 4.6.1. Alloc-Only model

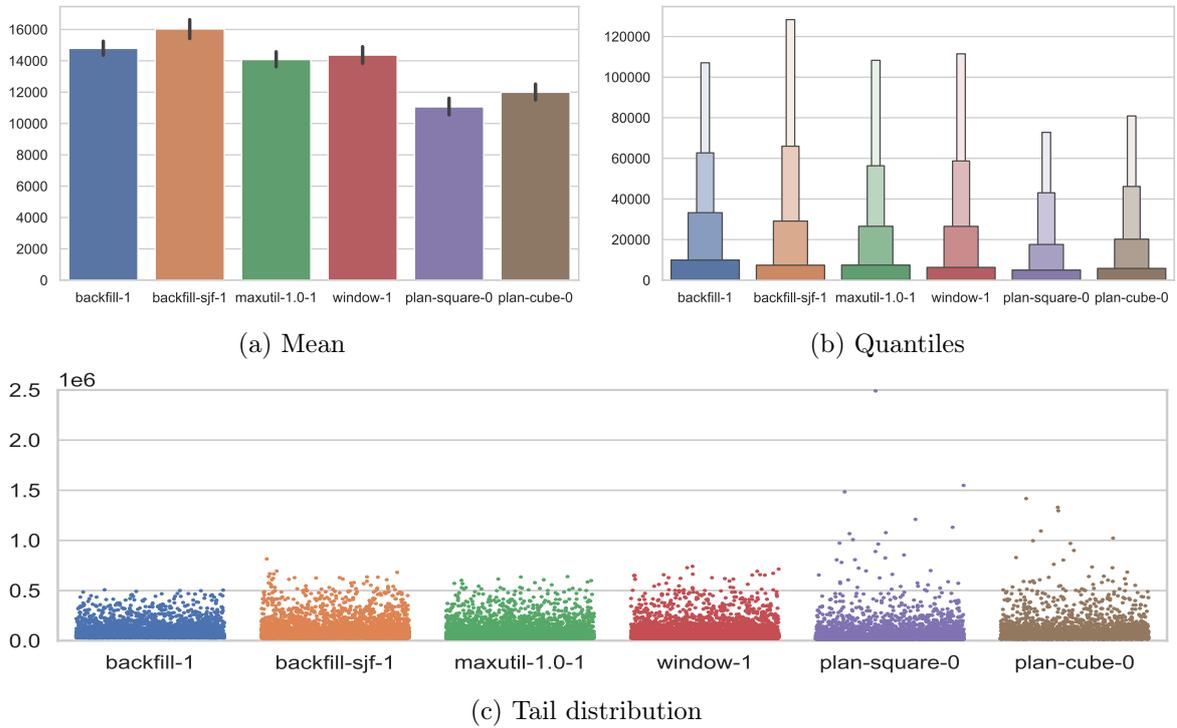

(a) Mean

(b) Quantiles

(c) Tail distribution

Figure 4.38: Waiting time

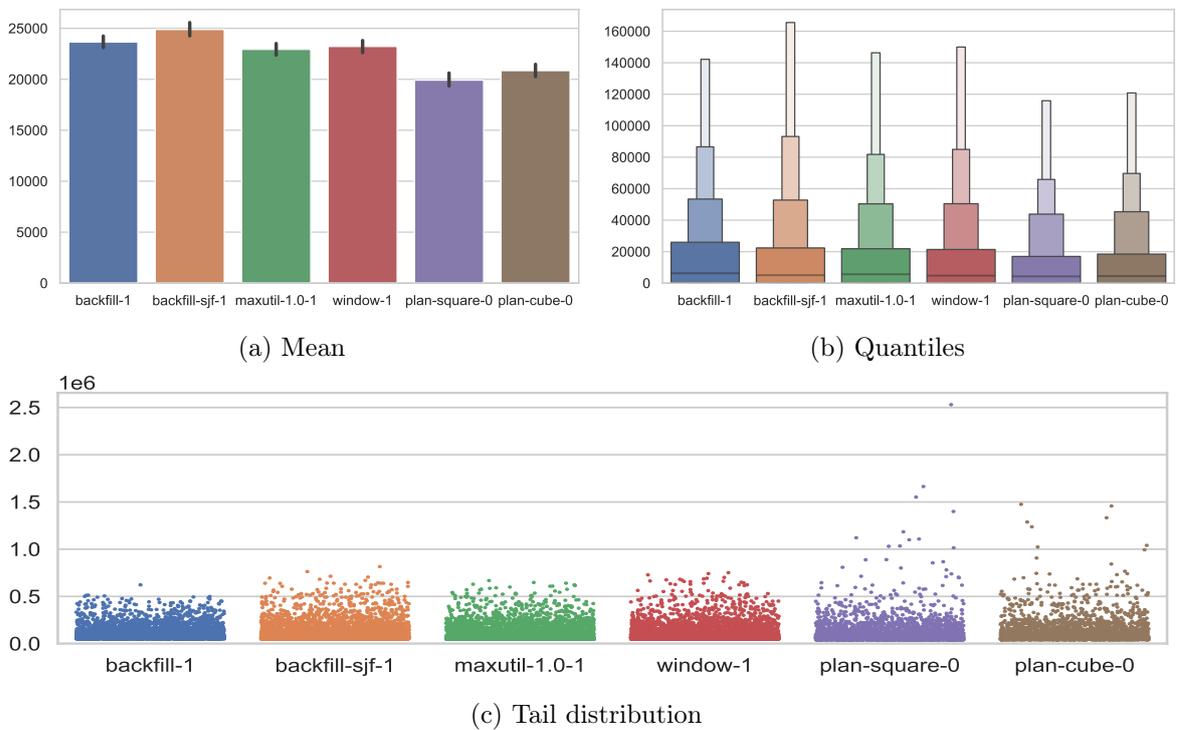

(a) Mean

(b) Quantiles

(c) Tail distribution

Figure 4.39: Turnaround time

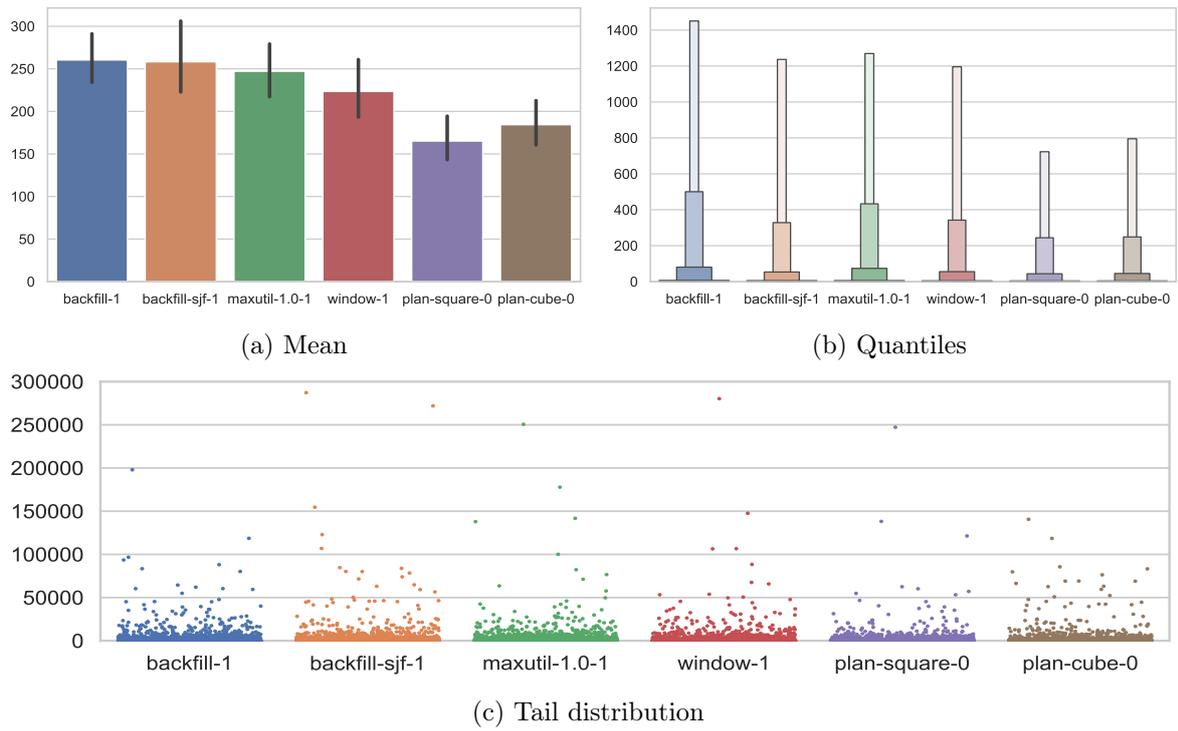

(a) Mean

(b) Quantiles

(c) Tail distribution

Figure 4.40: Slowdown

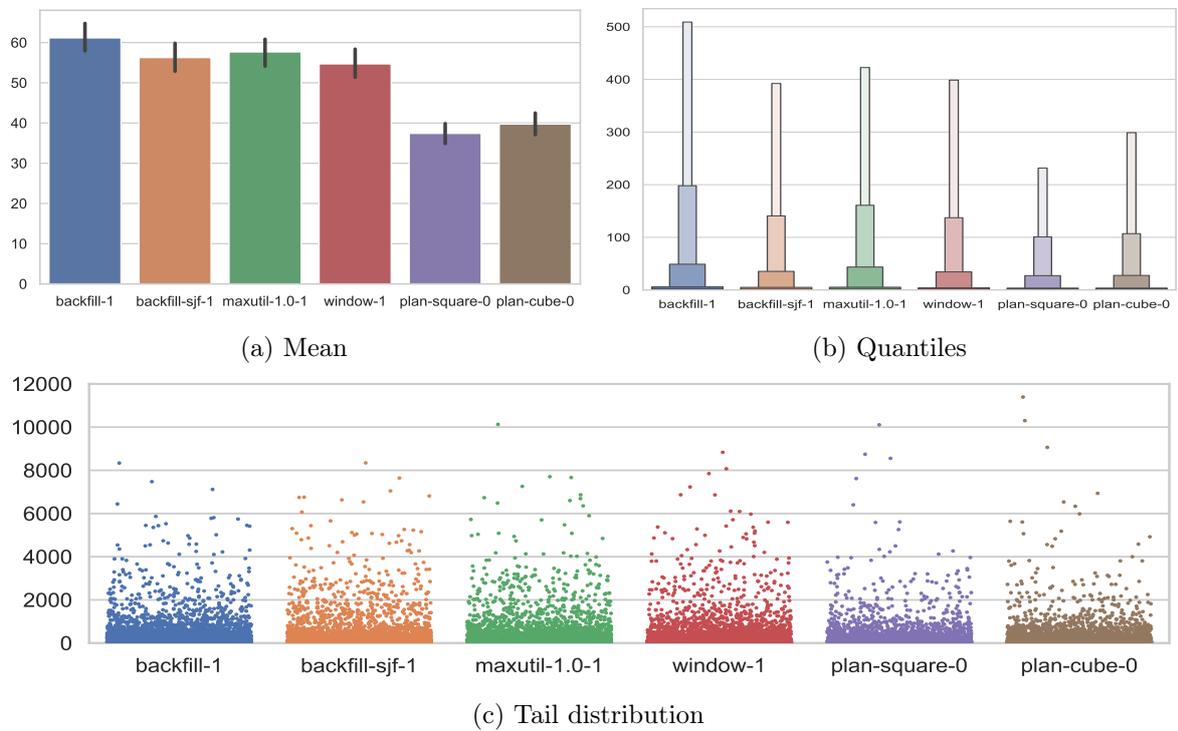

(a) Mean

(b) Quantiles

(c) Tail distribution

Figure 4.41: Bounded slowdown

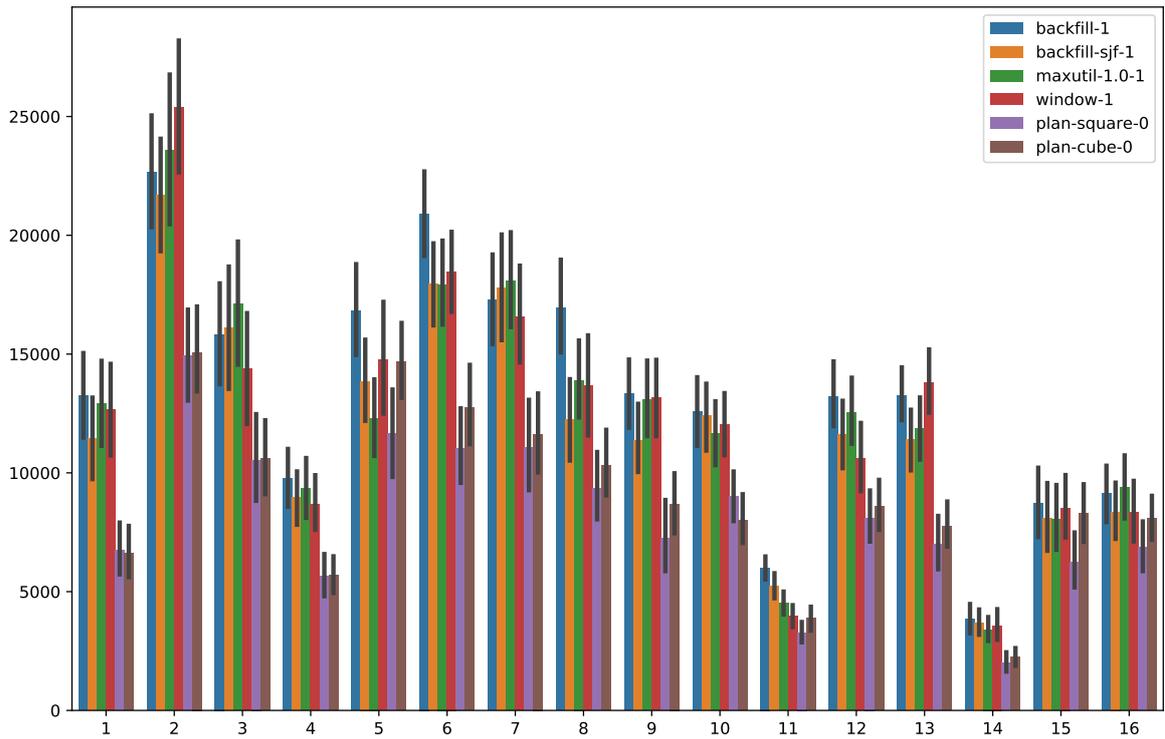

Figure 4.42: Mean waiting time of the split workload

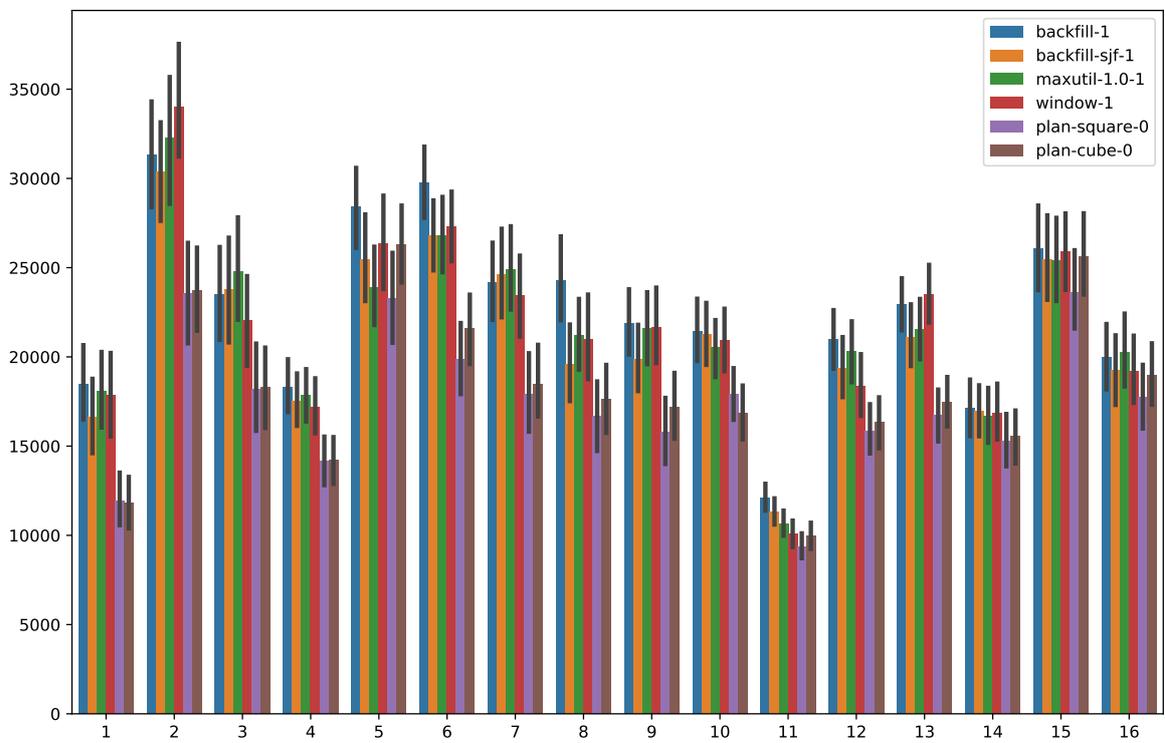

Figure 4.43: Mean turnaround time of the split workload

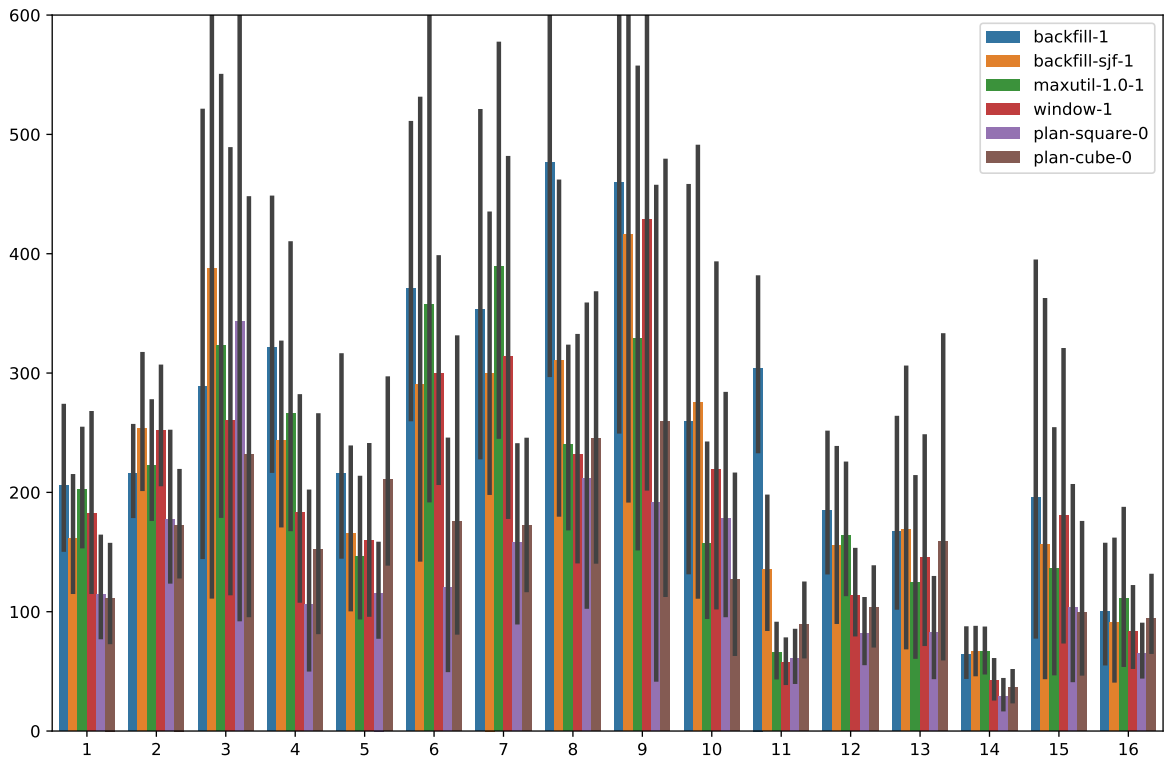

Figure 4.44: Mean slowdown of the split workload

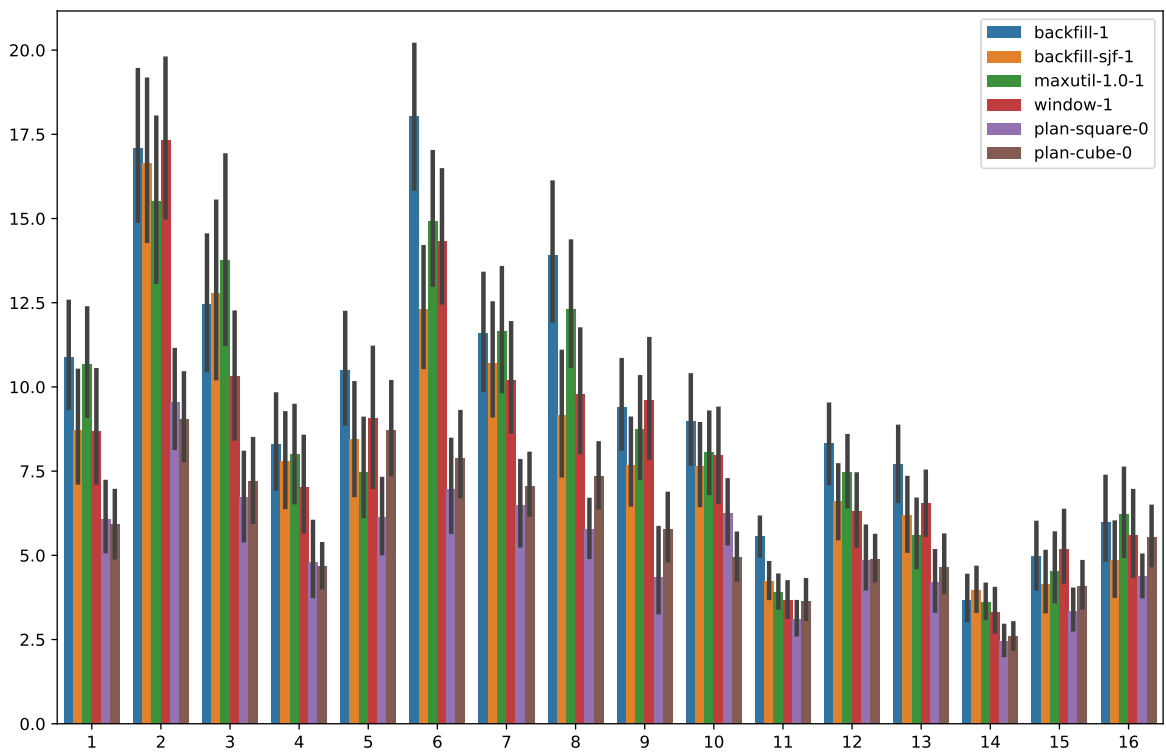

Figure 4.45: Mean bounded slowdown of the split workload

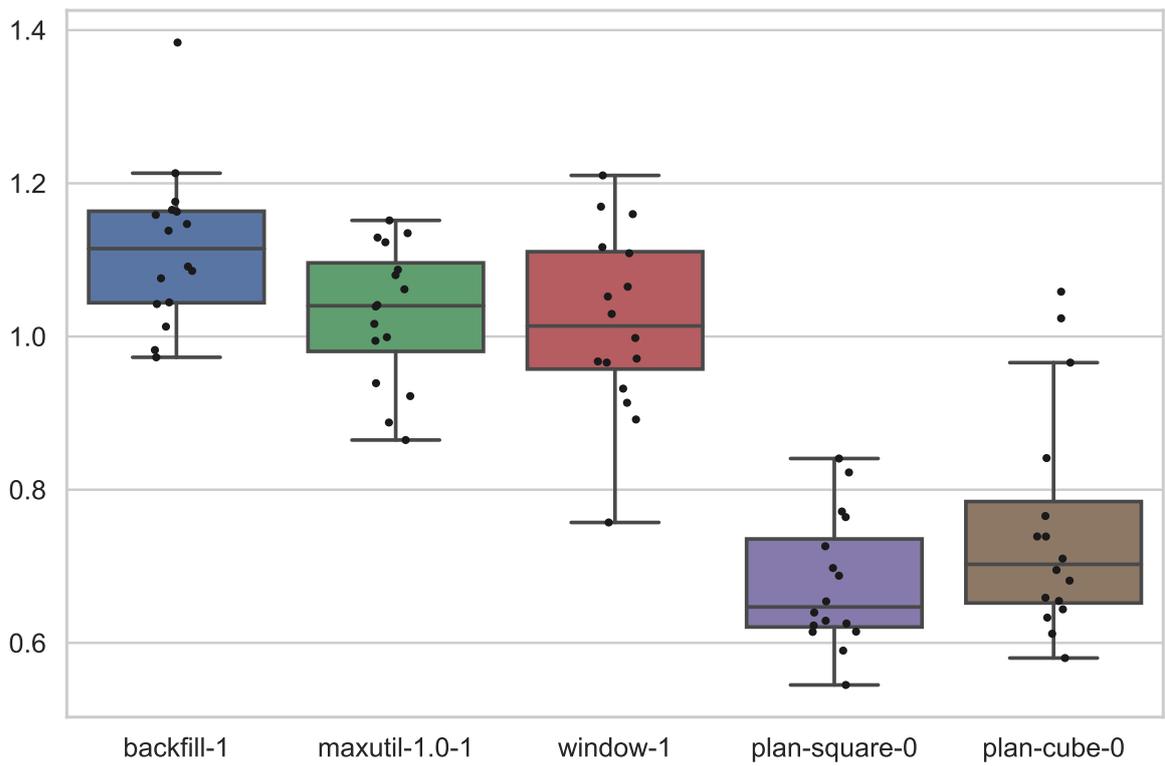

Figure 4.46: Mean waiting time for each part in the split workload (normalised to backfill-sjf-1)

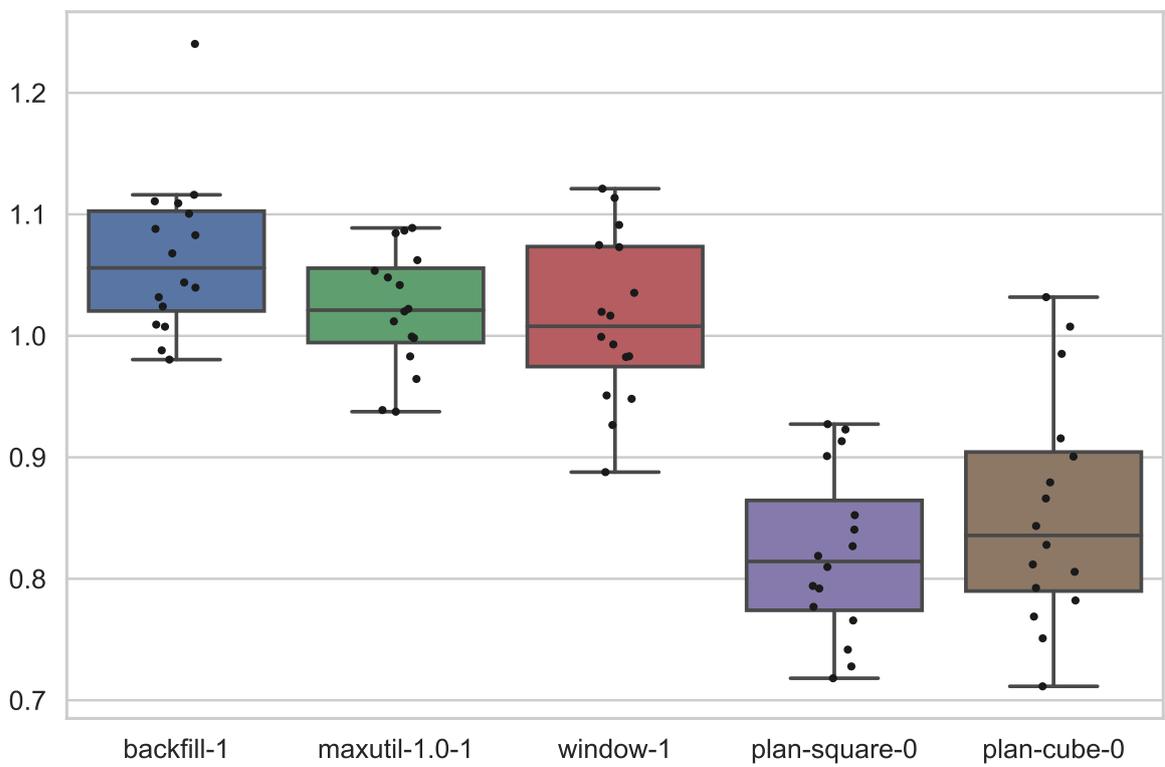

Figure 4.47: Mean turnaround time for each part in the split workload (normalised to backfill-sjf-1)

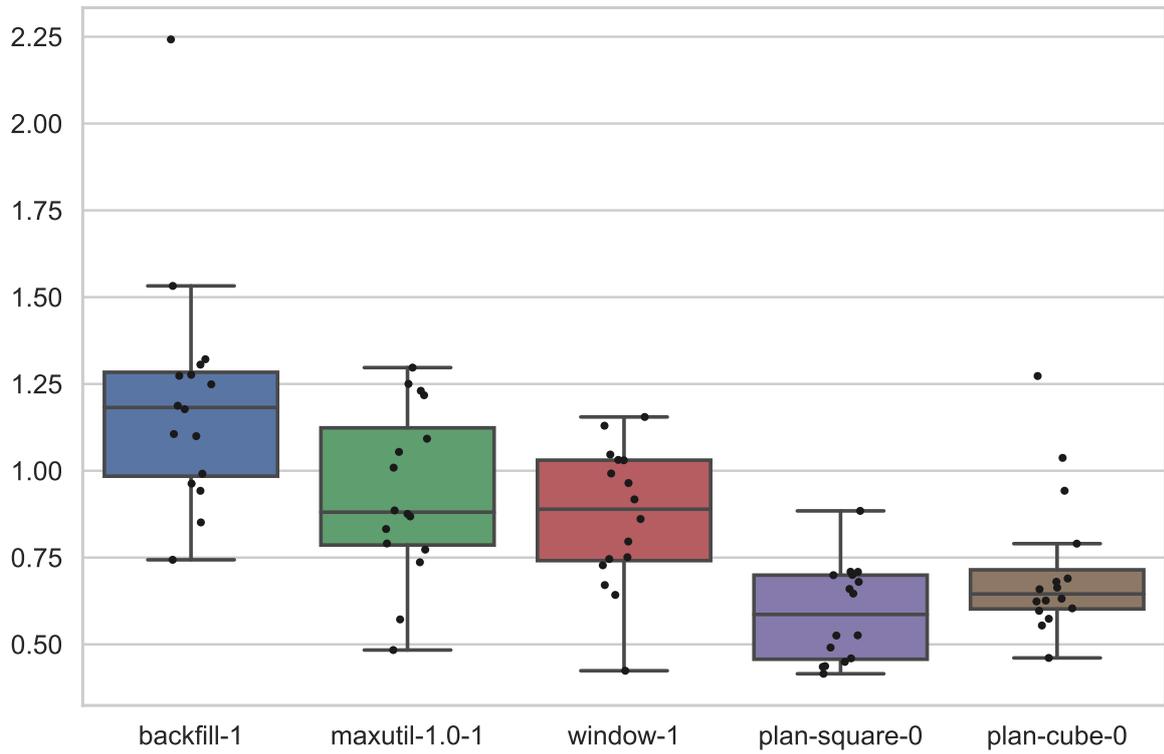

Figure 4.48: Mean slowdown for each part in the split workload (normalised to backfill-sjf-1)

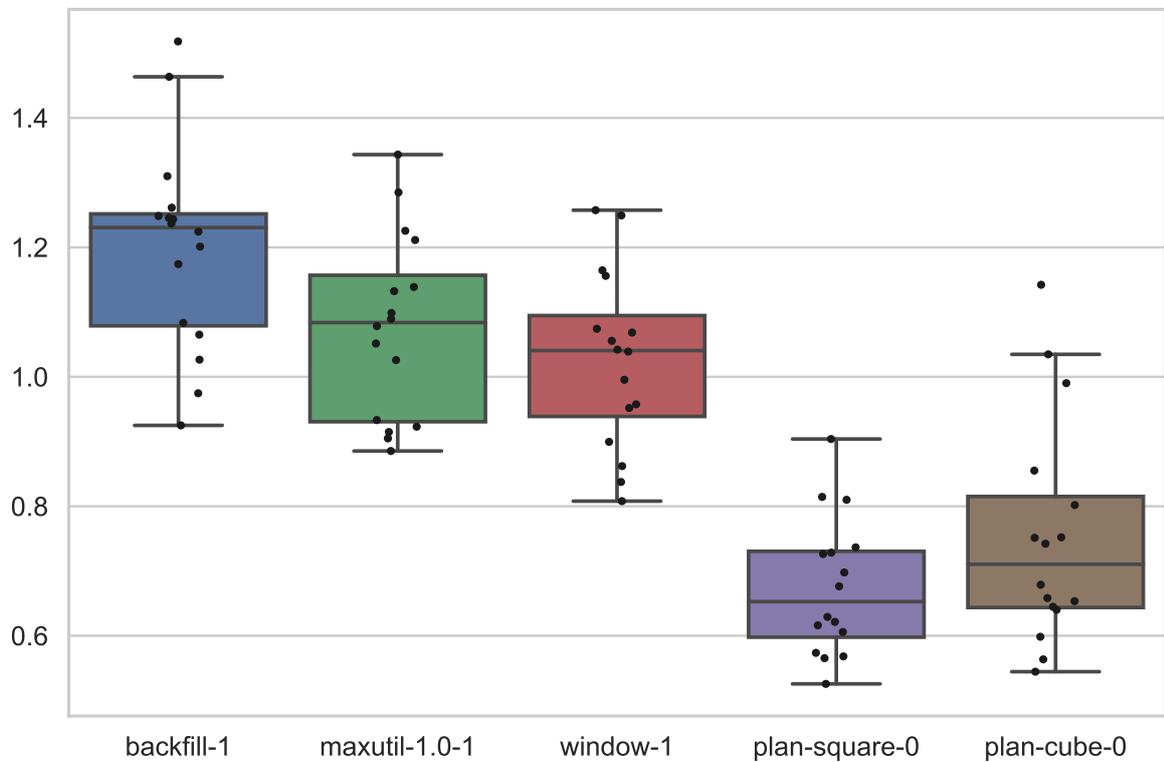

Figure 4.49: Mean bounded slowdown for each part in the split workload (normalised to backfill-sjf-1)

### 4.6.2. IO-Aware model

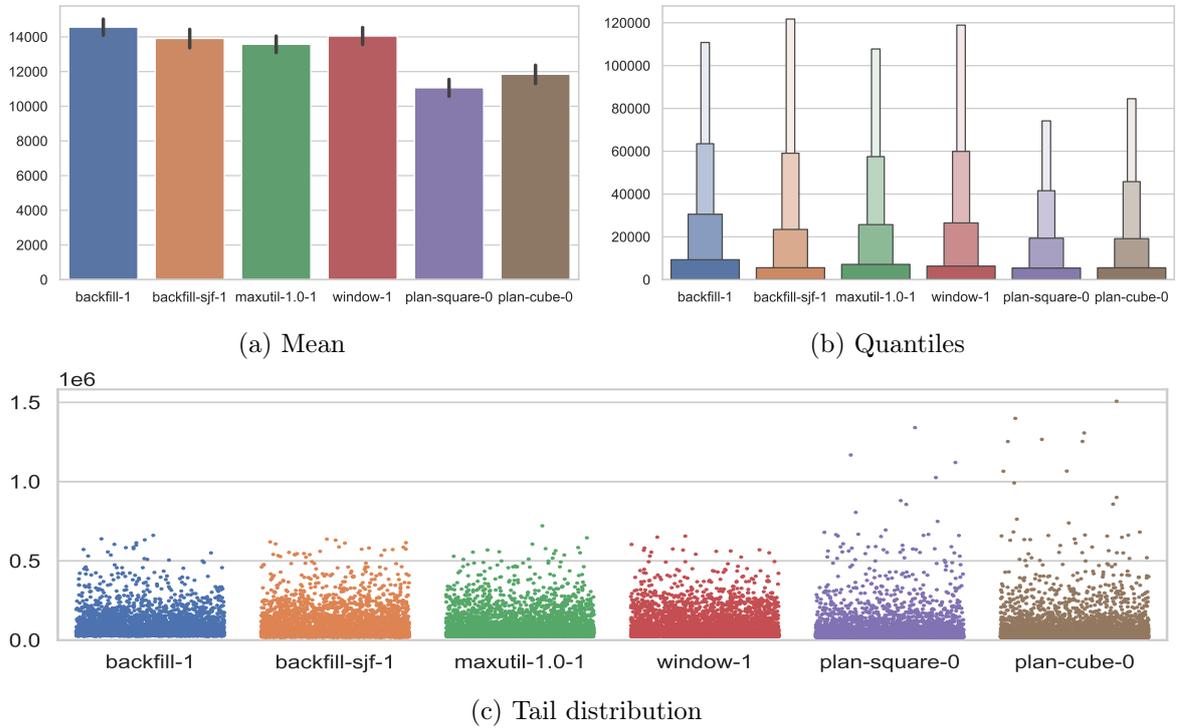

(a) Mean

(b) Quantiles

(c) Tail distribution

Figure 4.50: Waiting time

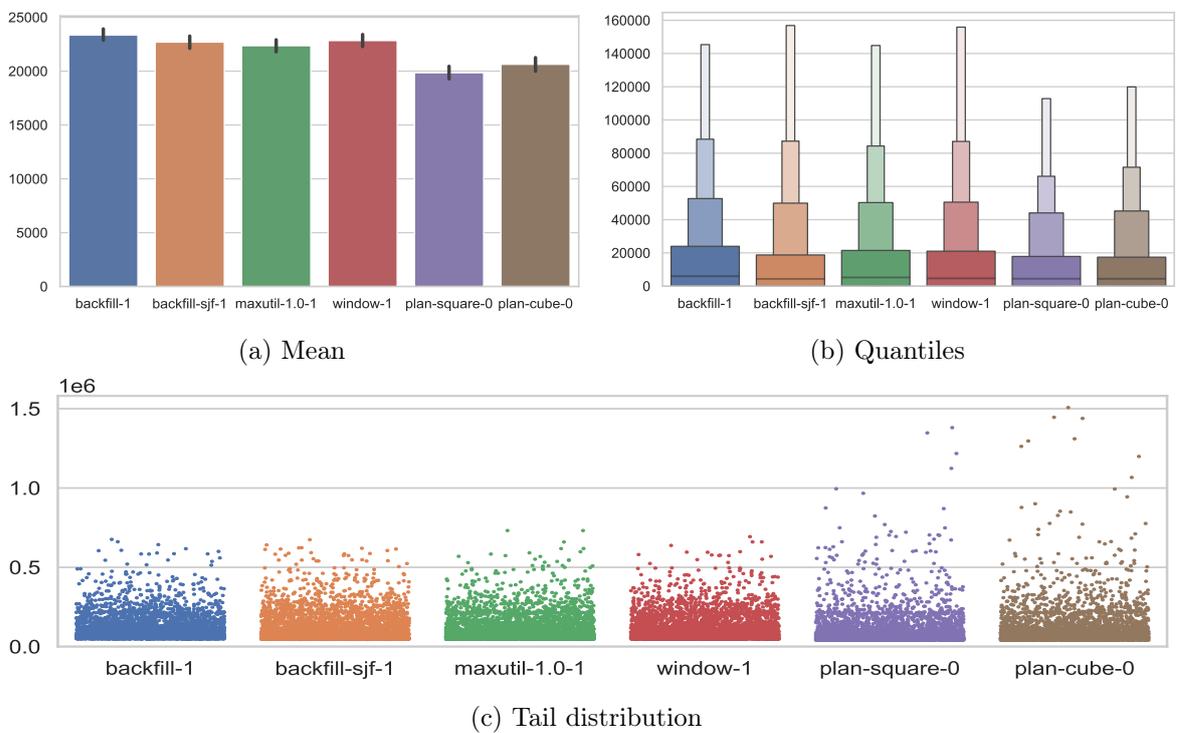

(a) Mean

(b) Quantiles

(c) Tail distribution

Figure 4.51: Turnaround time

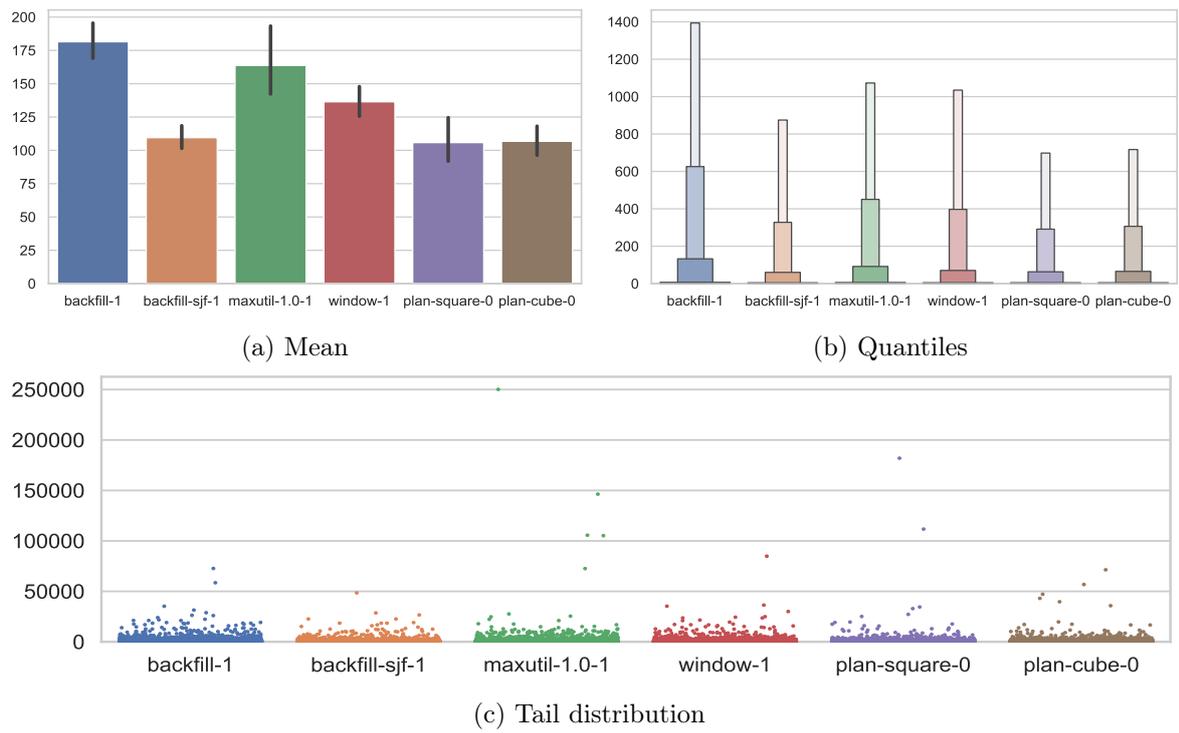

(a) Mean

(b) Quantiles

(c) Tail distribution

Figure 4.52: Slowdown

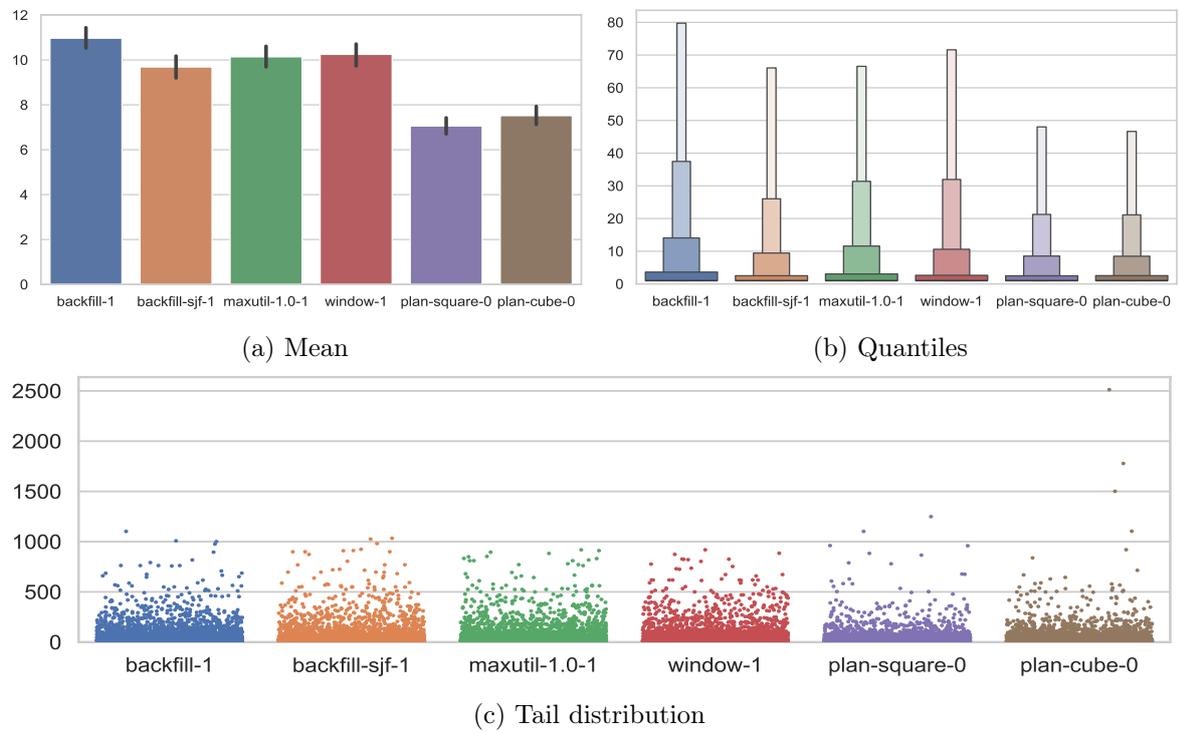

(a) Mean

(b) Quantiles

(c) Tail distribution

Figure 4.53: Bounded slowdown

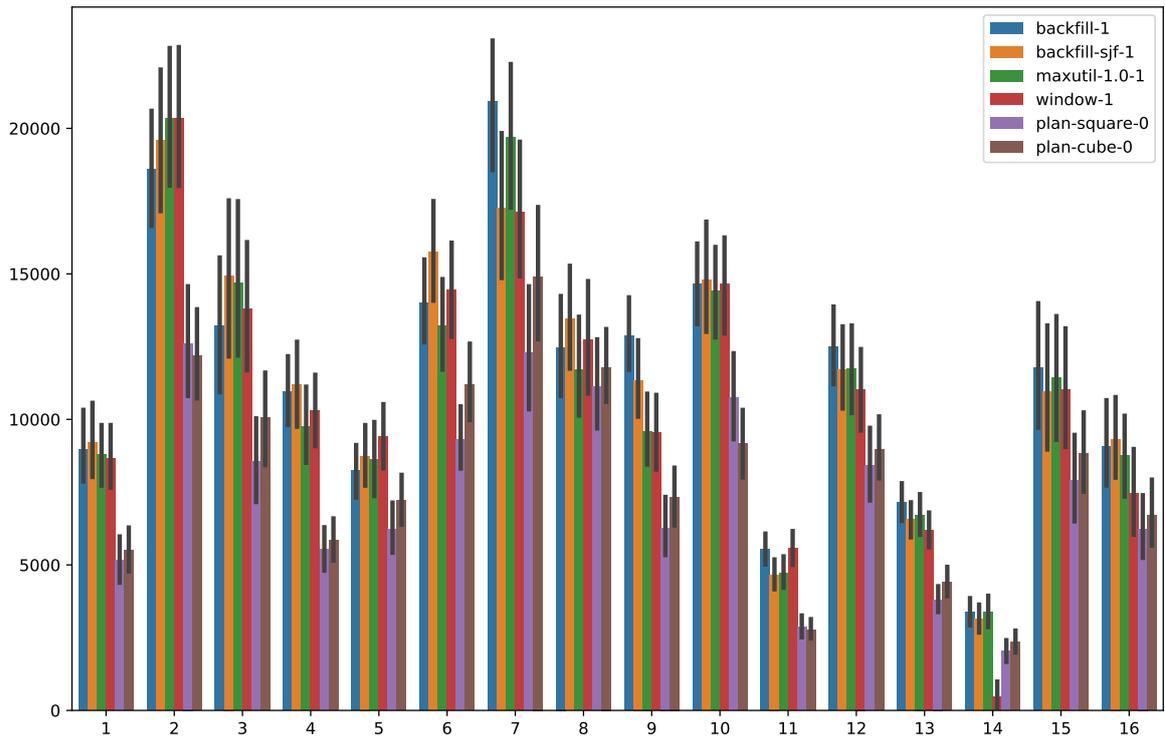

Figure 4.54: Mean waiting time of the split workload

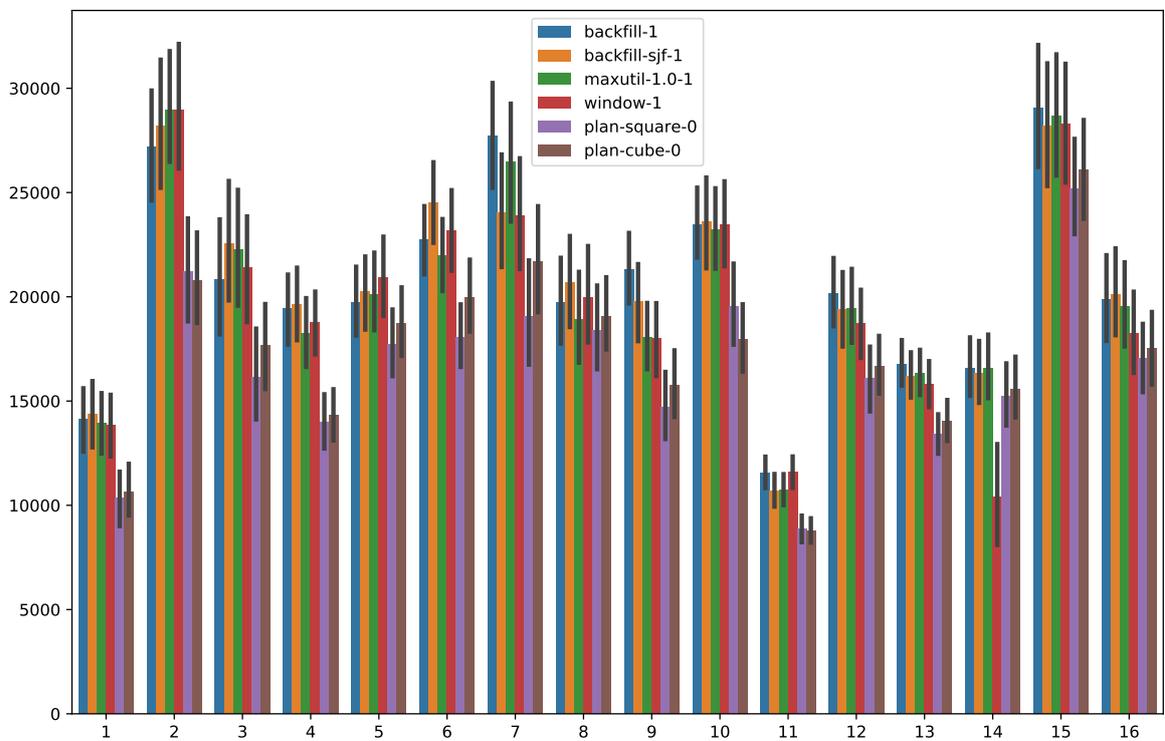

Figure 4.55: Mean turnaround time of the split workload

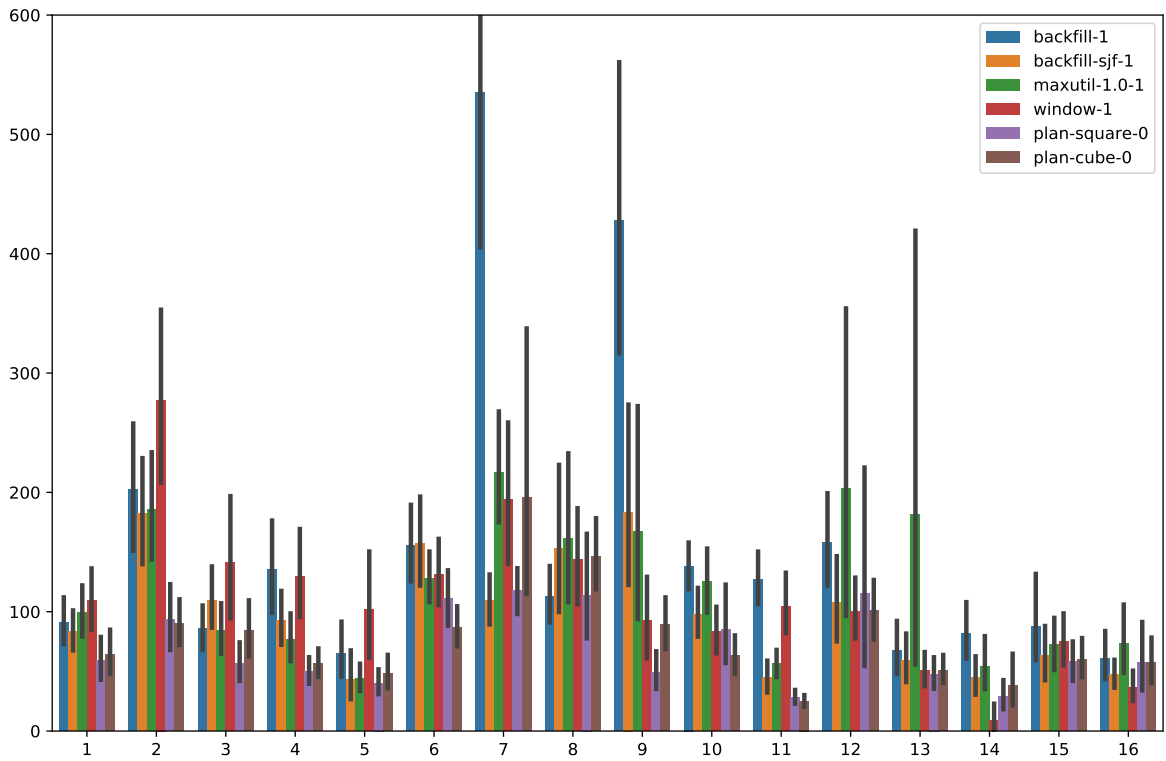

Figure 4.56: Mean slowdown of the split workload

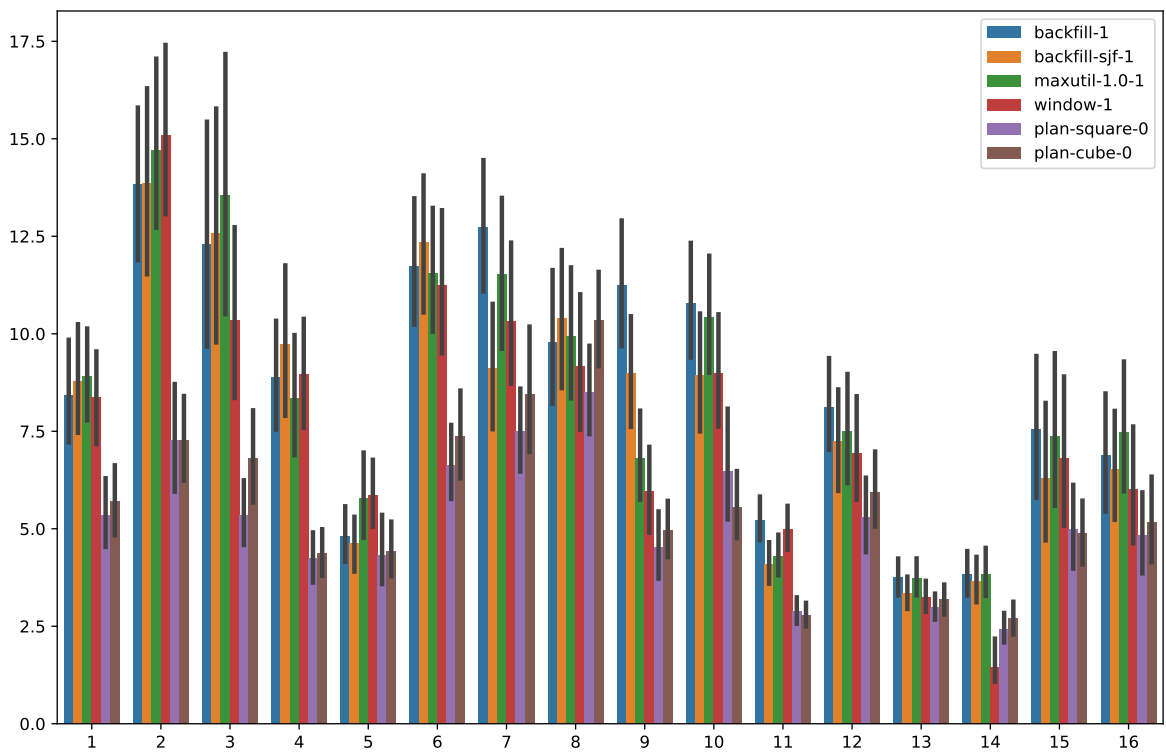

Figure 4.57: Mean bounded slowdown of the split workload

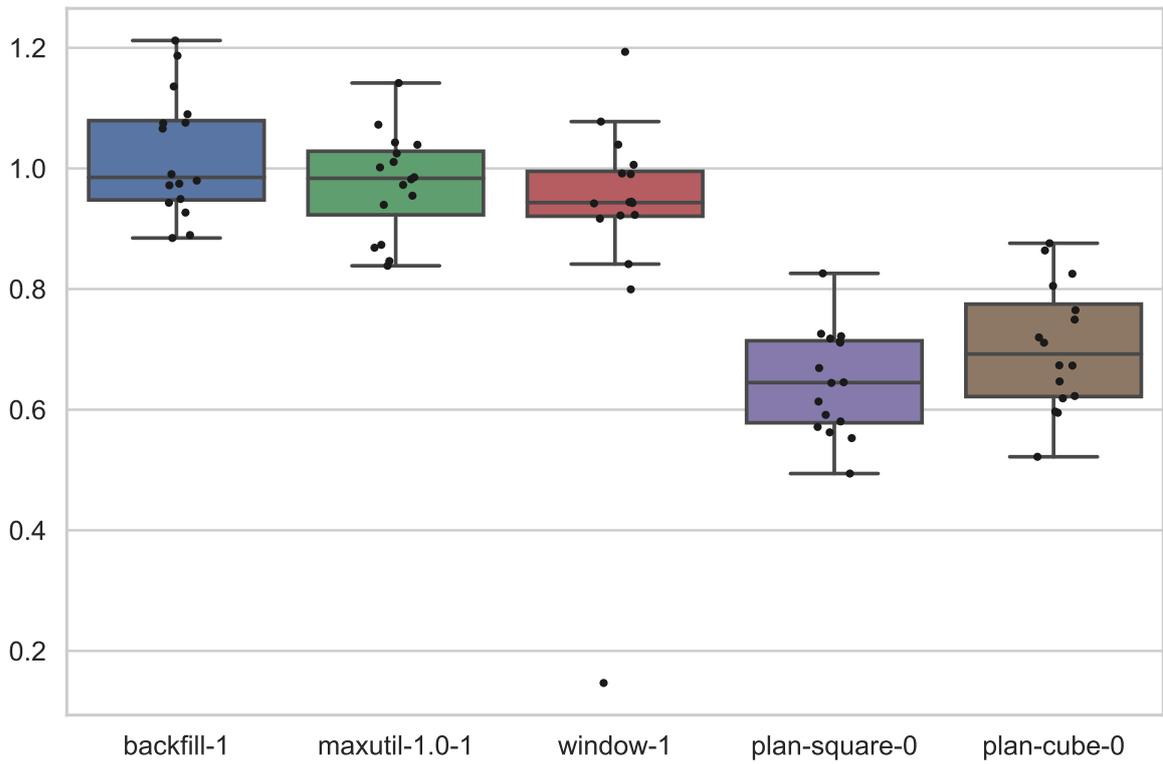

Figure 4.58: Mean waiting time for each part in the split workload (normalised to backfill-sjf-1)

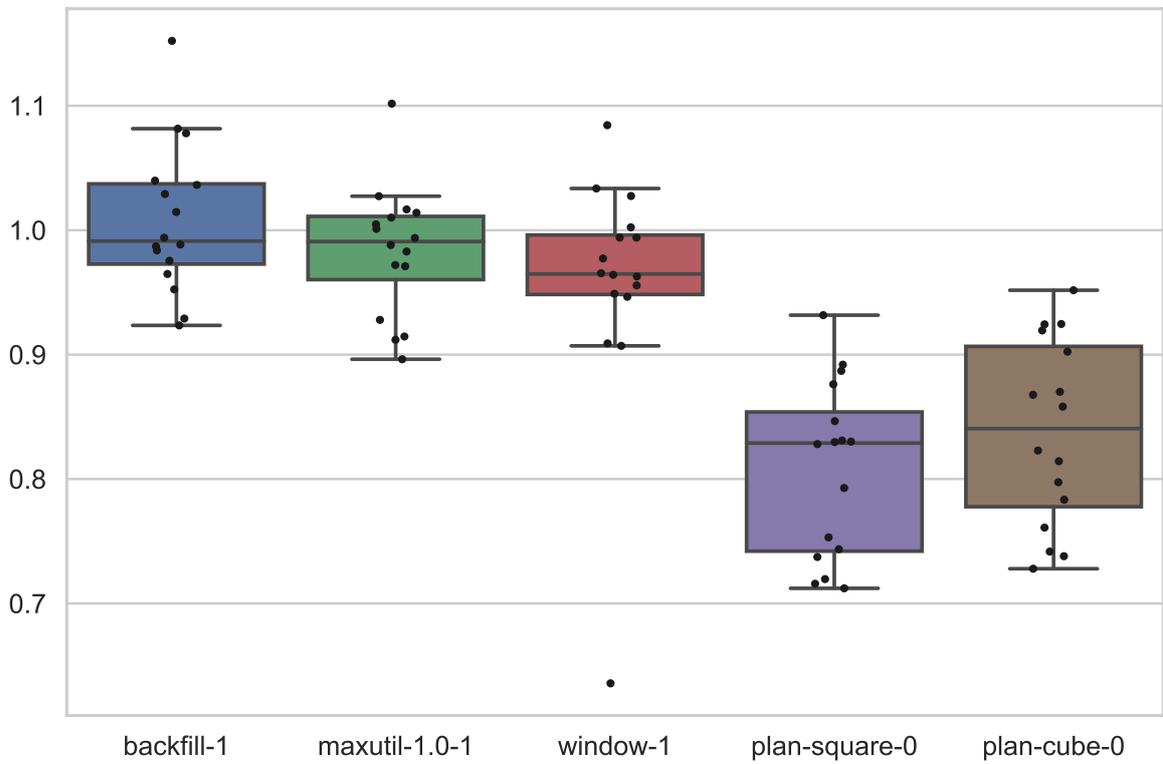

Figure 4.59: Mean turnaround time for each part in the split workload (normalised to backfill-sjf-1)

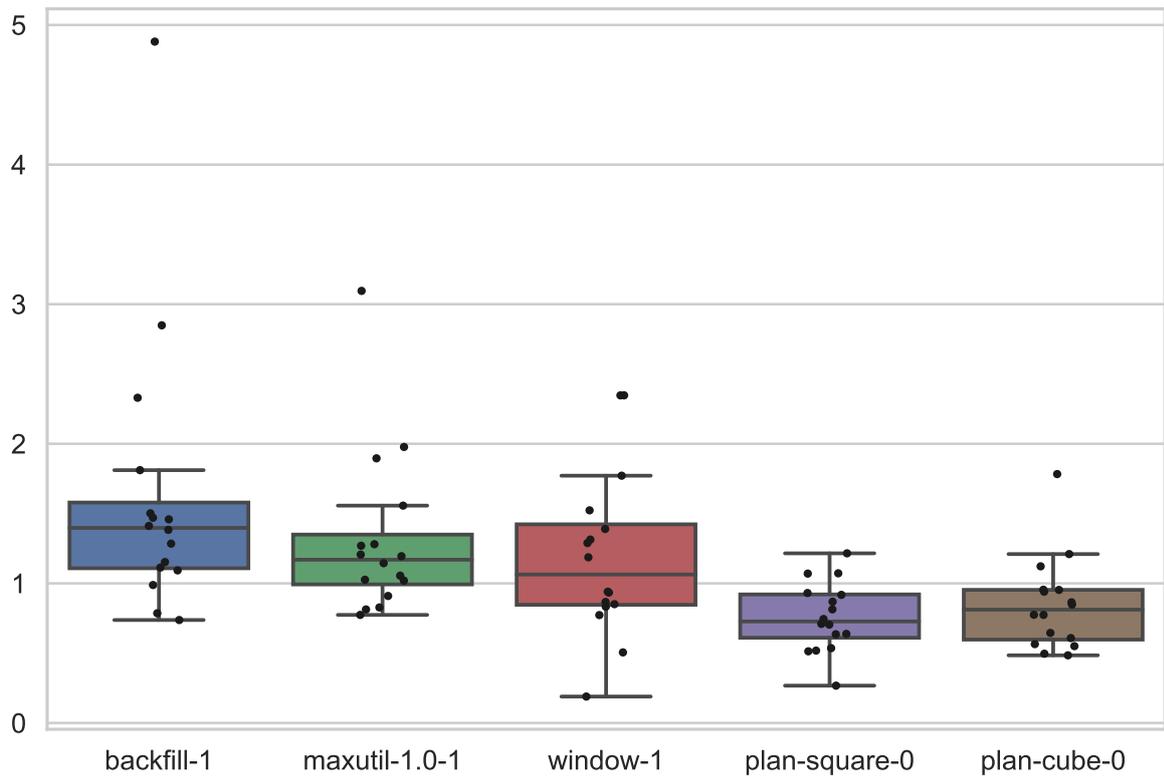

Figure 4.60: Mean slowdown for each part in the split workload (normalised to backfill-sjf-1)

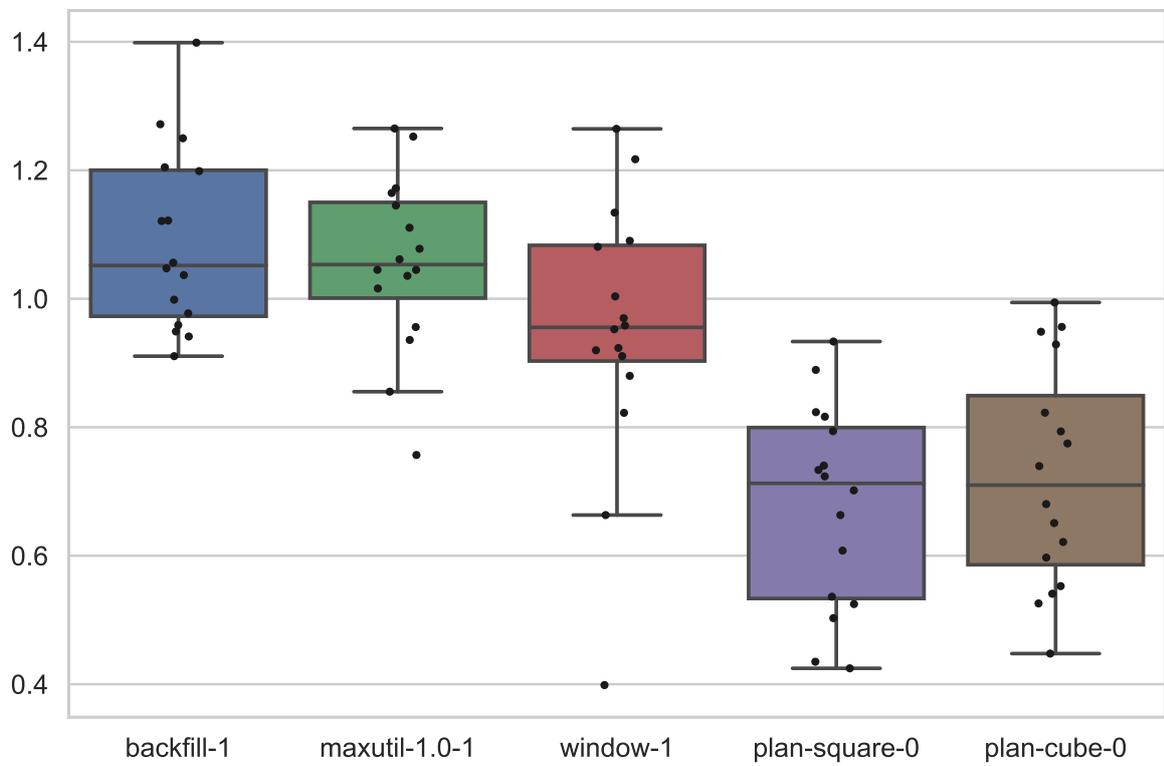

Figure 4.61: Mean bounded slowdown for each part in the split workload (normalised to backfill-sjf-1)

## 4.7. Backfilling reservation depth

In this section, we study the influence of the *reservation depth* parameter ($D$) on selected scheduling algorithms. We want to find whether increasing the number of jobs taken to reservations from the front of the waiting queue improves or worsen the efficiency of scheduling. Moreover, we want to find an optimal value of the *reservation depth* for each algorithm. We chose the canonical scheduling algorithms and the best performing plan-based policy. Namely, the selected algorithms are:

- Greedy FCFS filling without any reservations (filler, Algorithm 2)

- FCFS-backfilling with $D = 1$ (backfill-1, Algorithm 4)

- FCFS-backfilling with $D = 2$ (backfill-2, Algorithm 4)

- FCFS-backfilling with $D = 3$ (backfill-3, Algorithm 4)

- FCFS-backfilling with $D = 4$ (backfill-4, Algorithm 4)

- Greedy SJF filling without any reservations (filler-sjf)

- SJF-backfilling with $D = 1$ (backfill-sjf-1, Algorithm 5)

- SJF-backfilling with $D = 2$ (backfill-sjf-2, Algorithm 5)

- SJF-backfilling with $D = 3$ (backfill-sjf-3, Algorithm 5)

- SJF-backfilling with $D = 4$ (backfill-sjf-4, Algorithm 5)

- Plan-based minimising the sum of squared waiting time with $D = 0$ (plan-square-0, $\alpha = 2$, Algorithm 8)

- Plan-based minimising the sum of squared waiting time with $D = 1$ (plan-square-1, $\alpha = 2$, Algorithm 8)

- Plan-based minimising the sum of squared waiting time with $D = 2$ (plan-square-2, $\alpha = 2$, Algorithm 8)

- Plan-based minimising the sum of squared waiting time with $D = 3$ (plan-square-3, $\alpha = 2$, Algorithm 8)

Filler may be perceived as the FCFS-backfilling with $D = 0$. Analogously, filler-sjf can be viewed as SJF-backfilling with $D = 0$. We do not present plan-based scheduling with $D = 4$ due to its extraordinary extensive computation time. We evaluate the algorithms only the IO-Aware model.

For all presented metrics, we see that increasing the *reservation depth* above 1 deteriorate the scheduling performance for all algorithms. For FCFS and SJF backfilling, the results of the mean and all ordered statistics are monotonically increasing. The variants of those scheduling algorithms without reservations, filler and filler-sjf, also shows worse results than backfilling with $D = 1$.

For the plan-based policy, all results are monotonically increasing in terms of the mean values and quantiles.

The tail distributions plots indicate interesting results as well. At first, we want to point out that the tail distribution plots for waiting time, turnaround time and bounded slowdown were truncated. Specifically, for some jobs, filler-sjf resulted in values even an order of magnitude higher than the worst scores for scheduling policies with $D = 1$. Presenting all values for filler-sjf would dominate other algorithms and make them indistinctive. In general filler and filler-sjf have significantly worse tail distributions that all other policies. Excluding greedy filling algorithms, the dispersion of tail distributions is increasing with the $D$ parameter.

Based on our experiments for the full workload in the IO-Aware model, we conclude that the optimal *reservation depth* for canonical FCFS and SJF backfilling is 1. Whereas, for plan-based scheduling, which is minimising the sum of squared waiting time, no reservations results with the best outcome.

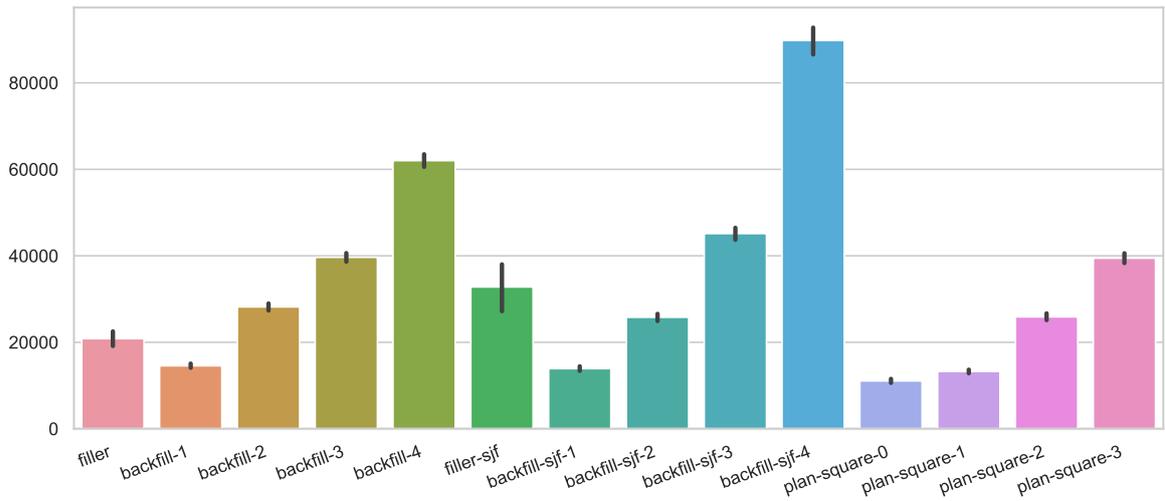

(a) Mean

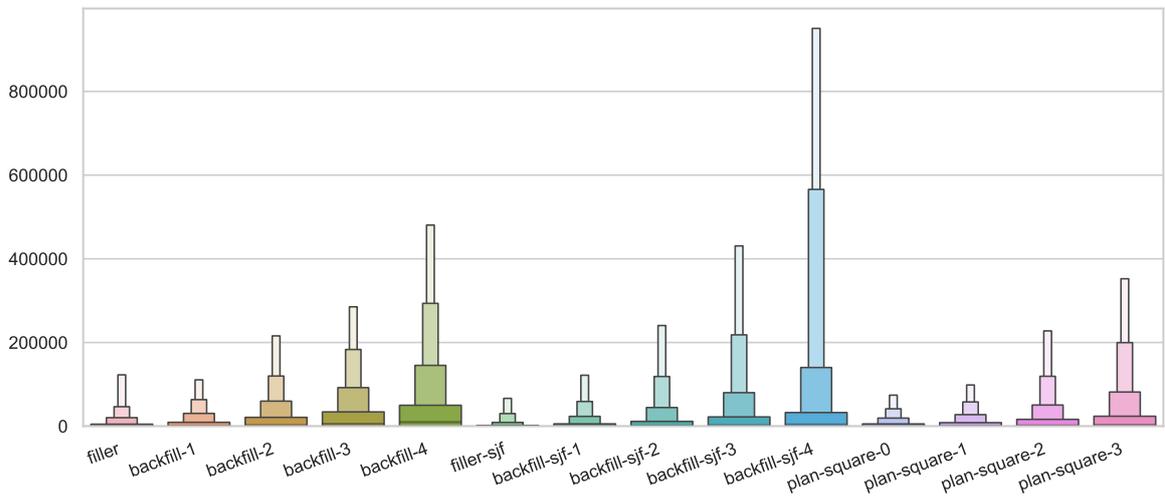

(b) Quantiles

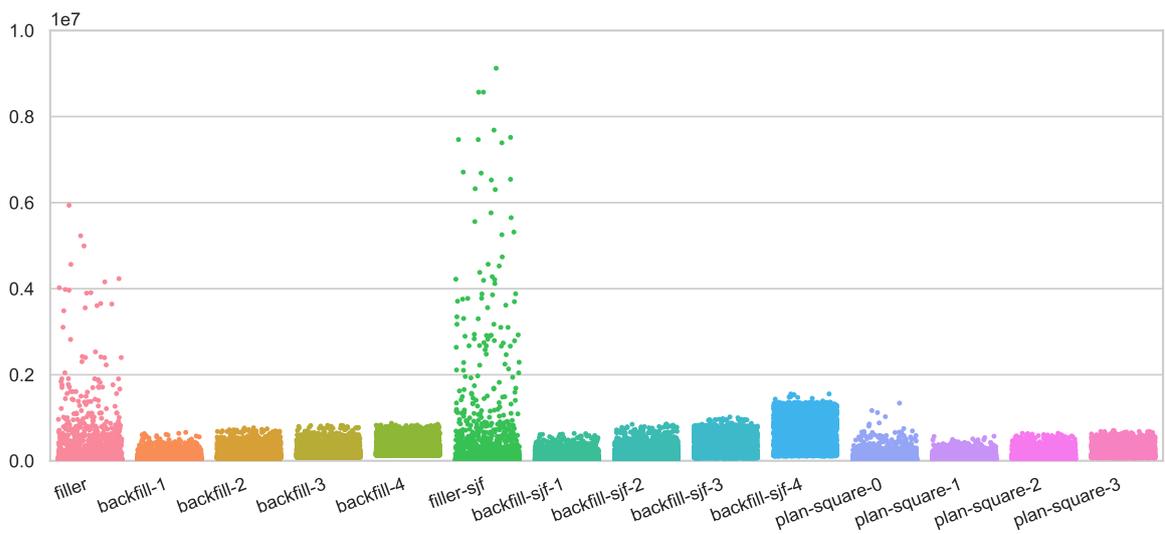

(c) Tail distribution

Figure 4.62: Waiting time

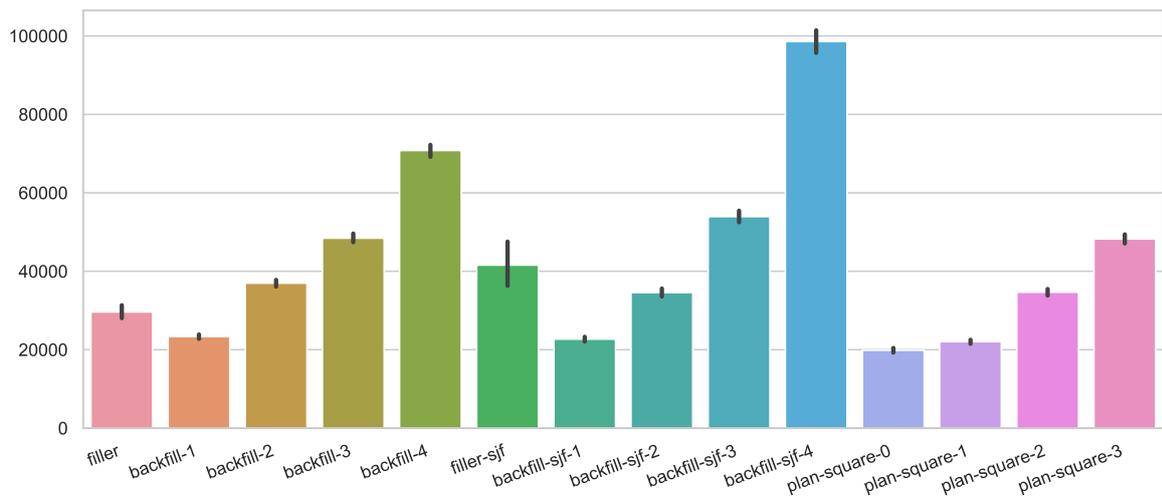

(a) Mean

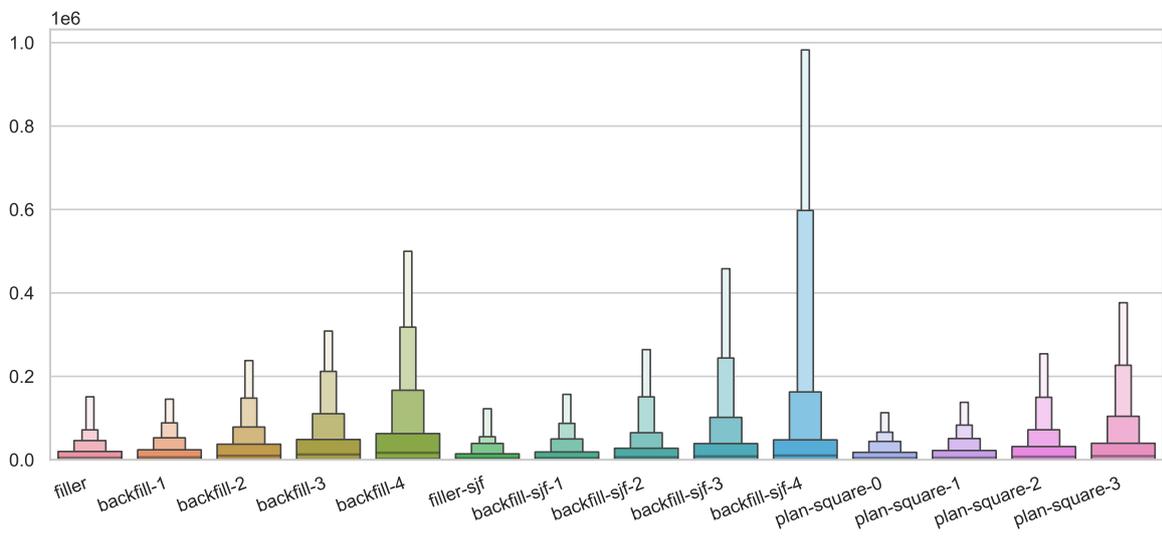

(b) Quantiles

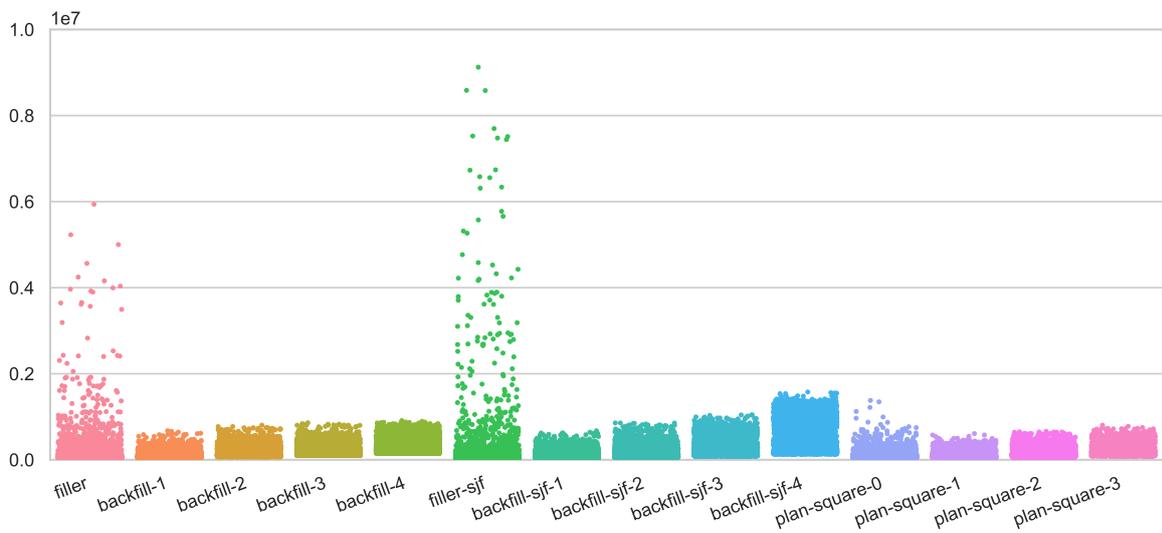

(c) Tail distribution

Figure 4.63: Turnaround time

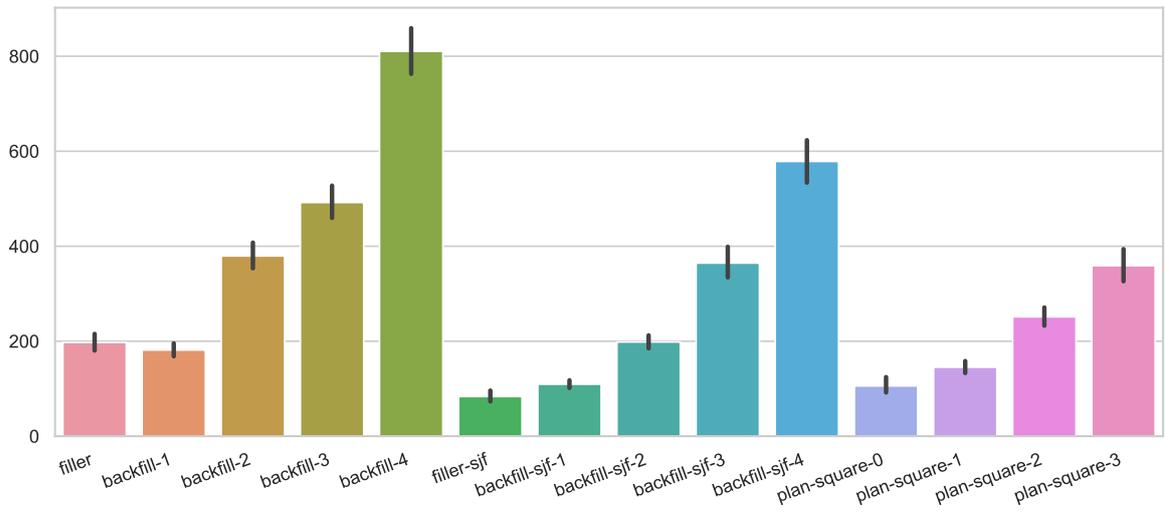

(a) Mean

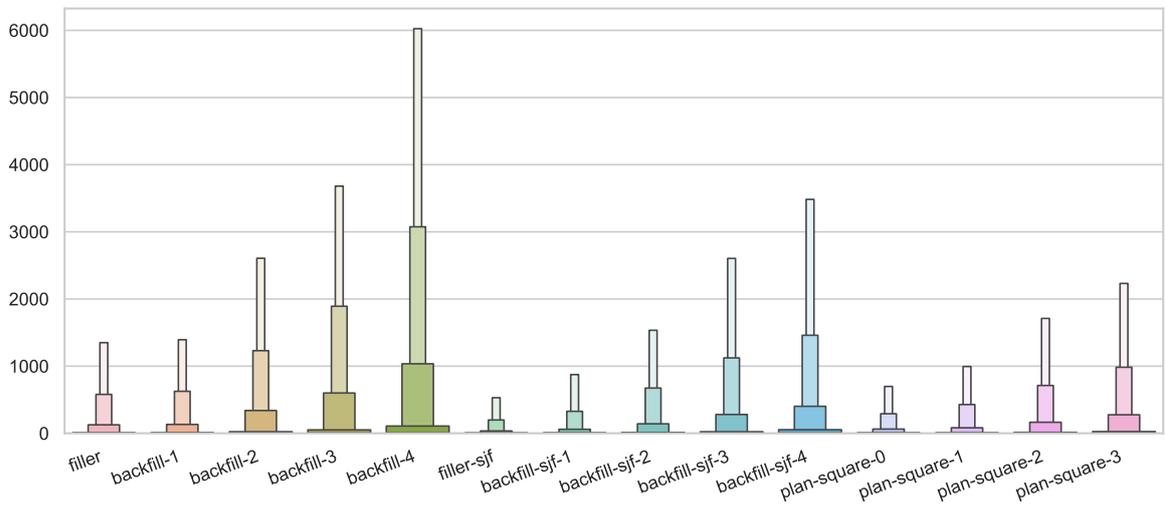

(b) Quantiles

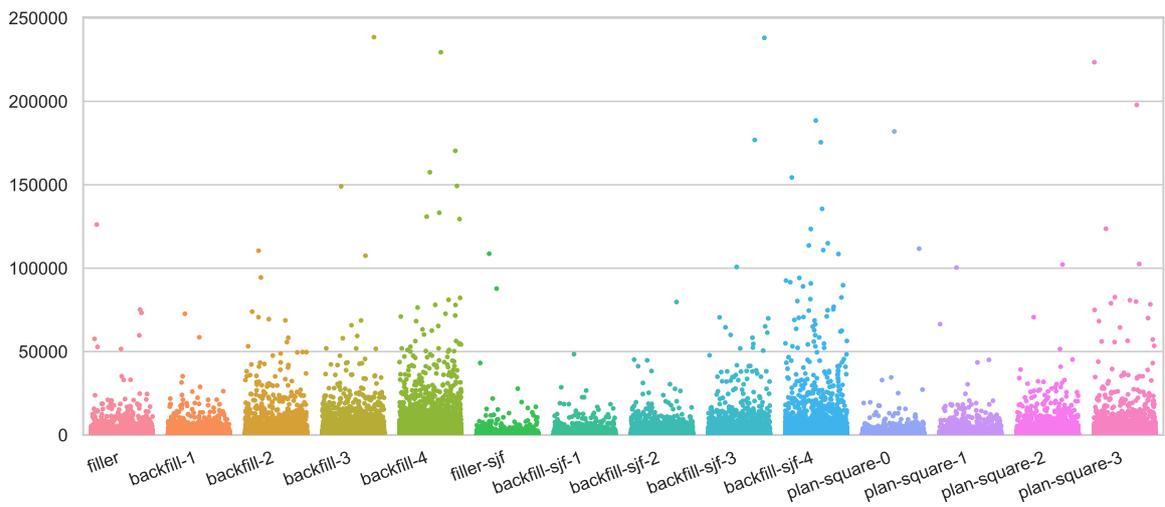

(c) Tail distribution

Figure 4.64: Slowdown

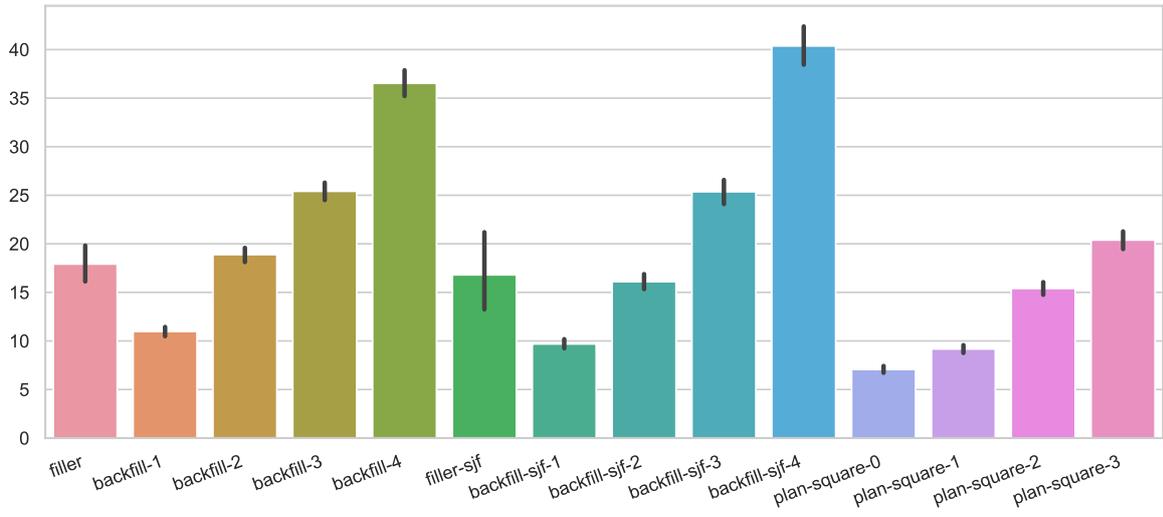

(a) Mean

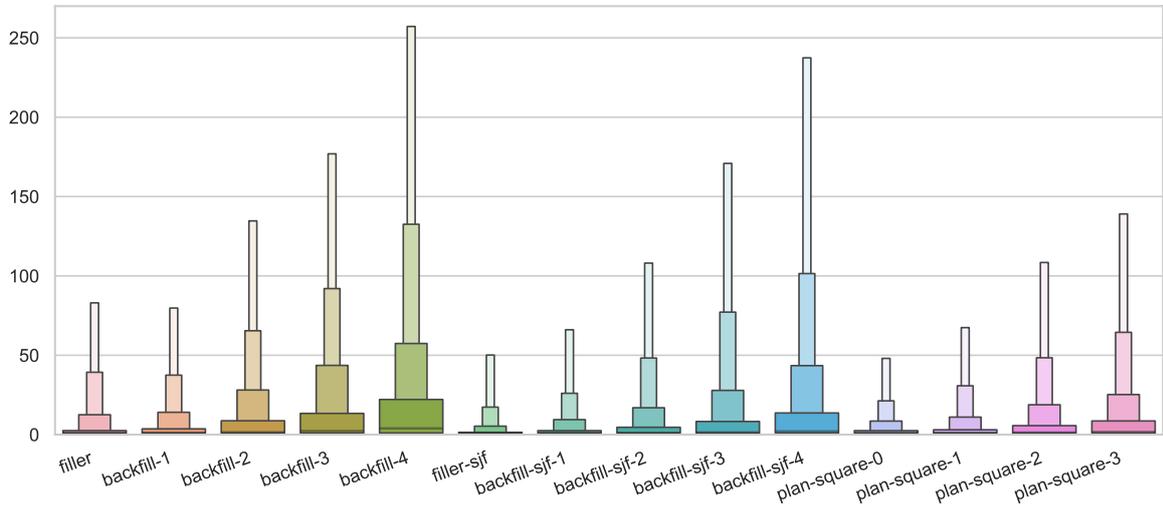

(b) Quantiles

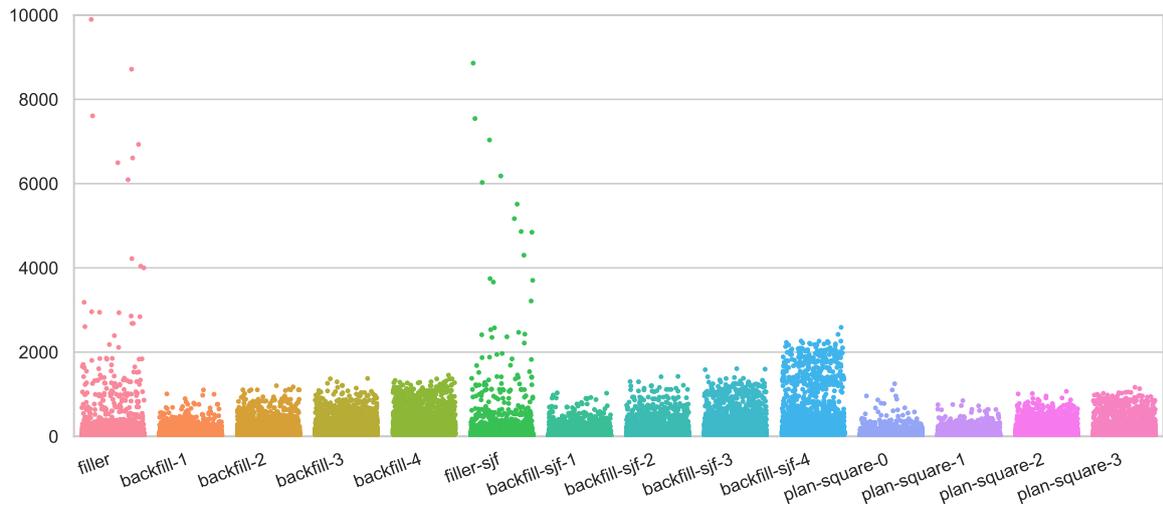

(c) Tail distribution

Figure 4.65: Bounded slowdown

# Chapter 5

# Conclusions

The ever increasing gap between compute and I/O performance in supercomputing platforms together with the development in persistence storage led to the idea of burst buffers—a fast persistence storage layer logically positioned between the random access main memory and parallel file system. Since the appearance of this concept, numerous supercomputers have been equipped with burst buffers exploiting various architectures such as compute-node–local burst buffers or remote shared burst buffers co-located with I/O nodes. In majority of cases, the novel NVMe Solid State Drives were selected as hardware implementation of the persistent storage. This new kind of technology sparked a novel branch in HPC research, which already brought several prominent publications on improving the software side of burst buffer capabilities.

Despite the development of real-world architectures as well as research concepts, Resource and Job Management Systems provide only marginal support for burst buffer utilisation. They usually allow to specify burst buffer allocation by users, but do not include storage reservations in backfilling during job scheduling. We presumed that this scheduling behaviour could potentially cause significant efficiency deterioration. That is the primary observation that motivated our research.

In this dissertation, we presented a detailed overview of burst buffer technology and online job scheduling concepts. Due to a lack of publicly available trace logs, we forged our workload with burst buffer requests based on analysis of main memory requests in logs from the Parallel Workloads Archive. We created two simulation models: one which performs job scheduling and simulates resource allocations and the other that extends it with I/O contention and I/O congestion effects. We proposed three new algorithms focused on improving scheduling performance by burst buffer awareness. Finally, we performed numerous experiments which led to several conclusions.

1. The lack of future burst buffer reservations in EASY-backfilling may drastically deteriorate the efficiency of job scheduling. In our simulations, we observed that backfilling without future burst buffer reservations resulted in mean waiting time comparable or even worse than First-Come-First-Served without any backfilling. Furthermore, backfilling future burst buffer reservations presented significantly more extensive distribution of job waiting times with a large number of outliers.

2. We also observed that the lack of future burst buffer reservations in EASY-backfilling could lead to starvation of medium-size and wide jobs. This situation happens when a wide job at the front of the waiting queue is selected for a future reservation in backfilling. This job needs to request not only a large number of processors but almost all available burst buffer capacity in a platform. Then, narrow-short jobs behind it could be backfilled at the front if only there are enough compute resource. Moreover, in online scheduling with a constant stream of submitted narrow-short jobs, the wide jobs can be almost arbitrary delayed.

3. Plan-based scheduling algorithms showed superior performance in burst-buffer–aware online job scheduling. The best plan-based algorithms improved the mean waiting time by 20% and mean bounded slowdown by 27% compared to Shortest-Jobs-First EASY-backfilling. Plan-based scheduling algorithms proved to be universally better in terms of summary statistics for all studied metrics. Their only downside is a slightly larger number of outliers compared to canonical backfilling algorithms. Furthermore, they are resource agnostic, that is they could be easily generalised to multi-resource scheduling for any kinds of resources. Plan-based scheduling algorithms are also agnostic from resource allocation techniques as they are based on optimising the order of jobs taken to scheduling.

4. The optimal value of the backfilling reservation depth is 1 for the canonical First-Come-First-Served and Shortest-Jobs-First backfilling. For all values above 1, we observed a monotonic decrease of scheduling efficiency in all investigated metrics. Removing resource reservations from backfilling deteriorates the performance as well by arbitrary delaying some jobs.

5. Mean is not a representative statistic for online job scheduling. We presented our results based on four metrics: waiting time, turnaround time, slowdown and bounded slowdown. For those metrics, we used different visualisation techniques such as mean, unaggregated values for a tail of distribution and boxenplots, which shows several ordered statistics. We observed that two schedules with almost identical mean may have completely different distributions.

6. The addition of I/O congestion and I/O contention effects did not significantly change the results of job scheduling compared to the simulation with resource allocations only. The only exception from this observation applied to Shortest-Jobs-First EASY-backfilling.

# Chapter 6

# Future work

Several research directions are arising from our work. We dedicate this chapter to discuss a few follow-up topics which in our opinion are especially worth to investigate. Additionally, we gathered various online resources in Appendix A that we found particularly useful in our research. We think they might occur to be helpful in derived works as well.

**Extension of the Standard Workload Format**   As we discussed in Section 2.5, there is no publicly available dataset containing a parallel job workload with real burst buffer requests. Existence of such a workload trace would be beneficial for the research community working on burst-buffer–aware and multi-resource job scheduling. A set of universal workloads will simplify new research initiative and make results reported in publications more comparable with each other. That was the case for the Parallel Workloads Archive maintained by D. Feitelson which we described in Section 1.11.

The Parallel Workloads Archive contains multiple workload logs from 1993 to 2018. However, the Standard Workload Format, to which all traces are converted, has not been updated since version 2.2 in 2006. Since then, supercomputers have evolved, and new concepts, such as burst buffers, were introduced.

Therefore, we propose to investigate which information are currently missing in the Standard Workload Format and update it with new fields. The extended format in our perception should contain information about new kinds of resources. Since 2006, supercomputers have been equipped with various accelerators such as GPUs or FPGAs, and new types of memory: burst buffers, High Bandwidth Memory and accelerator onboard memory.

The next step would be to collect execution traces from supercomputers. For this purpose, particularly useful might occur tools such as Darshan I/O (Section 2.5), Slurm logging (Section 1.6) and NVIDIA System Management Interface. The collected traced should be then combined and converted to the Standard Workload Format. The converted workloads may be analysed to discover various relations such as temporal locality, spatial locality, cross-correlation, self-similarity or long-range dependence.

**Burst-buffer–aware scheduling with reinforcement learning**   Novel publications in job scheduling usually exploit some sort of optimisation methods to improve the overall performance of scheduling. In this dissertation, we presented three such methods: simulated annealing, hill climbing and combinatorial optimisation. However, even more complex machine-learning–based approaches are possible.

Reinforcement learning approach for job scheduling has been already explored in [Mao+19], [Zha+20] and [BG20]. The first one focuses on scheduling for Big Data applications, while the other two are dedicated primarily to the HPC domain. Nonetheless, neither of them explores multi-layer storage systems and burst-buffer–aware scheduling.

We propose to merge research methodologies presented in this dissertation with reinforcement learning approaches explored in the above publications. Specifically, the scheduler might serve the role of an agent, and the simulator would be the reinforcement learning environment. We expect that a straightforward application of Q-learning should raise remarkable results.

**Implementation of plan-based scheduling in Slurm**  Plan-based scheduling algorithms was introduced in [Zhe+16], where they showed up to 40% improvement in mean waiting time and up to 30% in mean turnaround time. Our results indicated the superior performance of plan-based scheduling policies over all other algorithms for all studied metrics. As described in Chapter 5, they achieved over 20% improvement in mean waiting time and 27% in mean bounded slowdown. They are also not dependent on a concrete resource allocation scheme and can easily generalise to all kinds of shared resources.

Taking all advantages into consideration, we propose to implement the plan-based scheduling algorithm with the sum of squared waiting time minimisation objective in a real Resource and Job Management Systems, for instance, Slurm. To further improve the efficiency of scheduling, parallel simulated annealing may be applied as an optimisation method.

In order to test the implementation, a custom workload obtained by mixing existing HPC benchmarks can be created. Prominent candidates would be the LINPACK Benchmark (HPL) and HPCG Benchmark for compute-intensive job profiles, and IO500 Benchmark for I/O-intensive job profiles.

**Window-based scheduling**  A direct continuation of this research could be further development and testing of window-based combinatorial scheduling described in Section 3.3. As we indicated in Section 4.4, this scheduling algorithm showed promising results despite the limitation of the *window size* imposed by an issue in our programming dependencies. Rewiring our integer-linear program in a different code library should resolve the issue.

Another field of improvement for our methodologies would be creating a realistic networking model for MPI-like communication, as mentioned in Section 2.4.

Lastly, burst buffer requests in real RJMSs could be specified in two ways: either as an ephemeral which is existing for a lifetime of a given job or a persistent burst buffer reservation. In this dissertation, we explored only the first kind while we ignored the existence of persistent burst buffer reservations. However, it should be fairly easy to extend our research by treating persistent burst buffer reservations as jobs without requested compute resources.

# Appendix A

# Useful online resources

This is a collection of online resources that we found particularly useful during the research. We would like to share them as they might be very helpful in work on derived and related topics.

## A.1. This dissertation

☐ Code repository
`https://github.com/jankopanski/Burst-Buffer-Scheduling`

☐ Modified Pybatsim fork
`https://github.com/jankopanski/pybatsim`

## A.2. Data sources

☐ Parallel Workload Archive
Source of KTH-SP2-1996-2.1-cln and METACENTRUM-2013-3 workload logs.
`https://www.cs.huji.ac.il/labs/parallel/workload/`

☐ IO500 list
A ranking of top Parallel File Systems based on benchmarks.
`https://www.vi4io.org/io500/start`

☐ MetaCentrum hardware description
`https://metavo.metacentrum.cz/pbsmon2/hardware`

☐ Subset of MetaCentrum hardware description
`https://www.cerit-sc.cz/infrastructure/computing-capacity`

## A.3. Batsim

☐ Documentation
`https://batsim.readthedocs.io/en/latest/`

☐ Code repository
`https://gitlab.inria.fr/batsim/batsim`
`https://github.com/oar-team/batsim`

- ☐ Pybatsim
  Scheduler implementation in Python.
  `https://gitlab.inria.fr/batsim/pybatsim`

- ☐ Batsched
  Scheduler implementation in C++.
  `https://gitlab.inria.fr/batsim/batsched`

- ☐ Evalys
  Data analytics and visualisation library.
  `https://evalys.readthedocs.io/`
  `https://gitlab.inria.fr/batsim/evalys`

## A.4. SimGrid

- ☐ Website
  `https://simgrid.frama.io/`

- ☐ New documentation
  `https://simgrid.org/doc/latest/`

- ☐ Old documentation
  New documentation does not contain all information from the old documentation.
  `http://simgrid.gforge.inria.fr/simgrid/3.20/doc/`

- ☐ Platform tags description
  `http://simgrid.gforge.inria.fr/simgrid/3.20/doc/platform.html`

## A.5. Alternative simulators

- ☐ Alea
  Job scheduling simulator.
  `https://www.metacentrum.cz/cs/devel/simulator/`
  `https://github.com/aleasimulator/alea`

- ☐ CQSim
  Job scheduling simulator.
  `http://bluesky.cs.iit.edu/cqsim/`
  `https://github.com/SPEAR-IIT/CQSim`

- ☐ CODES
  Simulator of exascale storage architectures.
  `https://press3.mcs.anl.gov/codes/`
  `https://github.com/codes-org/codes`
  `https://xgitlab.cels.anl.gov/codes/codes`

## A.6. Modeling

- ☐ Dror G. Feitelson's Book *Workload Modeling for Computer Systems Performance Evaluation*
  `https://www.cs.huji.ac.il/~feit/wlmod/`

## A.7. Z3 solver

- ☐ Code repository
  `https://github.com/Z3Prover/z3`

- ☐ Programming Z3
  `https://theory.stanford.edu/~nikolaj/programmingz3.html`

- ☐ Dennis Yurichev's Book *SAT/SMT by Example*
  `https://sat-smt.codes/SAT_SMT_by_example.pdf`

- ☐ Python API documentation
  `https://z3prover.github.io/api/html/namespacez3py.html`

- ☐ SMT logics (tactics)
  `http://smtlib.cs.uiowa.edu/logics.shtml`